\pdfoutput=1
\documentclass{llncs}

\usepackage{amsmath}
\usepackage{amssymb}
\usepackage{booktabs}
\usepackage{etoolbox}
\usepackage{textcomp}
\usepackage{tikz}
\usepackage[normalem]{ulem}
\usepackage{xcolor}
\usepackage{xparse}
\usepackage{xspace}
\usepackage{paralist}
\usepackage{url}
\usepackage{array}
\usepackage{wrapfig}

\usetikzlibrary{calc,backgrounds,shapes,decorations.pathmorphing}

\begin{document}

\newcommand{\anonymoustool}{RSMC} %% Anonymized

\newcommand{\ifnotextended}[1]{}
\newcommand{\ifextended}[1]{#1}
\newcommand{\hide}[1]{}

\newcommand{\bjparagraph}[1]{\vspace*{4pt}\noindent{\bf #1 }}
\newcommand{\altparagraph}[1]{\vspace*{4pt}\noindent{\it #1 }}
%% \newcommand{\bjparagraph}[1]{\paragraph{#1}}

%% \titleformat{\subsection}[runin]
%%   {}
%%   {\thesubsection}
%%   {}
%%   {}

\def\by#1{\mathop{{\hbox{\setbox0=\hbox{$\scriptstyle{#1\quad}$}{$%
\mathrel{\mathop{\setbox1=\hbox to \wd0{\rightarrowfill}\ht1=3pt\dp1=-2pt\box1}\limits^{#1}}%
$}}}}}

%% Redefine Theorem, Lemma, Corollary, Definition without italics
\let\oldtheorem\theorem
\RenewDocumentCommand{\theorem}{o}{%
  \IfNoValueTF{#1}
  {\oldtheorem}
  {\oldtheorem[#1]}%
  \normalfont
}
\let\oldlemma\lemma
\RenewDocumentCommand{\lemma}{o}{%
  \IfNoValueTF{#1}
  {\oldlemma}
  {\oldlemma[#1]}%
  \normalfont
}
\let\oldcorollary\corollary
\RenewDocumentCommand{\corollary}{o}{%
  \IfNoValueTF{#1}
  {\oldcorollary}
  {\oldcorollary[#1]}%
  \normalfontÄ
}
\let\olddefinition\definition
\RenewDocumentCommand{\definition}{o}{%
  \IfNoValueTF{#1}
  {\olddefinition}
  {\olddefinition[#1]}%
  \normalfont
}
\newtheorem{assumption}{\textbf{\color{red}Assumption}}
\let\oldassumption\assumption
\RenewDocumentCommand{\assumption}{o}{%
  \IfNoValueTF{#1}
  {\oldassumption}
  {\oldassumption[#1]}%
  \normalfont
}
\newcommand{\Assumption}[1]{{\color{red}Assumption~\ref{#1}}}

\newcommand{\todo}[1]{\noindent\textbf{{\color{red}TODO:} #1}}
\newcommand{\mtodo}[1]{\marginpar{\fontsize{7pt}{8pt}\selectfont\todo{#1}}}
\newcommand{\carl}[1]{\noindent\textbf{{\color{orange}Carl:} #1}}
\newcommand{\td}[1]{\textcolor{blue}{[TODO: #1]}}

\newcommand{\rcommitbefore}{commit-before order}

\newcommand{\commitbefore}{commit-before order}
\newcommand{\fcommitbefore}{commit-before function}

\newcommand{\cbminuspo}[1]{\textrm{$\textsf{cb}^{-\textsf{po}}_{#1}$}\xspace}
\newcommand{\cb}[1]{\textrm{$\textsf{cb}_{#1}$}\xspace}
\newcommand{\fcb}{\textrm{$\textsf{cb}$}\xspace}
\newcommand{\cbzero}[1]{\textrm{$\textsf{cb}^0_{#1}$}\xspace}
\newcommand{\fcbzero}{\textrm{$\textsf{cb}^0$}\xspace}
\newcommand{\fcbpower}{\textrm{$\textsf{cb}^{\textrm{\scriptsize{\textsf{power}}}}$}\xspace}
\newcommand{\cbpower}[1]{\textrm{$\textsf{cb}^{\textrm{\scriptsize{\textsf{power}}}}_{#1}$}\xspace}
\newcommand{\fcbarm}{\textrm{$\textsf{cb}^{\textrm{\scriptsize{\textsf{arm}}}}$\textbf{\color{red}(REMOVE!)}}\xspace}
\newcommand{\cbarm}[1]{\textrm{$\textsf{cb}^{\textrm{\scriptsize{\textsf{arm}}}}_{#1}$\textbf{\color{red}(REMOVE!)}}\xspace}

\newcommand{\partition}{\mbox{\sl partition}}
\newcommand{\position}{\mbox{\sl position}}

\newcommand{\inltransition}[1]{\xrightarrow{#1}}

\newcommand{\cpath}{\rho}

\newcommand{\extend}{\mbox{\sl extend}}
\newcommand{\iinsert}{\mbox{\sl insert}}

\newcommand{\reach}{\mbox{\sl reach}}
\newcommand{\store}{\mbox{\sl W}}

\newcommand{\areg}{\textsf{r}\xspace}
\newcommand{\atid}{\textsf{t}\xspace}
\newcommand{\xvar}{\textsf{x}\xspace}
\newcommand{\yvar}{\textsf{y}\xspace}
\newcommand{\zvar}{\textsf{z}\xspace}
\newcommand{\tid}{\textsf{tid}\xspace}
\newcommand{\fetch}{\textsf{fetch}\xspace}

\newcommand{\aninstr}{\textsf{i}\xspace}
\newcommand{\instrof}[1]{\textsf{instr}(#1)}
\newcommand{\labelof}[1]{\textsf{label}(#1)}
\newcommand{\init}[1]{\textrm{$\textsf{init}_{#1}$}\xspace}
\newcommand{\anexec}{\textrm{$\pi$}\xspace} % An execution
\newcommand{\eventtype}{\textrm{$\mathbb{E}$}\xspace} % The set of all possible events
\newcommand{\parameventtype}{\textrm{$\mathbb{P}$}\xspace} % The set of all possible parameterized events
\newcommand{\events}{\textrm{$E$}\xspace} % The set of events in an execution
\newcommand{\eventtuple}{(\atid,n,l)}
\newcommand{\cbextends}[1]{\leq_{#1}}

\newcommand{\FLB}{\textit{FLB}}
\newcommand{\FLBmax}{\FLB(\textit{max})}

\newcommand{\ppo}{{\normalfont\textsf{ppo}}\xspace}

\newcommand{\fences}{{\normalfont\textsf{fences}}\xspace}
\newcommand{\fence}{{\normalfont\textsf{fence}}\xspace}

\newcommand{\prop}{{\normalfont\textsf{prop}}}

\newcommand{\axmodel}[2]{[\![ #1 ]\!]_{#2}^{\textsf{Ax}}}
\newcommand{\exmodel}[3]{[\![ #1 ]\!]_{#2,#3}^{\textsf{Ex}}}
\newcommand{\ltsof}[3]{TS_{#2,#3}^{#1}}
\newcommand{\extransitions}{\longrightarrow}

\newcommand{\anevent}{\textrm{$e$}\xspace}
\newcommand{\anexpr}{\textrm{$a$}\xspace}
\newcommand{\po}{{\normalfont\textsf{po}}\xspace}
\newcommand{\co}{{\normalfont\textsf{co}}\xspace}
\newcommand{\rf}{{\normalfont\textsf{rf}}\xspace}
\newcommand{\rfee}{{\normalfont\textsf{rfe}}\xspace}
\newcommand{\rfi}{{\normalfont\textsf{rfi}}\xspace}

\newcommand{\dpp}{{\normalfont\textsf{dp}}\xspace}

\newcommand{\hbb}{{\normalfont\textsf{hb}}\xspace}

\newcommand{\com}{{\normalfont\textsf{com}}\xspace}

\newcommand{\cc}{{\normalfont\textsf{cc}}\xspace}
\newcommand{\ic}{{\normalfont\textsf{ic}}\xspace}
\newcommand{\ii}{{\normalfont\textsf{ii}}\xspace}
\newcommand{\ci}{{\normalfont\textsf{ci}}\xspace}

\newcommand{\fr}{{\normalfont\textsf{fr}}\xspace}
\newcommand{\fre}{{\normalfont\textsf{fre}}\xspace}
\newcommand{\coe}{{\normalfont\textsf{coe}}\xspace}
\newcommand{\ctrl}{{\normalfont\textsf{ctrl}}\xspace}

\newcommand{\R}{\textrm{$\textsf{R}$}}
\newcommand{\W}{\textrm{$\textsf{W}$}}
\newcommand{\M}{\textrm{$\textsf{M}$}}
\newcommand{\RR}{\textrm{$\textsf{RR}$}}
\newcommand{\RW}{\textrm{$\textsf{RW}$}}
\newcommand{\RM}{\textrm{$\textsf{RM}$}}
\newcommand{\WR}{\textrm{$\textsf{WR}$}}
\newcommand{\WW}{\textrm{$\textsf{WW}$}}
\newcommand{\WM}{\textrm{$\textsf{WM}$}}
\newcommand{\MR}{\textrm{$\textsf{MR}$}}
\newcommand{\MW}{\textrm{$\textsf{MW}$}}
\newcommand{\MM}{\textrm{$\textsf{MM}$}}
\newcommand{\poloc}[1]{\textrm{$\textsf{po-loc}_{#1}$}\xspace}
\newcommand{\rdwstate}[1]{\textrm{$\textsf{rdw}_{#1}$}\xspace}
\newcommand{\detourstate}[1]{\textrm{$\textsf{detour}_{#1}$}\xspace}

\newcommand{\hbstate}[1]{\textrm{$\textsf{hb}_{#1}$}\xspace}

\newcommand{\comm}[1]{\textrm{$\textsf{com}_{#1}$}\xspace}
\newcommand{\val}[1]{\textrm{$\textrm{val}_{#1}$}\xspace}
\newcommand{\adeps}[1]{\textrm{$\textrm{adeps}_{#1}$}\xspace}
\newcommand{\addr}[1]{\textrm{$\textrm{address}_{#1}$}\xspace}
\newcommand{\data}[1]{\textrm{$\textrm{data}_{#1}$}\xspace}
\newcommand{\addrdep}[1]{\textrm{$\textsf{addr}_{#1}$}\xspace}
\newcommand{\datadep}[1]{\textrm{$\textsf{data}_{#1}$}\xspace}
\newcommand{\ffencedep}[1]{\textrm{$\textsf{sync}_{#1}$}\xspace}
\newcommand{\lwfencedep}[1]{\textrm{$\textsf{lwsync}_{#1}$}\xspace}
\newcommand{\lwsyncdep}[1]{\textrm{$\textsf{lwsync}_{#1}$}\xspace}

\newcommand{\ctrldep}[1]{\textrm{$\textsf{ctrl}_{#1}$}\xspace}
\newcommand{\sync}{\textsf{sync}\xspace}
\newcommand{\lwsync}{\textsf{lwsync}\xspace}
\newcommand{\isync}{\textsf{isync}\xspace}
\newcommand{\dmb}{\textsf{dmb}\xspace}
\newcommand{\dsb}{\textsf{dsb}\xspace}
\newcommand{\isb}{\textsf{isb}\xspace}
\newcommand{\ffence}{\textsf{sync}\xspace}
\newcommand{\lwfence}{\textsf{lwsync}\xspace}
\newcommand{\cfence}{\textsf{isync}\xspace}

\newcommand{\chracyc}{\textsf{cb-acyclic}}
\newcommand{\chracon}{\textsf{cb-consistent}}
\newcommand{\nodeadlock}{\textsf{no-cb-deadlock}}

\newcommand{\mmodel}{{\normalfont\textsc{M}}\xspace} % A generic memory model
\newcommand{\mmodelarm}{{\normalfont\textsc{M}}^{\textit{ARM}}\textbf{\color{red}(REMOVE!)}}
\newcommand{\mmodelpower}{{\normalfont\textsc{M}}^{\textit{POWER}}}
\newcommand{\emptyseq}{\langle \rangle}
\renewcommand{\emptyset}{\varnothing}
\newcommand{\states}{\textrm{$\mathbb{S}$}\xspace}
\newcommand{\astate}{\textrm{$\sigma$}\xspace}
\newcommand{\state}{\astate}
\newcommand{\initstate}{\astate_0}
\newcommand{\lbl}{\textit{lbl}}
\newcommand{\labels}{\textrm{$\mathbb{L}$}\xspace}
\newcommand{\fetched}{\textrm{$F$}\xspace}
\newcommand{\lblcur}{\textrm{$\lambda$}\xspace}
\newcommand{\exec}[1]{\textrm{$\textsf{exec}(#1)$}}
\newcommand{\arun}{\textrm{$\tau$}\xspace}
\newcommand{\aprog}{\textrm{$\mathcal{P}$}\xspace}
\newcommand{\lblnext}{\textrm{$\lambda_{\textsf{next}}$}\xspace}
\newcommand{\tids}{\mbox{\textbf{DEPRECATED (\texttt{$\backslash$tids})}}} %% Definition has been removed from semantics section
\newcommand{\lblinit}{\mbox{\textbf{DEPRECATED (\texttt{$\backslash$lblinit})}}} %% Definition has been removed from semantics section
\newcommand{\lblinstr}{\mbox{\textbf{DEPRECATED (\texttt{$\backslash$lblinstr})}}} %% Definition has been removed from semantics section
\newcommand{\vars}{\mbox{\textbf{DEPRECATED (\texttt{$\backslash$vars})}}} %% Definition has been removed from semantics section
\newcommand{\varinit}{\mbox{\textbf{DEPRECATED (\texttt{$\backslash$varinit})}}} %% Definition has been removed from semantics section
\newcommand{\addrofvar}[1]{\textrm{$\texttt{\&}{#1}$}\xspace}

\newcommand{\pair}[2]{\langle#1,#2\rangle}
\newcommand{\tuple}[1]{\langle #1 \rangle}
\newcommand{\set}[1]{\{ #1 \}}
\newcommand{\setcomp}[2]{\{ #1 \ \mid \ #2 \}}
\newcommand{\transpair}[2]{\pair{#1}{#2}}
\newcommand{\instantiate}[2]{{#1}_{#2}}
\renewcommand{\inst}[2]{{#1}[{#2}]}
\newcommand{\indepprefix}[1]{w_{#1}}
\newcommand{\varindepprefix}[1]{w_{#1}'}
\newcommand{\varvarindepprefix}[1]{w_{#1}''}
\newcommand{\exseq}{\tau}
\newcommand{\domof}[1]{\mbox{\it dom}(#1)}
\newcommand{\domofafter}[2]{\mbox{\it dom}_{[#1]}(#2)}
\newcommand{\stateafter}[1]{#1(\astate)}
\newcommand{\stateoafter}[1]{#1(\astate_0)}
\newcommand{\reaches}{\longrightarrow^*}

\newcommand{\exequiv}{\equiv}
\newcommand{\mtequiv}{\simeq}
\newcommand{\mtclass}[1]{[#1]_{\simeq}}
\newcommand{\prefix}{\leq}
\newcommand{\totorder}[1]{<_{#1}}
\newcommand{\functof}[1]{\mbox{\sl f}_{#1}}
\newcommand{\enabledof}[1]{\mbox{\sl enabled}(#1)}
\newcommand{\committable}{\mbox{\sl enabled}}
\newcommand{\fetchable}{\mbox{\sl fetchable}}

\newcommand{\hb}[1]{\rightarrow_{#1}}
\newcommand{\hbpar}[1]{\xdashrightarrow{}_{#1}}
\newcommand{\hbinst}[1]{\rightsquigarrow{}_{#1}}
\newcommand{\indepafter}[3]{#1 \!\! \models \!\! #2 \diamondsuit #3}
\newcommand{\notindepafter}[3]{#1 \!\! \not\models \!\! #2 \diamondsuit #3}
\newcommand{\notsucc}[2]{\mbox{\it notha}(#1,#2)}
\newcommand{\truncate}[2]{#1 \!\! \upharpoonright \!\! #2}
\newcommand{\mayrace}[1]{\lessdot_{#1}}
\newcommand{\mayinstrace}[1]{\precsim_{#1}}
\newcommand{\minsuff}[3]{\mbox{\sl MinSuff}_{#2}(#1,#3)}
\newcommand{\procof}[1]{\threadof{#1}}
\newcommand{\threadof}[1]{\tid(#1)}
\newcommand{\param}{p}
\newcommand{\qaram}{q}
\newcommand{\raram}{r}

\newcommand{\TRUE}{{\it true}}
\newcommand{\FALSE}{{\it false}}

%%%%%%%%%%%%%%%%%%%%%%%%%%%%%%%%%%%%%%%%%%%%%%%%%%%%%%%%%%%
%                                                         %
%              Notation that should be removed            %
%                                                         %
%%%%%%%%%%%%%%%%%%%%%%%%%%%%%%%%%%%%%%%%%%%%%%%%%%%%%%%%%%%

\newcommand{\firsttrans}[2]{\mbox{\it I}_{[#1]}(#2)}
\newcommand{\wfirsttrans}[2]{\mbox{\it WI}_{[#1]}(#2)}
\newcommand{\keyword}[1]{\mbox{\bf #1}}
\newcommand{\sleepset}{\mbox{\it Sleep}}
\newcommand{\finalsleep}{\mbox{\it Final\_sleep}}
\newcommand{\nextof}[2]{\mbox{\it next}_{[#1]}(#2)}
\newcommand{\ievent}{\inst{\anevent}{\param}}
\newcommand{\ieventq}{\inst{\anevent}{\qaram}}
\newcommand{\ieventr}{\inst{\anevent}{\raram}}
\newcommand{\ieventp}{\inst{\anevent'}{\param'}}
\newcommand{\ieventqp}{\inst{\anevent'}{\qaram'}}
\newcommand{\ieventdp}{\inst{\anevent''}{\param''}}
\newcommand{\ieventqdp}{\inst{\anevent''}{\qaram''}}
\newcommand{\varevent}{d}
\newcommand{\varievent}{\inst{\varevent}{\qaram}}
\newcommand{\blockevent}[1]{\varevent_{#1}}
\newcommand{\anievent}{e}
\newcommand{\anieventp}{\anievent'}
\newcommand{\anieventdp}{\anievent''}

%%%%%%%%%%%%%%%%%%%%%%%%%%%%%%%%%%%%%%%%%%%%%%%%%%%%%%%%%%%
%                                                         %
%          End of notation that should be removed         %
%                                                         %
%%%%%%%%%%%%%%%%%%%%%%%%%%%%%%%%%%%%%%%%%%%%%%%%%%%%%%%%%%%

%%%%  ARROWS

\makeatletter
\newcommand*{\da@rightarrow}{\mathchar"0\hexnumber@\symAMSa 4B }
\newcommand*{\da@leftarrow}{\mathchar"0\hexnumber@\symAMSa 4C }
\newcommand*{\xdashrightarrow}[2][]{%
  \mathrel{%
    \mathpalette{\da@xarrow{#1}{#2}{}\da@rightarrow{\,}{}}{}%
  }%
}
\newcommand{\xdashleftarrow}[2][]{%
  \mathrel{%
    \mathpalette{\da@xarrow{#1}{#2}\da@leftarrow{}{}{\,}}{}%
  }%
}
\newcommand*{\da@xarrow}[7]{%
  % #1: below
  % #2: above
  % #3: arrow left
  % #4: arrow right
  % #5: space left
  % #6: space right
  % #7: math style
  \sbox0{$\ifx#7\scriptstyle\scriptscriptstyle\else\scriptstyle\fi#5#1#6\m@th$}%
  \sbox2{$\ifx#7\scriptstyle\scriptscriptstyle\else\scriptstyle\fi#5#2#6\m@th$}%
  \sbox4{$#7\dabar@\m@th$}%
  \dimen@=\wd0 %
  \ifdim\wd2 >\dimen@
    \dimen@=\wd2 %
  \fi
  \count@=2 %
  \def\da@bars{\dabar@\dabar@}%
  \@whiledim\count@\wd4<\dimen@\do{%
    \advance\count@\@ne
    \expandafter\def\expandafter\da@bars\expandafter{%
      \da@bars
      \dabar@
    }%
  }%
  \mathrel{#3}%
  \mathrel{%
    \mathop{\da@bars}\limits
    \ifx\\#1\\%
    \else
      _{\copy0}%
    \fi
    \ifx\\#2\\%
    \else
      ^{\copy2}%
    \fi
  }%
  \mathrel{#4}%
}
\makeatother

%%%% END OF ARROWS

% Algorithm environment
% ---------------------
%
% Usage:
% \begin{algo}{Name}
%   \lin{First line}
%   \lin{Next line}
%   ...
% \end{algo}
%\newcounter{algolinctr}
\newlength{\algoinddepth}
\newlength{\algostdind}
\newenvironment{algo}[2]{
    \newcommand{\thename}{#1}
    \newcommand{\linctr}{#2}
    \newcommand{\commentcolor}{gray}
    \newcounter{\linctr}
    \setcounter{\linctr}{0}
    \setlength{\algoinddepth}{0pt}
    \setlength{\algostdind}{5pt}
    \newcommand{\icmt}[1]{{\color{\commentcolor}// ##1}} %% Inline comment
    \newcommand{\comment}[1]{\hphantom{MM: }\rule{\algoinddepth}{0pt}\icmt{##1}\\}
    \newcommand{\linunterm}[1]{\refstepcounter{\linctr}\hphantom{MM}\llap{\arabic{\linctr}}: \rule{\algoinddepth}{0pt}##1}
    \newcommand{\lnlabel}[1]{\addtocounter{\linctr}{-1}\refstepcounter{\linctr}\label{##1}}
    \newcommand{\lin}[1]{\linunterm{##1}\\}
    \newcommand{\incind}{\addtolength{\algoinddepth}{\algostdind}\global\algoinddepth=\algoinddepth}
    \newcommand{\decind}{\addtolength{\algoinddepth}{-\algostdind}\global\algoinddepth=\algoinddepth}
    \newcommand{\key}[1]{\textrm{\textbf{##1}}}
%%%%%%%%%%%%%%%%%%%%%%%
% Command definitions %
%%%%%%%%%%%%%%%%%%%%%%%
\newcommand{\dowhile}[2]{%
\lin{\key{do}\{\incind}%
##1%
\decind%
\lin{\}\key{while}(##2);}%
}
\newcommand{\while}[2]{%
\lin{\key{while}(##1)\{\incind}%
##2%
\decind%
\lin{\}}%
}
\newcommand{\whileoneline}[2]{%
\lin{\key{while}(##1)\{##2\}}%
}
\newcommand{\ifunterm}[3][]{%
\lin{\key{if}(##2)\{\ifstrempty{##1}{}{ \icmt{##1}}\incind}%
##3%
\decind%
\linunterm{\}}%
}
\newcommand{\ifterm}[3][]{\ifunterm[##1]{##2}{##3}\\}
\newcommand{\elseifunterm}[3][]{%
\key{else if}(##2)\{\ifstrempty{##1}{}{ \icmt{##1}}\incind\\%
##3%
\decind%
\linunterm{\}}%
}
\newcommand{\elseifterm}[3][]{\elseifunterm[##1]{##2}{##3}\\}
\newcommand{\elseterm}[2][]{%
\key{else}\{\ifstrempty{##1}{}{ \icmt{##1}}\incind\\%
##2%
\decind%
\lin{\}}%
}
\newcommand{\for}[2]{%
\lin{\key{for}(##1)\{\incind}%
##2%
\decind%
\lin{\}}%
}
%%%%%%%%%%%%%%%%%%%%%%%%%%%%%%%%%%%%%%%%%%%%%%%%%%%%%
% Command variations for tall condition expressions %
%%%%%%%%%%%%%%%%%%%%%%%%%%%%%%%%%%%%%%%%%%%%%%%%%%%%%
\newcommand{\whiletall}[2]{%
\lin{\key{while}$\left(\textrm{\texttt{##1}}\right)$\{\incind}%
##2%
\decind%
\lin{\}}%
}
\newcommand{\ifuntermtall}[2]{%
\lin{\key{if}$\left(\textrm{\texttt{##1}}\right)$\{\incind}%
##2%
\decind%
\linunterm{\}}%
}
\newcommand{\iftermtall}[2]{\ifuntermtall{##1}{##2}\\}
\newcommand{\elseifuntermtall}[2]{%
\key{else if}$\left(\textrm{\texttt{##1}}\right)$\{\incind\\%
##2%
\decind%
\linunterm{\}}%
}
\newcommand{\elseiftermtall}[2]{\elseifuntermtall{##1}{##2}\\}
\newcommand{\dowhiletall}[2]{%
\lin{\key{do}\{\incind}%
##1%
\decind%
\lin{\}\key{while}$\left(\textrm{\texttt{##2}}\right)$;}%
}
\newcommand{\fortall}[2]{%
\lin{\key{for}$\left(\textrm{\texttt{##1}}\right)$\{\incind}%
##2%
\decind%
\lin{\}}%
}
  \tt
  \begin{tabular}{@{}l@{}}
    {\bf \thename}\\
    \hline
}{
  \end{tabular}
  \normalfont
}

\sloppy

%% \special{papersize=8.5in,11in}
%% \setlength{\pdfpageheight}{\paperheight}
%% \setlength{\pdfpagewidth}{\paperwidth}

%% \conferenceinfo{CONF 'yy}{Month d--d, 20yy, City, ST, Country}
%% \copyrightyear{20yy}
%% \copyrightdata{978-1-nnnn-nnnn-n/yy/mm}
%% \doi{nnnnnnn.nnnnnnn}

% Uncomment one of the following two, if you are not going for the
% traditional copyright transfer agreement.

%\exclusivelicense                % ACM gets exclusive license to publish,
                                  % you retain copyright

%\permissiontopublish             % ACM gets nonexclusive license to publish
                                  % (paid open-access papers,
                                  % short abstracts)

%% \titlebanner{banner above paper title}        % These are ignored unless
%% \preprintfooter{Stateless Model Checking for POWER}   % 'preprint' option specified.

\title{Stateless Model Checking for POWER}

\author{}
\institute{}

\author{Parosh Aziz Abdulla \and Mohamed Faouzi Atig \and Bengt Jonsson \and Carl Leonardsson}

\institute{Dept.\ of Information Technology, Uppsala University, Sweden}

\date{}

\maketitle

%% \authorinfo{}{}{}
%% \authorinfo{Name1}
%%            {Affiliation1}
%%            {Email1}
%% \authorinfo{Name2\and Name3}
%%            {Affiliation2/3}
%%            {Email2/3}

\begin{abstract}
  We present the first framework for efficient application of
  stateless model checking (SMC) to programs running under the relaxed
  memory model of POWER.  The framework combines several
  contributions.
  The first contribution is that we develop a scheme for
  systematically deriving operational execution models from existing
  axiomatic ones. The scheme is such that the derived execution models
  are well suited for efficient SMC. We apply our scheme to the
  axiomatic model of POWER from~\cite{alglave2014herding}.
  Our main contribution is a technique for efficient SMC, called
  \emph{Relaxed Stateless Model Checking} (RSMC), which systematically
  explores the possible inequivalent executions of a program.  RSMC is
  suitable for execution models obtained using our scheme.  We prove
  that RSMC is sound and optimal for the POWER memory model, in the
  sense that each complete program behavior is explored exactly once.
  We show the feasibility of our technique by providing an
  implementation for programs written in C/pthreads.
\end{abstract}

%% \category{CR-number}{subcategory}{third-level}

%% % general terms are not compulsory anymore,
%% % you may leave them out
%% \terms
%% term1, term2

%% \keywords
%% keyword1, keyword2

\section{Introduction}

Verification and testing of concurrent programs is difficult,
since one must consider all the different ways in which parallel
threads can interact.
To make matters worse, current shared-memory multicore processors,
such as Intel's x86, IBM's POWER, and ARM,~\cite{x86-swdmanual-1-3,sparc-v9-manual,power-isa-v207,arm-v7ar-refman},
achieve higher performance by implementing \emph{relaxed memory models}
that allow threads to interact in even subtler ways than by
interleaving of their instructions, as would be the case in the model of
{\em sequential consistency} (SC)~\cite{Lamport:multiprocess:executes}.
Under the relaxed memory model of POWER, loads and stores to different
memory locations may be reordered by the hardware, and the accesses
may even be observed in different orders on different processor cores.

Stateless model checking (SMC)~\cite{Godefroid:popl97} is one
successful technique for verifying concurrent programs. It detects
violations of correctness by systematically exploring the set of
possible program executions.
Given a concurrent program which is terminating and threadwisely deterministic
(e.g., by fixing any input data to avoid data-nondeterminism),
a special runtime scheduler drives the SMC exploration by controlling
decisions that may affect subsequent computations,
so that the exploration covers all possible executions.
The technique is automatic, has no false positives,
can be applied directly to the program source code, and can easily
reproduce detected bugs. SMC has been successfully
implemented in tools, such as VeriSoft~\cite{Godefroid:verisoft-journal},
\textsc{Chess}~\cite{MQBBNN:chess}, Concuerror~\cite{Concuerror:ICST13},
rInspect~\cite{Zhang:pldi15}, and Nidhugg~\cite{tacas15:tso}.

However, SMC suffers from the state-space explosion problem, and
must therefore be equipped with techniques to reduce the number of
explored executions.
The most prominent one is
\emph{partial order reduction}~\cite{Valmari:reduced:state-space,Peled:representatives,Godefroid:thesis,CGMP:partialorder},
adapted to SMC as \emph{dynamic partial order reduction}
(DPOR)~\cite{abdulla2014optimal,FG:dpor,SeAg:haifa06,SKH:acsd12}.
DPOR addresses state-space explosion caused by the many possible ways to
schedule concurrent threads.
DPOR retains full behavior coverage, while reducing the number of
explored executions by exploiting that two schedules which induce the same
order between conflicting instructions will induce equivalent
executions.
DPOR has been adapted to the memory models TSO and
PSO~\cite{tacas15:tso,Zhang:pldi15}, by
introducing auxiliary threads that induce the reorderings allowed by
TSO and PSO, and using DPOR to counteract the
resulting increase in thread schedulings.

In spite of impressive progress in SMC techniques for SC, TSO, and PSO,
there is so far no effective technique for SMC
under more relaxed models, such as POWER.
A major reason is that POWER allows
more aggressive reorderings of instructions within each thread, as well as
looser synchronization between threads, making it significantly more complex
than SC, TSO, and PSO. Therefore, existing SMC techniques
for SC, TSO, and PSO can not be easily extended to POWER.

In this paper, we present the first SMC algorithm for programs running
under the POWER relaxed memory model. The technique is both
sound, in the sense that it guarantees to explore each
programmer-observable behavior at least once, and optimal, in the
sense that it does not explore the same complete behavior twice.
Our technique combines solutions to several major challenges.

The first challenge is to design an execution model
for POWER that is suitable for SMC. Existing execution models fall
into two categories.
Operational models,
such as~\cite{DM14,sarkar2011understanding,DBLP:conf/pldi/SarkarMOBSMAW12,DBLP:journals/corr/abs-1208-5915},
define behaviors as resulting from
sequences of small steps of an abstract processor.
Basing SMC on such a model would induce large numbers of executions
with equivalent programmer-observable
behavior, and it would be difficult to
prevent redundant exploration, even if DPOR techniques are employed.
Axiomatic models, such as~\cite{alglave2014herding,Mador-Haim:2012:AMM:2362216.2362263,AlglaveM11}, avoid such redundancy by
  being defined in terms of an
  abstract representation of programmer-observable behavior, due to
  Shasha and Snir~\cite{shasha1988efficient}, here called
  {\em Shasha-Snir traces}.
However, being axiomatic, they
judge whether an execution is allowed only after it has been completed.
Directly basing SMC on such a model would lead to much wasted exploration of
unallowed executions.
To address this challenge,
we have therefore developed a scheme for systematically deriving
execution models that are suitable for SMC. Our scheme
derives an execution model, in the form of a labeled transition system,
from an existing axiomatic model, defined in terms of Shasha-Snir traces.
Its states are partially constructed Shasha-Snir traces.
Each transition adds (``commits'') an instruction to the state,
and also equips the instruction with a parameter that determines how it is
inserted into the Shasha-Snir trace.
The parameter of a load is the store from which it reads its value.
The parameter of a store is its position in the coherence order
of stores to the same memory location.
The order in which instructions are added must
respect various dependencies between instructions, such that each
instruction makes sense at the time when it is added. For example,
when adding a store or a load instruction, earlier instructions that are needed to
compute which memory address it accesses must already have been added.
Our execution model
therefore takes as input a partial order, called {\em commit-before},
which constrains the order in which instructions can be added.
The commit-before order should be tuned to suit the given
axiomatic memory model.
We define a condition of {\em validity} for commit-before orders, under
which our derived execution model is equivalent to the
original axiomatic one, in that they generate the same sets of
Shasha-Snir traces.
We use our scheme to derive an execution model for POWER,
equivalent to the axiomatic model of~\cite{alglave2014herding}.

Having designed a suitable execution model, we address our
main challenge, which is to design an effective SMC algorithm that explores
all Shasha-Snir traces that can be generated by the execution model.
We address this challenge
by a novel exploration technique,
called {\em Relaxed Stateless Model Checking} (RSMC).
RSMC is suitable for execution models, in which
each instruction can be executed in many ways with
different effects on the program state, such as those derived using our execution model scheme.
The exploration by RSMC combines two mechanisms:
\begin{inparaenum}[(i)]
\item
RSMC  considers instructions one-by-one, respecting the
commit-before order, and explores the effects of each possible way in which
the instruction can be executed.
\item
  RSMC monitors the generated execution for
  data races from loads to subsequent stores,
  and initiates alternative explorations where instructions are reordered.
\end{inparaenum}
We define the property \emph{deadlock freedom} of execution models,
meaning intuitively that no run will block before being complete. We
prove that RSMC is sound for deadlock free execution models, and that
our execution model for POWER is indeed deadlock free.
We also prove that RSMC is optimal for POWER, in the sense that it
explores each \emph{complete} Shasha-Snir trace exactly once.
Similar to sleep set blocking for classical SMC/DPOR, it may happen
for RSMC that superfluous \emph{incomplete} Shasha-Snir traces are
explored. Our experiments indicate, however, that this is
rare.

To demonstrate the usefulness of our framework, we have implemented
RSMC in the stateless model checker Nidhugg~\cite{nidhugggithub}. For test cases written in C with pthreads, it
explores all Shasha-Snir traces allowed under the POWER memory model, up to some bounded length. We evaluate
our implementation on several challenging benchmarks. The results show that
RSMC efficiently explores the Shasha-Snir traces of a program, since
\begin{inparaenum}[(i)]
\item
  on most benchmarks, our implementation performs no superfluous exploration (as discussed above),
  and
\item
  the running times correlate to the number of Shasha-Snir traces of
  the program.
\end{inparaenum}
We show the competitiveness of our implementation by comparing with an
existing state of the art analysis tool for POWER:
\textsf{goto-instrument}~\cite{AlKNT13}.

\altparagraph{Outline.}
The next section presents our derivation of execution models.
Section~\ref{sec:dpor} presents our RSMC algorithm, and
Section~\ref{sec:experiments} presents our implementation and experiments.
\ifextended{
Proofs of all theorems, and formal definitions, are provided in the
appendix. Our implementation is available at~\cite{nidhugggithub}.}%
\ifnotextended{
Proofs of all theorems, and formal definitions, are provided in our
technical report~\cite{rsmctechreport}. Our implementation is
available at~\cite{nidhugggithub}.
\mtodo{Tech report}
}

  \newcommand{\gramvar}[1]{\langle#1\rangle}
  \newcommand{\gramlit}[1]{\textrm{\textquotesingle}\texttt{#1}\textrm{\textquotesingle}}

\begin{figure}
  \centering
  \begin{math}
    \begin{array}{r@{\;}l@{\;}l}
      \gramvar{prog} & ::= & \gramvar{varinit}^* \; \gramvar{thrd}^+\\
      \gramvar{varinit} & ::= & \gramvar{var} \;\gramlit{=}\; \mathbb{Z} \\
      \gramvar{thrd} &:=& \gramlit{thread} \; \gramvar{tid}  \;\gramlit{:} \;  \gramvar{linstr}^+\\
      \gramvar{linstr} & ::= & \gramvar{label} \;\gramlit{:} \; \gramvar{instr} \;\gramlit{;} \\
      \gramvar{instr} & ::= &
      \gramvar{reg} \;\gramlit{:=}\; \gramvar{expr} \;| \hfill\textrm{// register assignment}\\
      & & \gramlit{if} \; \gramvar{expr} \; \gramlit{goto} \; \gramvar{label} \; | \hfill\textrm{// conditional branch}\\
      & & \gramvar{reg} \;\gramlit{:=}\; \gramlit{[} \;\gramvar{expr}\; \gramlit{]} \;| \hfill\textrm{// memory load}\\
      & & \gramlit{[}\; \gramvar{expr} \;\gramlit{]} \; \gramlit{:=} \; \gramvar{expr} \;| \hfill\textrm{// memory store}\\
      & & \gramlit{sync} \;|\; \gramlit{lwsync} \;|\; \gramlit{isync} \hfill\textrm{// fences}\\
      \gramvar{expr} & ::= & \textrm{(arithmetic expression over literals and registers)}\\
    \end{array}
  \end{math}
  \caption{The grammar of concurrent programs}\label{fig:lang:grammar}
\end{figure}

\section{Execution Model for Relaxed Memory Models}
\label{sec:model}

\bjparagraph{POWER --- a Brief Glimpse.}
\label{subsec:power-glimpse}
The programmer-observable behavior of POWER multiprocessors emerges from
a combination of many features, including out-of-order and speculative
execution, various buffers, and caches. POWER provides
significantly weaker ordering guarantees than, e.g., SC and TSO.

We consider programs consisting of a number of threads, each of which
runs a deterministic code, built as a sequence of assembly instructions.
The grammar of our assumed language is given in Fig.~\ref{fig:lang:grammar}.
The threads access a shared memory, which is a mapping from  addresses to values.
A program may start by declaring named global variables with specific initial values.
Instructions include register assignments
and conditional branches with the usual semantics.
A load
\mbox{\textquotesingle\texttt{\areg:=[{\anexpr}]}\textquotesingle}
loads the value from the memory address given by the
arithmetic expression $\anexpr$ into the register $\areg$.
A store
\mbox{\textquotesingle\texttt{[{$\anexpr_0$}]:=$\anexpr_1$}\textquotesingle{}}
stores the value of the expression $\anexpr_1$ to the
memory location addressed by the evaluation of $\anexpr_0$.
For a global variable $\xvar$, we use $\xvar$ as syntactic
sugar for \texttt{[\&$\xvar$]}, where \texttt{\&\xvar} is the address
of \xvar.
The instructions \ffence, \lwfence, {\cfence} are fences (or memory
barriers), which are special instructions preventing some memory
ordering relaxations.
Each instruction is given a label, which is assumed to be unique.

As an example, consider the program in Fig.~\ref{fig:prog:LB:data}. It
consists of two threads $P$ and $Q$, and has two zero-initialized
memory locations $\xvar$ and $\yvar$. The thread $P$ loads the value
of $\xvar$, and stores that value plus one to $\yvar$. The thread $Q$
is similar, but always stores the value 1, regardless of the loaded
value.
Under the SC or TSO memory models, at least one of the loads
\texttt{L0} and \texttt{L2} is guaranteed to load the initial value 0
from memory. However, under POWER the order between the load
\texttt{L2} and the store \texttt{L3} is not maintained. Then it is
possible for $P$ to load the value 1 into $\areg_0$, and for $Q$ to
load 2 into $\areg_1$.
Inserting a \ffence between \texttt{L2} and \texttt{L3} would prevent
such a behavior.

\begin{figure}[t]
  \centering
  \begin{tabular}{@{}c@{\;\;\;}c@{}}
    \mbox{
      \texttt{
        \begin{tabular}{ll}
          \multicolumn{2}{l}{\xvar = 0 \qquad \yvar = 0}\\
          thread $P$:                     & thread $Q$:\\
          L0: $\areg_0$ := $\xvar$;   & L2: $\areg_1$ := $\yvar$;\\
          L1: $\yvar$ := $\areg_0$+1; & L3: $\xvar$ := 1;\\
        \end{tabular}
      }
    }
    &
    \mbox{
      \begin{tikzpicture}[baseline=-15pt]
        \node (l0) at (0,0) [anchor=west] {\texttt{L0: $\areg_0$ := $\xvar$}};
        \node (l1) at ($(l0.west)+(0,-1)$) [anchor=west] {\texttt{L1: $\yvar$ := $\areg_0$+1}};
        \node (l2) at (4,0) [anchor=west] {\texttt{L2: $\areg_1$ := $\yvar$}};
        \node (l3) at ($(l2.west)+(0,-1)$) [anchor=west] {\texttt{L3: $\xvar$ := 1}};
        \draw[->] (l0) -- node [left] {\textsf{po,data}} (l0 |- l1.north);
        \draw[->] (l2) -- node [right] {\textsf{po}} (l2 |- l3.north);
        \draw[->,out=0,in=180] (l1) to node [right] {\textsf{rf}} (l2);
        \draw[->,out=180,in=0] (l3) to node [left] {\textsf{rf}} (l0);
      \end{tikzpicture}
    }
    \\
  \end{tabular}
  \caption{Left: An example program: LB+data. Right: A trace of the same program.}
  \label{fig:prog:LB:data}
  \label{fig:exec:LB:data:final:state}
\end{figure}
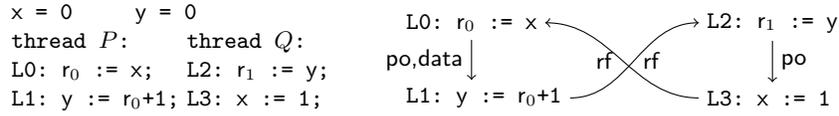

\begin{figure}[t]
  \centering
  \begin{tabular}{l@{\;\;}c@{\;\;}l}
    Event & Parameter & Semantic Meaning\\
    \hline
    \texttt{\phantom{MM}L3: $\xvar$ := 1} & 0 & First in coherence order for $\xvar$\\
    \texttt{L0: $\areg_0$ := $\xvar$} & \texttt{L3} & Read value 1 from \texttt{L3}\\
    \texttt{L1: $\yvar$ := $\areg_0$+1} & 0 & First in coherence order for $\yvar$\\
    \texttt{\phantom{MM}L2: $\areg_1$ := $\yvar$} & \texttt{L1} & Read value 2 from \texttt{L1}\\
  \end{tabular}
  \caption{The run
    $\inst{\texttt{L3}}{0}.\inst{\texttt{L0}}{\texttt{L3}}.\inst{\texttt{L1}}{0}.\inst{\texttt{L2}}{\texttt{L1}}$,
    of the program in Fig.~\ref{fig:prog:LB:data} (left), leading to
    the complete state corresponding to the trace given in
    Fig.~\ref{fig:exec:LB:data:final:state} (right). Here we use the
    labels \texttt{L0}-\texttt{L3} as shorthands for the corresponding
    events.}\label{fig:run:LB:data}
\end{figure}

\bjparagraph{Axiomatic Memory Models.}
Axiomatic memory models, of the form in~\cite{alglave2014herding}, operate
on an abstract representation of observable program behavior,
introduced by Shasha and Snir~\cite{shasha1988efficient},
here called {\em traces}.
A trace is a directed graph,
in which vertices are executed instructions (called {\em events}),
and edges capture dependencies between them.
More precisely,
a {\em trace} $\anexec$ is a
quadruple $(\events,\po,\co,\rf)$ where \events is a set of {\em events},
and \po, \co, and \rf are relations over \events
\footnote{\cite{alglave2014herding} uses the term ``execution'' to denote what we call ``trace.''}.
An {\em event} is a tuple $\eventtuple$ where
$\atid$ is an identifier for the
executing thread, $l$ is the unique label of the instruction,
and $n$ is a natural number which disambiguates instructions.
Let $\eventtype$ denote the set of all possible events.
For an event $\anevent=\eventtuple$, let $\tid(\anevent)$ denote
$\atid$ and let $\instrof{\anevent}$ denote the instruction labelled
$l$ in the program code.
The relation \po (for ``program order'') totally orders all events executed by
the same thread.
The relation \co (for ``coherence order'') totally orders all stores to the
same memory location.
The relation \rf (for ``read-from'') contains the pairs $(\anevent,\anevent')$
such that $\anevent$ is a store and
$\anevent'$ is a load which gets its value from
$\anevent$.
For simplicity,
we assume that the initial value of each memory address $\xvar$ is assigned by
a special {\em initializer instruction} $\init{\xvar}$, which is first in the coherence
order for that address.
A trace is a {\em complete trace of the program $\aprog$}
if the program order over the committed events of each thread makes
up a path from the first instruction in the code of the thread,
to the last instruction, respecting the evaluation of conditional branches.
Fig.~\ref{fig:exec:LB:data:final:state} shows the complete trace
corresponding to the behavior described in
the beginning of this section, in which
each thread loads the value stored by the other thread.

An axiomatic memory model \mmodel (following the
framework~\cite{alglave2014herding})
is defined as a predicate \mmodel over traces \anexec, such
that $\mmodel(\anexec)$ holds precisely when \anexec is an
allowed trace under the model.
Deciding whether $\mmodel(\anexec)$ holds involves checking%
\begin{inparaenum}[(i)]
\item that the trace is internally consistent, defined in the natural
  way (e.g., the relation \co relates precisely events that access the
  same memory location), and
\item
that various combinations of
relations that are derived from the trace are acyclic or irreflexive.
\end{inparaenum}
Which specific relations need to be acyclic depends on the memory
model.

We define the {\em axiomatic semantics under $\mmodel$} as a
mapping from programs $\aprog$ to their denotations
$\axmodel{\aprog}{\mmodel}$,
where $\axmodel{\aprog}{\mmodel}$
is the set of complete traces $\anexec$
of $\aprog$ such that $\mmodel(\anexec)$ holds.
In the following, we assume that the axiomatic memory model for
POWER, here denoted $\mmodelpower$,
is defined as in~\cite{alglave2014herding}.
The interested reader is encouraged to read the details
in~\cite{alglave2014herding}, but the high-level understanding given
above should be enough to understand the remainder of this text.

\bjparagraph{Deriving an Execution Model.}
\label{sec:semantics}
Let an axiomatic model $\mmodel$ be given, in the style
of~\cite{alglave2014herding}. We will derive an equivalent execution model in the
form of a transition system.

\altparagraph{States.} States of our execution model are traces,
augmented with a set of
fetched events.
A state $\state$ is a tuple of the form
$(\lblcur,\fetched,\events, \po,\co,\rf)$ where $\lblcur(\atid)$ is
a label in the code of $\atid$ for each thread $\atid$,
$\fetched\subseteq\eventtype$ is a set of events, and
$(\events,\po|_{\events},\co,\rf)$ is a trace such that
$\events\subseteq\fetched$. (Here
$\po|_{\events}$ is the restriction of $\po$ to $\events$.)
For a state $\astate=(\lblcur,\fetched,\events,\po,\co,\rf)$, we let
$\exec{\astate}$ denote the trace $(\events,\po|_{\events},\co,\rf)$.
Intuitively, $\fetched$ is the set of all currently fetched events and
$\events$ is the set of events that have been committed. The function
$\lblcur$ gives the label of the next instruction to fetch for each
thread. The relation $\po$ is the program order between all fetched
events. The relations $\co$ and $\rf$ are defined for committed events
(i.e., events in \events) only.
The set of all possible states is denoted $\states$. The initial state
$\astate_0\in\states$ is defined as
$\astate_0=(\lblcur_0,\events_0,\events_0,\emptyset,\emptyset,\emptyset)$
where $\lblcur_0$ is the function providing the initial label of each
thread, and $\events_0$ is the set of initializer events
for all memory locations.

\altparagraph{Commit-Before.}
The order in which events can be committed -- effectively a
linearization of the trace -- is restricted by a
\emph{commit-before order}. It is a parameter of our execution model
which can be tuned to suit the given axiomatic model.
Formally, a commit-before order is defined by a {\em \fcommitbefore}
$\fcb$, which associates with each state
$\astate=(\lblcur,\fetched,\events, \po,\co,\rf)$, a
\emph{\rcommitbefore} $\cb{\astate}\subseteq\fetched\times\fetched$,
which is a partial order on the set of fetched events.  For each state
$\astate$, the \rcommitbefore\ $\cb{\astate}$ induces a predicate
$\committable_\astate$ over the set of fetched events
$\anevent\in\fetched$ such that $\committable_{\astate}(\anevent)$
holds if and only if $\anevent\not\in\events$ and the set
$\{\anevent'\in\fetched\,|\,(\anevent',\anevent)\in\cb{\astate}\}$ is
included in $\events$. Intuitively, $\anevent$ can be committed only
if all the events it depends on have already been committed.
Later in this section, we define requirements on commit-before
functions, which are necessary for the execution model and for the RSMC
algorithm respectively.

\altparagraph{Transitions.}  The transition relation between states is
given by a set of rules, in Fig.~\ref{fig:oper:sem}.
The function $\val{\astate}(\anevent,\anexpr)$ denotes the
value taken by the arithmetic expression $\anexpr$, when evaluated at
the event $\anevent$ in the state $\astate$. The value is computed in
the natural way, respecting data-flow.%
\ifextended{(Formal definition given in Appendix~\ref{app:semantics}.)}
\ifnotextended{(Formal definition in the technical report~\cite{rsmctechreport}.)}
For example, in the state $\astate$ corresponding to the trace given
in Fig.~\ref{fig:exec:LB:data:final:state}, where $\anevent$ is the
event corresponding to label \texttt{L1}, we would have
$\val{\astate}(\anevent,\textrm{\texttt{$\areg_0$+1}})=2$.  The
function $\addr{\state}(\anevent)$ associates with each load or store
event $\anevent$ the memory location accessed.
For a label $l$, let $\lblnext(l)$ denote the next label following $l$
in the program code.
Finally, for a state $\astate$ with coherence order $\co$ and a store
$\anevent$ to some memory location $\xvar$, we let
$\extend_{\astate}(\anevent)$ denote the set of coherence orders
$\co'$ which result from inserting $\anevent$ anywhere in the
total order of stores to $\xvar$ in $\co$.
For each such order $\co'$, we let $\position_{\co'}(\anevent)$ denote
the position of $\anevent$ in the total order:
I.e. $\position_{\co'}(\anevent)$ is the number of (non-initializer)
events $\anevent'$ which precede $\anevent$ in $\co'$.

\begin{figure}[t]
  \centering
  \begin{math}
    \begin{array}{@{}c@{}}
      \begin{array}{@{}c@{}}
        \fetched_{\atid}= \{\anevent'' \in\fetched | \tid(\anevent'') = \atid\} \;\;\;\;
        \anevent = (\atid,|\fetched_{\atid}|,\lblcur(\atid))\\
        \not\exists \anevent',\anexpr,l \; . \; \anevent'\in\fetched\setminus\events \wedge \tid(\anevent')=\atid \wedge \instrof{\anevent'} = (\texttt{if }\anexpr\texttt{ goto }l)\\
        \hline
        \astate \xrightarrow{\FLB}
        (\lblcur[\atid\hookleftarrow\lblnext(\lblcur(\atid))],
        \fetched\cup\{\anevent\},
        \events,
        \po\cup (\fetched_{\atid}\times\{\anevent\}),
        \co,
        \rf
        )\\
      \end{array}
      \begin{array}{@{}c@{}}
        \\
        \\
        \textrm{FETCH}\\
        \\
      \end{array}
      \\
      \\
      \begin{array}{@{}c@{}}
        \instrof{\anevent}=(\texttt{if }\anexpr\texttt{ goto }l) \;\;\;\; \atid=\tid(\anevent) \\
        \val{\astate}(\anevent,\anexpr)\in\mathbb{Z}\setminus\{0\} \;\;\;\;
        \committable_{\astate}(\anevent)
        \\
        \hline
        \astate\xrightarrow{\FLB}
        (\lblcur[\atid\hookleftarrow l],\fetched,\events\cup\{\anevent\},\po,\co,\rf)\\
      \end{array}
      \begin{array}{@{}c@{}}
        \\
        \\
        \textrm{BRT}\\
        \\
      \end{array}
      \hspace{5pt}
      \begin{array}{@{}c@{}}
        \instrof{\anevent}\in\{\ffence,\lwfence,\cfence,\areg\texttt{:=}\anexpr\}  \\
     \committable_{\astate}(\anevent)
        \\
        \hline
        \astate\xrightarrow{\FLB}
        (\lblcur,\fetched,\events\cup\{\anevent\},\po,\co,\rf)\\
      \end{array}
      \begin{array}{@{}c@{}}
        \\
        \\
        \textrm{LOC}\\
        \\
      \end{array}
      \\
      \\
      \begin{array}{@{}c@{}}
        \instrof{\anevent}=(\texttt{if }\anexpr\texttt{ goto }l)  \\
        \val{\astate}(\anevent,\anexpr)=0 \;\;\;\; \committable_{\astate}(\anevent)
        \\
        \hline
        \astate\xrightarrow{\FLB}
        (\lblcur,\fetched,\events\cup\{\anevent\},\po,\co,\rf)\\
      \end{array}
      \begin{array}{@{}c@{}}
        \\
        \\
        \textrm{BRF}\\
        \\
      \end{array}
      \hspace{5pt}
      \begin{array}{c}
        \instrof{\anevent}=(\texttt{[}a\texttt{]:=}\anexpr')
         \;\;\; \committable_{\astate}(\anevent)\;\;\; \mmodel(\exec{\astate'})
         \\
         \astate' = (\lblcur,\fetched,\events\cup\{\anevent\},\po,\co',\rf)  \;\;
         \co' \in \extend_{\state}(\anevent)\\
        \hline
        \astate\xrightarrow{\inst{\anevent}{\position_{\co'}(\anevent)}}
        \astate'\\
      \end{array}
      \begin{array}{@{}c@{}}
        \\
        \\
        \textrm{ST}\\
        \vphantom{\astate\xrightarrow{\inst{\anevent}{\position_{\co'}(\anevent)}}}\\
      \end{array}
      \\
      \\
      \begin{array}{@{}c@{}}
        \instrof{\anevent}=(\areg\texttt{:=[}\anexpr\texttt{]})
                \;\;\;\; \committable_{\astate}(\anevent)
         \;\;\;
        \anevent_w\in\events \;\;\;
        \instrof{\anevent_w} = (\texttt{[}\anexpr'\texttt{]:=}\anexpr'')
        \\ \addr{\astate}(\anevent_w) = \addr{\astate}(\anevent) \;\;\;
        \astate' = (\lblcur,\fetched,\events\cup\{\anevent\},\po,\co,\rf\cup\{(\anevent_w,\anevent)\}) \;\;\;
        \mmodel(\exec{\astate'})\\
        \hline
        \astate\xrightarrow{\inst{\anevent}{\anevent_w}}
        \astate'\\
      \end{array}
      \begin{array}{@{}c}
        \\
        \\
        \textrm{LD}\\
        \\
      \end{array}
    \end{array}
  \end{math}
  \caption{Execution model of programs under the
    memory model \mmodel. Here
    $\astate = (\lblcur,\fetched,\events,\po,\co,\rf)$.
  }
  \label{fig:oper:sem}
\end{figure}

The intuition behind the rules in Fig.~\ref{fig:oper:sem} is that
events are committed non-deterministically out of order, but
respecting the constraints induced by the commit-before order.  When a
memory access (load or store) is committed, a non-deterministic choice
is made about its effect. If the event is a store, it is
non-deterministically inserted somewhere in the coherence order. If
the event is a load, we non-deterministically pick the store from
which to read. Thus, when committed, each memory access event
$\anevent$ is parameterized by a choice $\param$: the coherence
position for a store, and the source store for a load.
We call $\inst{\anevent}{\param}$ a \emph{parameterized event}, and
let $\parameventtype$ denote the set of all possible parameterized
events.
A transition committing a memory access is only enabled if the
resulting state is allowed by the memory model \mmodel.
Transitions are labelled with $\FLB$ when an event is fetched or a
local event is committed, or with $\inst{\anevent}{\param}$ when a
memory access event $\anevent$ is committed with parameter $\param$.

We illustrate this intuition for the program in
Fig.~\ref{fig:prog:LB:data} (left). The trace in
Fig.~\ref{fig:exec:LB:data:final:state} (right) can be produced by
committing the instructions (events) in the order \texttt{L3},
\texttt{L0}, \texttt{L1}, \texttt{L2}. For the load \texttt{L0}, we
can then choose the already performed \texttt{L3} as the store from
which it reads, and for the load \texttt{L2}, we can choose to read
from the store \texttt{L1}.  Each of the two stores \texttt{L3} and
\texttt{L1} can only be inserted at one place in their respective
coherence orders, since the program has only one store to each memory
location. We show the resulting sequence of committed events in
Fig.~\ref{fig:run:LB:data}: the first column shows the sequence of
events in the order they are committed, the second column is the
parameter assigned to the event, and the third column explains the
parameter. Note that other traces can be obtained by choosing
different values of parameters. For instance, the load \texttt{L2} can
also read from the initial value, which would generate a different
trace.

Next we explain each of the rules:
The rule FETCH allows to fetch the next instruction according to the
control flow of the program code. The first two requirements identify
the next instruction. To fetch an event, all preceding branch events
must already be committed. Therefore events are never fetched along a
control flow path that is not taken. We point out that this
restriction does not prevent our execution model from capturing the
observable effects of speculative execution (formally ensured by
Theorem~\ref{thm:equiv:oper:axiom}).

The rules LOC, BRT and BRF describe how to commit non-memory access
events.

When a store event is committed by the ST rule, it is inserted
non-deterministically at some position $n=\position_{\co'}(\anevent)$
in the coherence order. The guard $\mmodel(\exec{\astate'})$ ensures
that the resulting state is allowed by the axiomatic memory model.

The rule LD describes how to commit a load event $\anevent$. It is
similar to the ST rule. For a load we non-deterministically choose a
source store $\anevent_w$, from which the value can be read. As before,
the guard $\mmodel(\exec{\astate'})$ ensures that the resulting state
is allowed.

Given two states $\state, \state' \in \states$, we use
$\state\xrightarrow{\FLBmax} \, \state'$ to denote that
$\state{\xrightarrow{\FLB}}^{*} \state'$ and there is no state
$\state''\in\states$ with $\state'\xrightarrow{\FLB}\state''$.
A {\em run} $\exseq$ from some state $\state$ is a sequence of parameterized events $\inst{\anevent_1}{\param_1}.\inst{\anevent_2}{\param_2}.\cdots{}.\inst{\anevent_k}{\param_k}$ such that
{\small
  \begin{math}
  \state \inltransition{\FLBmax} \,\state_1 \inltransition{\inst{\anevent_1}{\param_1}}
  \state'_1\, \inltransition{\FLBmax}
  \cdots  \inltransition{\inst{\anevent_k}{\param_k}} \state'_k \inltransition{\FLBmax} \state_{k+1}
\end{math}
}
for some states $\state_1, \state'_1,\ldots, \state'_k,\state_{k+1}\in \states$.
We write $\inst{\anevent}{\param} \in \exseq$ to denote that the
parameterized event $\inst{\anevent}{\param}$ appears in
$\exseq$. Observe that the sequence $\exseq$ leads to a uniquely
determined state $\state_{k+1}$, which we denote $\stateafter{\exseq}$.
A run $\exseq$, from the initial state $\initstate$, is {\em complete}
iff the reached trace $\exec{\stateoafter{\tau}}$ is complete.
Fig.~\ref{fig:run:LB:data} shows an example complete run of the
program in Fig.~\ref{fig:prog:LB:data} (left).

In summary, our execution model represents a program
$\aprog$ as a labeled transition system
$\ltsof{\aprog}{\mmodel}{\fcb} = (\states,\initstate,\longrightarrow)$,
where
$\states$ is the set of states,
$\initstate$ is the initial state, and
$\extransitions \ \subseteq \states \times (\parameventtype \cup \set{\FLB}) \times \states$ is the transition relation.
We define the {\em execution semantics under $\mmodel$ and $\fcb$} as a
mapping, which maps each program $\aprog$ to its denotation
$\exmodel{\aprog}{\mmodel}{\fcb}$, which is
the set of complete runs $\arun$ induced by
$\ltsof{\aprog}{\mmodel}{\fcb}$.

\bjparagraph{Validity and Deadlock Freedom.}
Here, we define validity and deadlock freedom for memory models and
commit-before functions. Validity is necessary for the correct
operation of our execution model
(Theorem~\ref{thm:equiv:oper:axiom}). Deadlock freedom is necessary
for soundness of the RSMC algorithm
(Theorem~\ref{thm:soundness2}). First, we introduce some auxiliary
notions.

We say that a state
$\astate'=(\lblcur',\fetched',\events',\po',\co',\rf')$ is a
{\em $\fcb$-extension} of a state
$\astate=(\lblcur,\fetched,\events,\po,\co,\rf)$,
denoted $\astate\cbextends{\fcb}\astate'$, if
$\astate'$
can be obtained from $\astate$ by fetching in program order or committing events in $\fcb$ order.
Formally $\astate\cbextends{\fcb}\astate'$ if
$\po=\po'|_{\fetched}$, $\co=\co'|_{\events}$, $\rf=\rf'|_{\events}$,
$\fetched$ is a $\po'$-closed subset of $\fetched'$, and
$\events$ is a $\cb{\astate'}$-closed subset of $\events'$.
More precisely, the condition on $\fetched$ means that
for any events $\anevent,\anevent' \in \fetched'$, we have
$\left[\anevent'\in\fetched \land (\anevent,\anevent')\in\po'\right] \Rightarrow \anevent\in\fetched$.
The condition on $\events$ is analogous.

We say that $\fcb$ is \emph{monotonic} w.r.t. $\mmodel$ if whenever
$\astate\cbextends{\fcb}\astate'$, then
\begin{inparaenum}[(i)]
\item
  $\mmodel(\exec{\astate'}) \Rightarrow \mmodel(\exec{\astate})$,
\item
  $\cb{\astate} \subseteq \cb{\astate'}$, and
\item
  for all $\anevent\in\fetched$ such that either
  $\anevent\in\events$ or
  $\left(\committable_{\astate}(\anevent)\wedge\anevent\not\in\events'\right)$, we have
  $(\anevent',\anevent)\in\cb{\astate}\Leftrightarrow(\anevent',\anevent)\in\cb{\astate'}$
  for all $\anevent'\in\fetched'$.
\end{inparaenum}
Conditions (i) and (ii) are natural monotonicity requirements on
\mmodel and \fcb. Condition (iii) says that while an
event is committed or enabled, its $\fcb$-predecessors do not change.

A state $\astate$ induces a number of relations over its fetched
(possibly committed) events. Following~\cite{alglave2014herding}, we
let $\addrdep{\astate}$, $\datadep{\astate}$, $\ctrldep{\astate}$,
denote respectively address dependency, data dependency and control
dependency. Similarly, $\poloc{\astate}$ is the subset of $\po$ that
relates memory accesses to the same memory location. Lastly,
$\ffencedep{\astate}$ and $\lwsyncdep{\astate}$ relate events that are
separated in program order by respectively a $\ffence$ or
$\lwfence$. The formal definitions can be found
in~\cite{alglave2014herding},%
\ifextended{as well as in Appendix~\ref{app:semantics}.}
\ifnotextended{and in our technical report~\cite{rsmctechreport}.}
We can now define a weakest reasonable commit-before function
$\fcbzero$, capturing natural dependencies:
\[
\cbzero{\state} = (\addrdep{\state} \cup \datadep{\state} \cup \ctrldep{\state}\cup\rf)^+ \ \ ,
\]
where $R^+$ denotes the transitive (but not reflexive) closure of $R$.

We say that a commit-before function $\fcb$ is {\em valid} w.r.t.\ a memory model
\mmodel if $\fcb$ is monotonic w.r.t. $\mmodel$, and for all states
$\astate$ such that $\mmodel(\exec{\astate})$ we have that
$\cb{\astate}$ is acyclic and $\cbzero{\astate}\subseteq\cb{\astate}$.

\begin{theorem}[Equivalence with Axiomatic Model]\label{thm:equiv:oper:axiom}
  Let  $\fcb$ be a commit-before function
  valid w.r.t. a memory model \mmodel.
  Then $\axmodel{\aprog}{\mmodel}=\setcomp{\exec{\stateoafter{\exseq}}}{\exseq \in \exmodel{\aprog}{\mmodel}{\fcb}}$.
\ifnotextended{\qed}
\end{theorem}

\begin{figure}[t]
  \centering
  \begin{tabular}{ccc}
    Program & Blocked run $\arun$ & Blocked state $\astate$\\
    \hline
    \mbox{
      \texttt{
        \begin{tabular}{l@{\;\;\;}l}
          $\xvar$ = 0 & $\yvar$ = 0\\
          thread P: & thread Q:\\
          L0: $\areg_0$:=$\yvar$; & L3: $\xvar$:=3;\\
          L1: $\xvar$:=$\areg_0$; & L4: \ffence;\\
          L2: $\xvar$:=2;         & L5: $\yvar$:=1;\\
        \end{tabular}
      }
    }
    &
    \mbox{
      \texttt{
        \begin{tabular}{l}
          \phantom{MM}$\inst{\texttt{L3}}{0}$\\
          \phantom{MM}$\inst{\texttt{L5}}{0}$\\
          $\inst{\texttt{L0}}{\texttt{L5}}$\\
          $\inst{\texttt{L2}}{0}$\\
          \texttt{\color{red}(L1 blocked)}
        \end{tabular}
      }
    }
    &
    \begin{tikzpicture}[baseline=-25pt]
      \node (l0) at (0,0) {\texttt{L0: $\areg_0$:=$\yvar$}};
      \node (l1) at ($(l0.west)+(0,-0.8)$) [anchor=west] {\texttt{\color{red}L1: $\xvar$:=$\areg_0$}};
      \node (l2) at ($(l1.west)+(0,-0.8)$) [anchor=west] {\texttt{L2: $\xvar$:=2}};
      \node (l3) at ($(l0.east)+(1,0)$) [anchor=west] {\texttt{L3: $\xvar$:=3}};
      \node (l4) at ($(l3.west)+(0,-0.8)$) [anchor=west] {\texttt{L4: \ffence}};
      \node (l5) at ($(l4.west)+(0,-0.8)$) [anchor=west] {\texttt{L5: $\yvar$:=1}};
      \draw[->,draw=red] (l0) -- node [left] {\textsf{\color{red}data}} (l0 |- l1.north);
      \draw[->,draw=red] (l1) -- node [left] {\textsf{\color{red}po-loc}} (l1 |- l2.north);
      \draw[->] (l3) -- node [right] {\textsf{\ffence}} (l3 |- l4.north);
      \draw[->] (l4) -- node [right] {\textsf{\ffence}} (l4 |- l5.north);
      \draw[->,out=0,in=180] (l2) to node [left] {\textsf{\vphantom{f}co}} (l3);
      \draw[->,out=180,in=0] (l5) to node [right] {\textsf{rf}} (l0);
    \end{tikzpicture}
    \\
  \end{tabular}
  \caption{If the weak commit-before function $\fcbzero$ is used, the
    POWER semantics may deadlock.
    When the program above (left) is executed according to the run
    $\arun$ (center) we reach a state $\astate$ (right) where
    \texttt{L0}, \texttt{L2}, \texttt{L3}-\texttt{L5} are successfully
    committed. However, any attempt to commit \texttt{L1} will close a
    cycle in the relation
    $\co;\ffencedep{\astate};\rf;\datadep{\astate};\poloc{\astate}$,
    which is forbidden under POWER.
    This blocking behavior is prevented when the stronger
    commit-before function $\fcbpower$ is used, since it requires
    \texttt{L1} and \texttt{L2} to be committed in program order.
  }{}\label{fig:cb0:deadlock}
\end{figure}

The commit-before function
$\fcbzero$ is valid w.r.t. $\mmodelpower$,
implying (by Theorem~\ref{thm:equiv:oper:axiom}) that
$\exmodel{\aprog}{\mmodelpower}{\fcbzero}$ is a faithful execution model
for POWER.
However, $\fcbzero$ is not strong enough to prevent blocking runs in
the execution model for POWER. I.e., it is possible, with $\fcbzero$,
to create an incomplete run, which cannot be completed.
Any such blocking is undesirable for SMC, since it corresponds to
wasted exploration.
Fig.~\ref{fig:cb0:deadlock} shows an example of how the POWER
semantics may deadlock when based on $\fcbzero$.

We say that a memory model $\mmodel$ and a commit before function
$\fcb$ are \emph{deadlock free} if for all runs $\arun$ from
$\initstate$ and memory access events $\anevent$ such that
$\committable_{\stateoafter{\arun}}(\anevent)$ there exists a
parameter $\param$ such that $\arun.\inst{\anevent}{\param}$ is a run
from $\initstate$. I.e., it is impossible to reach a state where some
event is enabled, but has no parameter with which it can be committed.

\bjparagraph{Commit-Before Order for POWER.}\label{sec:cb-power}
We will now define a stronger commit before function for
POWER, which is both valid and deadlock free:

\nopagebreak

\[
\cbpower{\state} = (\cbzero{\state} \cup (\addrdep{\state};\po) \cup \poloc{\state} \cup \ffencedep{\state} \cup \lwsyncdep{\state})^+
\]

\begin{theorem}\label{thm:power:valid}
  $\fcbpower$ is valid w.r.t. $\mmodelpower$.
\end{theorem}

\begin{theorem}\label{thm:power:no:deadlock}
  $\mmodelpower$ and $\fcbpower$ are deadlock free.
\end{theorem}

\section{The RSMC Algorithm}\label{sec:dpor}

Having derived an execution model, we address the challenge of
defining an SMC algorithm, which explores all allowed traces of a
program in an efficient manner. Since each trace can be generated by
many equivalent runs, we must, just as in standard SMC for SC, develop
techniques for reducing the number of explored runs, while still
guaranteeing coverage of all traces.
Our RSMC algorithm is designed to do this in the context of semantics
like the one defined above, in which instructions can be committed
with several different parameters, each yielding different results.

Our exploration technique basically combines two mechanisms:
\begin{enumerate}[(i)]
\item
  In each state, RSMC considers an instruction $\anevent$,
  whose $\fcb$-predecessors have
  already been committed. For each possible parameter value $p$ of $\anevent$ in
  the current state, RSMC extends the state by $\anevent[p]$ and
  continues the exploration recursively from the new state.
\item
  RSMC monitors generated runs to detect read-write conflicts (or
  ``races''), i.e., the occurrence of a load and a subsequent store to
  the same memory location, such that the load would be able to read
  from the store if they were committed in the reverse order. For each
  such conflict, RSMC starts an alternative exploration, in which the
  load is preceded by the store, so that the load can read from the
  store.
\end{enumerate}
Mechanism (ii) is analogous to the detection and reversal of races in
conventional DPOR, with the difference that RSMC need only detect conflicts
in which a load is followed by a store.
A race in the opposite direction (store followed by load) does not
induce reordering by mechanism (ii). This is because our execution
model allows the load to read from any of the already committed stores
to the same memory location, without any reordering. An anlogous
observation applies to occurrences of several stores to the same
memory location.

\begin{figure}
  \begin{center}
    \texttt{
      \begin{tabular}{l@{\;\;\;}c@{\;\;\;}l}
        \textrm{Instruction} & \textrm{Parameter} & \textrm{Semantic Meaning} \\
        \hline
        L0: $\areg_0$ := $\xvar$ & $\init{\xvar}$ & \textrm{(read  initial value)}\\
        L1: $\yvar$ := $\areg_0$+1 & 0 & \textrm{(first in coherence of \yvar)}\\
        \rule{10pt}{0pt}L2: $\areg_1$ := $\yvar$ & $\init{\yvar}$ & \textrm{(read initial value)}\\
        \rule{10pt}{0pt}L3: $\xvar$ := 1 & 0 & \textrm{(first in coherence of \xvar)}\\
      \end{tabular}
    }
  \end{center}
  \caption{The first explored run of the program in Fig.~\ref{fig:prog:LB:data}}\label{fig:exec:LB:data:ex0}
\end{figure}

We illustrate the basic idea of RSMC on the program in
Fig.~\ref{fig:prog:LB:data} (left). As usual in SMC, we start by running the
program under an arbitrary schedule, subject to the constraints
imposed by the commit-before order $\fcb$. For each instruction, we
explore the effects of each parameter value which is allowed by the
memory model. Let us assume that we initially explore the instructions
in the order \texttt{L0}, \texttt{L1}, \texttt{L2}, \texttt{L3}. For
this schedule, there is only one possible parameter for \texttt{L0},
\texttt{L1}, and \texttt{L3}, whereas \texttt{L2} can read either from
the initial value or from \texttt{L1}.
Let us assume that it reads the initial value. This gives us the first
run, shown in Fig.~\ref{fig:exec:LB:data:ex0}. The second run is
produced by changing the parameter for \texttt{L2}, and let it read
the value 1 written by \texttt{L1}.

During the exploration of the first two runs, the RSMC algorithm also
detects a race between the load \texttt{L0} and the store
\texttt{L3}. An important observation is that \texttt{L3} is not
ordered after \texttt{L0} by the commit-before order, implying that
their order can be reversed. Reversing the order between \texttt{L0}
and \texttt{L3} would allow \texttt{L0} to read from
\texttt{L3}. Therefore, RSMC initiates an exploration where the load
\texttt{L0} is preceded by \texttt{L3} and reads from it. (If
\texttt{L3} would have been preceded by other events that enable
\texttt{L3}, these would be executed before \texttt{L3}.)  After the
sequence \texttt{L3[0]}.\texttt{L0[L3]}, RSMC is free to choose the
order in which the remaining instructions are considered. Assume that
the order \texttt{L1}, \texttt{L2} is chosen. In this case, the load
\texttt{L2} can read from either the initial value or from
\texttt{L1}. In the latter case, we obtain the run in
Fig.~\ref{fig:run:LB:data}, corresponding to the trace in
Fig.~\ref{fig:exec:LB:data:final:state} (right).

After this, there are no more unexplored parameter choices, and so the
RSMC algorithm terminates, having explored four runs corresponding to
the four possible traces.

In the following section, we will provide a more detailed look at the
RSMC algorithm, and see formally how this exploration is carried out.

\subsection{Algorithm Description}

\newcommand{\newbnc}{\textrm{\texttt{S}}}
\newcommand{\newalgo}{\textrm{\mbox{\textbf{Explore}}}}
\newcommand{\newalgorun}{\textrm{\mbox{\textbf{Traverse}}}}
\newcommand{\newalgodetectrace}{\textrm{\mbox{\textbf{DetectRace}}}}
\newcommand{\prefixmap}{\texttt{P}\xspace}
\newcommand{\contmap}{\texttt{Q}\xspace}
\newcommand{\cutrun}{\textrm{\textsf{cut}}}
\newcommand{\dporstar}{\textrm{\texttt{*}}}
\newcommand{\normalizerun}{\textrm{\textsf{normalize}}}

In this section, we present
our algorithm, RSMC, for SMC under POWER.
We prove soundness of RSMC, and optimality w.r.t. explored \emph{complete} traces.

\newlength{\dhatheight}
\newlength{\dhatwidth}
\newcommand{\dhat}[1]{%
\settoheight{\dhatheight}{\ensuremath{\hat{#1}}}%
\settowidth{\dhatwidth}{\ensuremath{\hat{#1}}}%
\addtolength{\dhatheight}{-1pt}%
\hat{\phantom{\rule{\dhatwidth}{\dhatheight}}}%
\llap{$\hat{#1}$}%
}

The RSMC algorithm is shown in Fig.~\ref{fig:algo:dpor}. It uses
the recursive procedure $\newalgo$, which takes
parameters $\arun$ and $\astate$ such that $\astate = \stateoafter{\arun}$.
$\newalgo$ will explore all states that can be reached by complete runs extending $\arun$.

First, on line~\ref{ln:dpor2:flb:start}, we fetch instructions and
commit all local instructions as far as possible from $\astate$. The
order of these operations makes no difference. Then we turn to memory
accesses.
If the run is not yet terminated, we select an enabled event
$\anevent$ on line~\ref{ln:dpor2:select:e:disarm}.

\newcommand{\scripttext}[1]{{\textrm{\fontsize{7pt}{8pt}\selectfont$#1$}}}

\begin{figure}[t]
  {\fontsize{8pt}{8pt}\selectfont
    \begin{tabular}{@{}m{.5\linewidth}m{.49\linewidth}@{}}
      \texttt{\color{gray}// $\prefixmap$[$\anevent$] holds a run}\linebreak
      \texttt{\color{gray}// preceding the load event $\anevent$.}\linebreak
      \texttt{\mbox{\textbf{global} \prefixmap = $\lambda\anevent.\emptyseq$}}\linebreak
      \texttt{\color{gray}// $\contmap$[$\anevent$] holds a set of continuations}\linebreak
      \texttt{\color{gray}// leading to the execution of the}\linebreak
      \texttt{\color{gray}// load event $\anevent$ after $\prefixmap$[$\anevent$].}\linebreak
      \texttt{\mbox{\textbf{global} \contmap = $\lambda\anevent.\emptyset$}}\linebreak
      \linebreak
      \begin{algo}{$\newalgo(\arun,\astate)$}{exploretwolinctr}
        \comment{Fetch \& commit local greedily.}
        \whileoneline{$\exists \astate' . \astate\xrightarrow{\scripttext{FLB}}\astate'$\label{ln:dpor2:flb:start}}{%
          $\astate$ := $\astate'$;%
        }\lnlabel{ln:dpor2:flb:end}

        \comment{Find committable memory access $\anevent$.}
        \ifterm{$\exists\anevent.\committable_{\astate}(\anevent)$\label{ln:dpor2:select:e:disarm}}{
          \ifunterm{$\anevent$ is a store\label{ln:dpor2:store:begin}}{
            \comment{Explore all ways to execute $\anevent$.}
            \lin{$\newbnc$ := $\{(n,\astate') | \astate\xrightarrow{\scripttext{\inst{\anevent}{n}}}\astate'\}$;\label{ln:dpor2:store:comp:S}}
            \for{$(n,\astate')\in\newbnc$\label{ln:dpor2:store:rec:loop:begin}}{
              \lin{$\newalgo(\arun.\inst{\anevent}{n},\astate')$;\label{ln:dpor2:store:rec:call}}
            }
            \lin{$\newalgodetectrace(\arun,\astate,\anevent)$;\label{ln:dpor2:store:call:detect:race}}%
            \lnlabel{ln:dpor2:store:end}%
          }\elseterm[$\anevent$ is a load\label{ln:dpor2:load:begin}]{
            \lin{$\prefixmap$[$\anevent$] := $\arun$;\label{ln:dpor2:load:store:prefix}}
            \comment{Explore all ways to execute $\anevent$.}
            \lin{$\newbnc$ := $\{(\anevent_{\scripttext{w}},\astate') | \astate\xrightarrow{\scripttext{\inst{\anevent}{\anevent_{\scripttext{w}}}}}\astate'\}$;\label{ln:dpor2:load:comp:S}}
            \for{$(\anevent_{\scripttext{w}},\astate')\in\newbnc$\label{ln:dpor2:load:rec:loop:begin}}{
              \lin{$\newalgo(\tau.\inst{\anevent}{\anevent_{\scripttext{w}}},\astate')$;\label{ln:dpor2:load:rec:call}}
            }\lnlabel{ln:dpor2:load:rec:loop:end}
            \comment{Handle R -> W races.}
            \lin{explored = $\emptyset$;\label{ln:dpor2:load:cont:begin}}
            \while{$\exists\arun'\in\contmap$[$\anevent$]$\setminus$explored}{
              \lin{explored := explored$\cup\{\arun'\}$;\label{ln:dpor2:add:explored}}
              \lin{$\newalgorun(\arun,\astate,\arun')$;\label{ln:dpor2:call:explore:run}}
            }\lnlabel{ln:dpor2:load:cont:end}%
          }\lnlabel{ln:dpor2:load:end}%
        }
      \end{algo}
      &
      \begin{algo}{$\newalgodetectrace(\arun,\astate,\anevent)$}{exploretwodetectracelinctr}
        \fortall{\begin{tabular}{l}$\inst{\anevent_{\scripttext{r}}}{\anevent_{\scripttext{w}}}\in\arun$ s.t. \\
            \hspace{7pt}$\anevent_{\scripttext{r}}$ is a load $\wedge\;(\anevent_{\scripttext{r}},\anevent)\not\in \cb{\astate}$\\
            \hspace{7pt}$\wedge\;\addr{{\scripttext{\astate}}}(\anevent_{\scripttext{r}}) = \addr{{\scripttext{\astate}}}(\anevent)$\\
          \end{tabular}\label{ln:dpor2:detect:conflict:begin}}{
          \comment{Compute postfix after $\prefixmap$[$\anevent_{\scripttext{r}}$].}
          \lin{$\arun'$ := the $\arun'$ s.t. $\arun = \textrm{\texttt{$\prefixmap$[$\anevent_{\scripttext{r}}$]}}.\arun'$;\label{ln:dpor2:cut:branch:begin}}
          \comment{Remove events not cb-before $\anevent$.}
          \lin{$\arun''$ := $\normalizerun(\cutrun(\arun',\anevent,\astate),\cb{{\scripttext{\astate}}})$;\label{ln:dpor2:cut:branch:end}}
          \comment{Construct new continuation.}
          \lin{$\arun'''$ := $\arun''.\inst{\anevent}{\dporstar}.\inst{\anevent_{\scripttext{r}}}{\anevent}$;}
          \comment{Add to \contmap, to explore later.}
          \lin{$\contmap$[$\anevent_{\scripttext{r}}$] := $\contmap$[$\anevent_{\scripttext{r}}$]$\cup\{\arun'''\}$;\label{ln:dpor2:store:add:branch}}
        }\lnlabel{ln:dpor2:detect:conflict:end}
      \end{algo}

      \begin{algo}{$\newalgorun(\arun,\astate,\arun')$}{exploretworunlinctr}
        \ifunterm{$\arun' = \emptyseq$}{
          \lin{$\newalgo(\arun,\astate)$;\label{ln:dpor2run:return:to:dpor2}}
        }\elseterm{
          \comment{Fetch \& commit local greedily.}
          \whileoneline{$\exists \astate' . \astate\xrightarrow{\scripttext{FLB}}\astate'$}{%
            $\astate$ := $\astate'$;%
          }
          \lin{$\inst{\anevent}{\param}.\arun''$ := $\arun'$; \icmt{Get first event.}}
          \ifunterm{$\param = \dporstar$\label{ln:dpor2run:handle:star:begin}}{
            \comment{Explore all ways to execute $\anevent$.}
            \lin{$\newbnc$ := $\{(n,\astate') | \astate\xrightarrow{\scripttext{\inst{\anevent}{n}}}\astate'\}$;}
            \for{$(n,\astate')\in\newbnc$}{
              \lin{$\newalgorun(\arun.\inst{\anevent}{n},\astate',\arun'')$;}
            }\lnlabel{ln:dpor2run:handle:star:end}%
          }\elseifunterm{$\exists \astate' . \astate\xrightarrow{\scripttext{\inst{\anevent}{\param}}}\astate'$}{
            \lin{$\newalgorun(\arun.\inst{\anevent}{\param},\astate',\arun'')$;}
          }\elseterm{\lnlabel{ln:dpor2run:fail:commit}
            \comment{Only happens when the final}
            \comment{load in $\arun'$ does not accept its}
            \comment{parameter. Stop exploring.}
          }
        }
      \end{algo}
      \\
    \end{tabular}
  }
  \caption{An algorithm to explore all traces of a given program. The initial call is $\newalgo(\emptyseq,\initstate)$.}\label{fig:algo:dpor2}\label{fig:algo:dpor}\label{fig:algo:dpor2run}
\end{figure}

If the chosen event $\anevent$ is a store
(lines~\ref{ln:dpor2:store:begin}-\ref{ln:dpor2:store:end}), we first
collect, on line~\ref{ln:dpor2:store:comp:S}, all parameters for
$\anevent$ which are allowed by the memory model. For each of them, we
recursively explore all of its continuations on
line~\ref{ln:dpor2:store:rec:call}. I.e., for each coherence position
$n$ that is allowed for $\anevent$ by the memory model, we explore the
continuation of $\arun$ obtained by committing $\anevent[n]$.
Finally, we call
$\newalgodetectrace$. We will return shortly to a discourse of that
mechanism.

If $\anevent$ is a load
(lines~\ref{ln:dpor2:load:begin}-\ref{ln:dpor2:load:end}), we proceed
in a similar manner.
Line~\ref{ln:dpor2:load:store:prefix} is related to
$\newalgodetectrace$, and discussed later.
On line~\ref{ln:dpor2:load:comp:S} we compute all
allowed parameters for the load $\anevent$. They are (some of the) stores in $\arun$ which access the same address as
$\anevent$. On line~\ref{ln:dpor2:load:rec:call},
we make one recursive call to \newalgo{} per allowed parameter.
The structure of this exploration is illustrated in the two branches
from $\astate_1$ to $\astate_2$ and $\astate_5$ in
Fig.~\ref{fig:dpor2:rec:calls}(a).

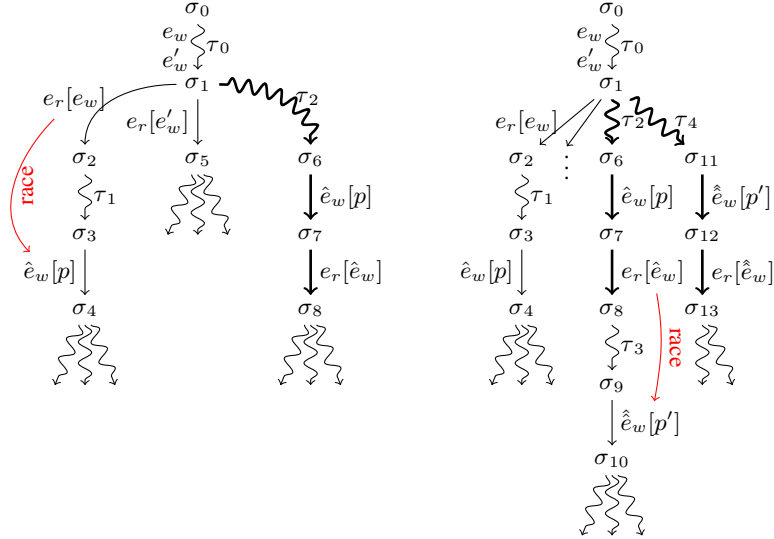
\begin{figure}[t]
  \begin{center}
    \begin{tabular}{@{}p{.45\linewidth}@{\;\;}p{.45\linewidth}@{}}
      \begin{center}
        \begin{tikzpicture}
          \node (s0) at (0,0) [] {$\initstate$};
          \node (s1) at ($(s0)+(0,-1)$) [] {$\astate_1$};
          \draw[decorate,decoration={snake,post length=1pt},segment aspect=0, segment length=8pt,->] (s0) -- node [right] {$\arun_0$} node [left] {\begin{tabular}{@{}l@{}}\small{$\anevent_w$}\\\small{$\anevent_w'$}\end{tabular}} (s1);
          \node (s2a) at ($(s1)+(-1.5,-1)$) [] {$\astate_2$};
          \node (s2c) at ($(s1)+(0,-1)$) [] {$\astate_5$};
          \node (s3a) at ($(s2a)+(0,-1)$) [] {$\astate_3$};
          \node (s4a) at ($(s3a)+(0,-1)$) [] {$\astate_4$};
          \draw[->,out=180,in=90] (s1) to node [left,name=er] {\small{$\inst{\anevent_r}{\anevent_w}$}} (s2a);
          \draw[->] (s1) to node [left] {\small{$\inst{\anevent_r}{\anevent_w'}$}} (s2c);
          \draw[decorate,decoration={snake,post length=1pt},segment aspect=0, segment length=8pt,->]
          (s2a) -- node [right] {$\arun_1$} (s3a);
          \draw[->] (s3a) to node [left,name=ew] {\small{$\inst{\hat{\anevent}_w}{\param}$}} (s4a);
          \draw[decorate,decoration={snake,post length=1pt},segment aspect=0, segment length=8pt,->]
          (s4a) -- ($(s4a)+(-.4,-1)$);
          \draw[decorate,decoration={snake,post length=1pt},segment aspect=0, segment length=8pt,->]
          (s4a) -- ($(s4a)+(0,-1)$);
          \draw[decorate,decoration={snake,post length=1pt},segment aspect=0, segment length=8pt,->]
          (s4a) -- ($(s4a)+(.4,-1)$);
          \node (s2b) at ($(s1)+(1.5,-1)$) [] {$\astate_6$};
          \node (s3b) at ($(s2b)+(0,-1)$) [] {$\astate_7$};
          \node (s4b) at ($(s3b)+(0,-1)$) [] {$\astate_8$};
          \draw[decorate,decoration={snake,pre length=3pt,post length=1pt},segment aspect=0, segment length=6pt,->,line width=1pt,out=0,in=90]
          (s1) to node [right] {\hspace{2pt}$\arun_2$} (s2b);
          \draw[->,line width=1pt] (s2b) -- node [right] {\small{$\inst{\hat{\anevent}_w}{\param}$}} (s3b);
          \draw[->,line width=1pt] (s3b) -- node [right] {\small{$\inst{\anevent_r}{\hat{\anevent}_w}$}} (s4b);
          \draw[decorate,decoration={snake,post length=1pt},segment aspect=0, segment length=8pt,->]
          (s4b) -- ($(s4b)+(-.4,-1)$);
          \draw[decorate,decoration={snake,post length=1pt},segment aspect=0, segment length=8pt,->]
          (s4b) -- ($(s4b)+(0,-1)$);
          \draw[decorate,decoration={snake,post length=1pt},segment aspect=0, segment length=8pt,->]
          (s4b) -- ($(s4b)+(.4,-1)$);
          \draw[decorate,decoration={snake,post length=1pt},segment aspect=0, segment length=8pt,->]
          (s2c) -- ($(s2c)+(-.4,-1)$);
          \draw[decorate,decoration={snake,post length=1pt},segment aspect=0, segment length=8pt,->]
          (s2c) -- ($(s2c)+(0,-1)$);
          \draw[decorate,decoration={snake,post length=1pt},segment aspect=0, segment length=8pt,->]
          (s2c) -- ($(s2c)+(.4,-1)$);
          \draw[->,out=225,in=135,draw=red] (er) to node [below,rotate=90] {\small{\color{red}race}} (ew);
        \end{tikzpicture}
      \end{center}
      &
      \begin{center}
        \begin{tikzpicture}
          \node (s0) at (0,0) [] {$\initstate$};
          \node (s1) at ($(s0)+(0,-1)$) [] {$\astate_1$};
          \draw[decorate,decoration={snake,post length=1pt},segment aspect=0, segment length=8pt,->] (s0) -- node [right] {$\arun_0$} node [left] {\begin{tabular}{@{}l@{}}\small{$\anevent_w$}\\\small{$\anevent_w'$}\end{tabular}} (s1);
          \node (s2a) at ($(s1)+(-1.2,-1)$) [] {$\astate_2$};
          \node (s2d) at ($(s1)+(-.6,-1)$) [] {$\vdots$};
          \draw[->] (s1) -- ($(s2d.north)+(0,-.2)$);
          \node (s3a) at ($(s2a)+(0,-1)$) [] {$\astate_3$};
          \node (s4a) at ($(s3a)+(0,-1)$) [] {$\astate_4$};
          \draw[->] (s1) to node [left,name=er] {\small{$\inst{\anevent_r}{\anevent_w}$}} (s2a);
          \draw[decorate,decoration={snake,post length=1pt},segment aspect=0, segment length=8pt,->]
          (s2a) -- node [right] {$\arun_1$} (s3a);
          \draw[->] (s3a) to node [left,name=ew] {\small{$\inst{\hat{\anevent}_w}{\param}$}} (s4a);
          \draw[decorate,decoration={snake,post length=1pt},segment aspect=0, segment length=8pt,->]
          (s4a) -- ($(s4a)+(-.4,-1)$);
          \draw[decorate,decoration={snake,post length=1pt},segment aspect=0, segment length=8pt,->]
          (s4a) -- ($(s4a)+(0,-1)$);
          \draw[decorate,decoration={snake,post length=1pt},segment aspect=0, segment length=8pt,->]
          (s4a) -- ($(s4a)+(.4,-1)$);
          \node (s2b) at ($(s1)+(0,-1)$) [] {$\astate_6$};
          \node (s3b) at ($(s2b)+(0,-1)$) [] {$\astate_7$};
          \node (s4b) at ($(s3b)+(0,-1)$) [] {$\astate_8$};
          \draw[decorate,decoration={snake,post length=1pt},segment aspect=0, segment length=7pt,->,line width=1pt]
          (s1) -- node [right] {$\arun_2$} (s2b);
          \draw[->,line width=1pt] (s2b) -- node [right] {\small{$\inst{\hat{\anevent}_w}{\param}$}} (s3b);
          \draw[->,line width=1pt] (s3b) -- node [right,name=er2] {\small{$\inst{\anevent_r}{\hat{\anevent}_w}$}} (s4b);
          \node (s5b) at ($(s4b)+(0,-1)$) [] {$\astate_9$};
          \node (s6b) at ($(s5b)+(0,-1)$) [] {$\astate_{10}$};
          \draw[decorate,decoration={snake,post length=1pt},segment aspect=0, segment length=8pt,->]
          (s4b) -- node [right] {$\arun_3$} (s5b);
          \draw[->] (s5b) -- node [right,name=ewpp] {\small{$\inst{\dhat{\anevent}_w}{\param'}$}} (s6b);
          \draw[decorate,decoration={snake,post length=1pt},segment aspect=0, segment length=8pt,->]
          (s6b) -- ($(s6b)+(-.4,-1)$);
          \draw[decorate,decoration={snake,post length=1pt},segment aspect=0, segment length=8pt,->]
          (s6b) -- ($(s6b)+(0,-1)$);
          \draw[decorate,decoration={snake,post length=1pt},segment aspect=0, segment length=8pt,->]
          (s6b) -- ($(s6b)+(.4,-1)$);
          \node (s2c) at ($(s1)+(1.2,-1)$) [] {$\astate_{11}$};
          \node (s3c) at ($(s2c)+(0,-1)$) [] {$\astate_{12}$};
          \node (s4c) at ($(s3c)+(0,-1)$) [] {$\astate_{13}$};
          \draw[decorate,decoration={snake,post length=1pt},segment aspect=0, segment length=6.5pt,->,line width=1pt]
          (s1) -- node [right] {\hspace{3pt}$\arun_4$} (s2c);
          \draw[->,line width=1pt] (s2c) -- node [right] {\small{$\inst{\dhat{\anevent}_w}{\param'}$}} (s3c);
          \draw[->,line width=1pt] (s3c) -- node [right] {\small{$\inst{\anevent_r}{\dhat{\anevent}_w}$}} (s4c);
          \draw[decorate,decoration={snake,post length=1pt},segment aspect=0, segment length=8pt,->]
          (s4c) -- ($(s4c)+(0,-1)$);
          \draw[decorate,decoration={snake,post length=1pt},segment aspect=0, segment length=8pt,->]
          (s4c) -- ($(s4c)+(.4,-1)$);
          \draw[->,out=280,in=80,draw=red] (er2) to node [above,rotate=270] {\small{\color{red}race}} (ewpp);
        \end{tikzpicture}
      \end{center}\\
      (a)
      A new branch
      $\arun_2.\inst{\hat{\anevent}_w}{\dporstar}.\inst{\anevent_r}{\hat{\anevent}_w}$
      is added to \texttt{$\contmap$[$\anevent_r$]} and later
      explored, starting from $\astate_1$. $\arun_2$ is a
      restriction of $\arun_1$, containing only events that are
      $\cb{\astate_4}$-before
      $\hat{\anevent}_w$.
      %% \label{fig:dpor2:load:rec:calls:2:1}
      &
      (b)
      Another read-write race is detected,
      starting from the leaf of a branch explored by
      \newalgorun{}.
      The new branch
      $\arun_4.\inst{\dhat{\anevent}_w}{\dporstar}.\inst{\anevent_r}{\dhat{\anevent}_w}$
      is added at $\astate_1$, not at
      $\astate_7$. %% \label{fig:dpor2:load:rec:calls:2:2}
    \end{tabular}
  \end{center}
  \caption{How $\newalgo$ applies event parameters, and
    introduces new branches.
    Thin arrows indicate exploration performed directly by \newalgo{}.
    Bold arrows indicate traversal by
    \newalgorun{}.}\label{fig:dpor2:rec:calls}
\end{figure}

Notice in the above that both for stores and loads, the available
parameters are determined entirely by $\arun$, i.e. by the events that
precede $\anevent$ in the run. In the case of stores, the parameters
are coherence positions between the earlier stores occurring in
$\arun$. In the case of loads, the parameters are the earlier stores
occurring in $\arun$. For stores, this way of exploring is
sufficient. But for loads it is necessary to also consider parameters
which appear later than the load in a run. Consider the example in
Fig.~\ref{fig:dpor2:rec:calls}(a).
During the recursive exploration of a run from $\initstate$ to
$\astate_4$ we encounter a new store $\hat{\anevent}_w$, which is in a
race with $\anevent_r$.
If the load
$\anevent_r$ and the store $\hat{\anevent}_w$ access the same memory
location, and $\anevent_r$ does not precede $\hat{\anevent}_w$ in the
$\fcb$-order, they could appear in the opposite order in a run (with
$\hat{\anevent}_w$ preceding $\anevent_r$), and $\hat{\anevent}_w$
could be an allowed parameter for the load $\anevent_r$.
This read-write race is detected on
line~\ref{ln:dpor2:detect:conflict:begin} in the function
$\newalgodetectrace$, when it is called from
line~\ref{ln:dpor2:store:call:detect:race} in $\newalgo$ when the
store $\hat{\anevent}_w$ is being explored.
We must then ensure that some run is explored where $\hat{\anevent}_w$
is committed before $\anevent_r$ so that $\hat{\anevent}_w$ can be
considered as a parameter for $\anevent_r$. Such a run must include
all events that are before $\hat{\anevent}_w$ in $\fcb$-order, so that
$\hat{\anevent}_w$ can be committed. We construct $\arun_2$, which is
a template for a new run, including precisely the events in $\arun_1$
which are $\fcb$-before the store $\hat{\anevent}_w$. The run template
$\arun_2$ can be explored from the state $\astate_1$ (the state where
$\anevent_r$ was previously committed) and will then lead to a state
where $\hat{\anevent}_w$ can be committed. The run template $\arun_2$
is computed from the complete run in $\newalgodetectrace$ on
lines~\ref{ln:dpor2:cut:branch:begin}-\ref{ln:dpor2:cut:branch:end}.
This is done by first removing (at
line~\ref{ln:dpor2:cut:branch:begin}) the prefix $\arun_0$ which
precedes $\anevent_r$ (stored in \texttt{$\prefixmap$[$\anevent_r$]}
on line~\ref{ln:dpor2:load:store:prefix} in $\newalgo$). Thereafter
(at line~\ref{ln:dpor2:cut:branch:end}) events that are not
$\fcb$-before $\hat{\anevent}_w$ are removed using the function
$\cutrun$ (here, $\cutrun(\arun,\anevent,\astate)$ restricts $\arun$
to the events which are $\cb{\astate}$-before $\anevent$),
and the resulting run is normalized. The function $\normalizerun$
normalizes a run by imposing a predefined order on the events which
are not ordered by $\fcb$. This is done to avoid unnecessarily
exploring two equivalent run templates.
\ifextended{(Formal definitions in Appendix~\ref{app:aux:dpor2}.)}
The run template
$\arun_2.\inst{\hat{\anevent}_w}{\dporstar}.\inst{\anevent_r}{\hat{\anevent}_w}$
is then stored on line~\ref{ln:dpor2:store:add:branch} in the set
\texttt{$\contmap$[$\anevent_r$]}, to ensure that it is explored
later. Here we use the special pseudo-parameter $\dporstar$ to indicate that
every allowed parameter for $\hat{\anevent}_w$ should be explored (See
lines~\ref{ln:dpor2run:handle:star:begin}-\ref{ln:dpor2run:handle:star:end}
in $\newalgorun$.).

All of the run templates collected in \texttt{$\contmap$[$\anevent_r$]} are
explored from the same call to $\newalgo(\arun_0,\astate_1)$ where
$\anevent_r$ was originally committed. This is done on
lines~\ref{ln:dpor2:load:cont:begin}-\ref{ln:dpor2:load:cont:end}. The
new branch is shown in Fig.~\ref{fig:dpor2:rec:calls}(a) in the run
from $\astate_0$ to $\astate_8$. Notice on
line~\ref{ln:dpor2:call:explore:run} that the new branch is explored
by the function \newalgorun{}, rather than by \newalgo{} itself. This
has the effect that $\arun_2$ is traversed, with each event using the
parameter given in $\arun_2$, until
$\inst{\anevent_r}{\hat{\anevent}_w}$ is committed. The traversal by
\newalgorun{} is marked with bold arrows in
Fig.~\ref{fig:dpor2:rec:calls}. If the memory model does not allow
$\anevent_r$ to be committed with the parameter $\hat{\anevent}_w$,
then the exploration of this branch terminates on
line~\ref{ln:dpor2run:fail:commit} in \newalgorun{}. Otherwise, the
exploration continues using \newalgo{}, as soon as $\anevent_r$ has
been committed (line~\ref{ln:dpor2run:return:to:dpor2} in
\newalgorun{}).

Let us now consider the situation in
Fig.~\ref{fig:dpor2:rec:calls}(b) in the run from
$\initstate$ to $\astate_{10}$. Here
$\arun_2.\inst{\hat{\anevent}_w}{\dporstar}.\inst{\anevent_r}{\hat{\anevent}_w}$,
is explored as described above. Then \newalgo{} continues the exploration, and a
read-write race is discovered from
$\anevent_r$ to $\dhat{\anevent}_w$. From earlier DPOR algorithms such as
e.g. \cite{FG:dpor}, one might expect that this case is handled by
exploring a new branch of the form
$\arun_2.\inst{\hat{\anevent}_w}{\param}.\arun_3'.\inst{\dhat{\anevent}_w}{\param'}.\inst{\anevent_r}{\dhat{\anevent}_w}$,
where $\anevent_r$ is simply delayed after $\astate_7$ until
$\dhat{\anevent}_w$ has been committed.
Our algorithm handles the case differently, as shown in the run from
$\initstate$ to $\astate_{13}$.
Notice that \texttt{$\prefixmap$[$\anevent_r$]} can be used to
identify the position in the run where $\anevent_r$ was last committed
by \newalgo{} (as opposed to by \newalgorun{}), i.e., $\astate_1$ in
Fig.~\ref{fig:dpor2:rec:calls}(b). We start the new branch from that
position ($\astate_1$), rather than from the position where
$\anevent_r$ was committed when the race was detected (i.e.,
$\astate_7$). The new branch $\arun_4$ is constructed when the race is
detected on
lines~\ref{ln:dpor2:cut:branch:begin}-\ref{ln:dpor2:cut:branch:end} in
\newalgodetectrace{}, by restricting the sub-run
$\arun_2.\inst{\hat{\anevent}_w}{\param}.\inst{\anevent_r}{\hat{\anevent}_w}.\arun_3$
to events that $\fcb$-precede the store $\dhat{\anevent}_w$.

The reason for returning all the way up to $\astate_1$, rather than
starting the new branch at $\astate_7$, is to avoid exploring multiple
runs corresponding to the same trace. This could otherwise happen
when the same race is detected in multiple runs.
To see this happen, let us consider the program given in
Fig.~\ref{fig:redundant:branches:prog}. A part of its exploration
tree is given in Fig.~\ref{fig:redundant:branches:tree}.
In the interest of brevity, when describing the exploration of the
program runs, we will ignore some runs which would be explored by the
algorithm, but which have no impact on the point of the example.
Throughout this example, we will use the labels \texttt{L0}, \texttt{L1},
and \texttt{L2} to identify the events corresponding to the labelled
instructions.
We assume that in the first run to be explored (the path from
$\initstate$ to $\astate_3$ in
Fig.~\ref{fig:redundant:branches:tree}), the load at \texttt{L0} is
committed first (loading the initial value of \xvar), then the stores
at \texttt{L1} and \texttt{L2}. There are two read-write races
in this run, from \texttt{L0} to \texttt{L1} and to \texttt{L2}. When
the races are detected, the branches \texttt{L1[\dporstar].L0[L1]} and
\texttt{L2[\dporstar].L0[L2]} will be added to
$\contmap[\textrm{\texttt{L0}}]$. These branches are later explored,
and appear in Fig.~\ref{fig:redundant:branches:tree} as the paths
from $\initstate$ to $\astate_6$ and from $\initstate$ to $\astate_9$ respectively.
In the run ending in $\astate_9$, we discover the race from
\texttt{L0} to \texttt{L1} again.
This indicates that a run should be explored where \texttt{L0} reads
from \texttt{L1}.
If we were to continue exploration from $\astate_7$ by delaying
\texttt{L0} until \texttt{L1} has been committed, we would follow the
path from $\astate_7$ to $\astate_{11}$ in
Fig.~\ref{fig:redundant:branches:tree}. In $\astate_{11}$, we have
successfully reversed the race between \texttt{L0} and
\texttt{L1}. However, the trace of $\astate_{11}$ turns out to be
identical to the one we already explored in $\astate_6$.
Hence, by exploring in this manner, we would end up exploring
redundant runs. The \newalgo{} algorithm avoids this redundancy by
exploring in the different manner described above: When the race from
\texttt{L0} to \texttt{L1} is discovered at $\astate_9$, we consider
the entire sub-run \texttt{L2[0].L0[L2].L1[1]} from $\initstate$, and
construct the new sub-run \texttt{L1[\dporstar].L0[L1]} by removing
all events that are not $\fcb$-before \texttt{L1}, generalizing the
parameter to \texttt{L1}, and by appending \texttt{L0[L1]} to the
result. The new branch \texttt{L1[\dporstar].L2[L1]} is added to
\texttt{$\contmap$[L0]}. But \texttt{$\contmap$[L0]} already contains
the branch \texttt{L1[\dporstar].L2[L1]} which was added at the
beginning of the exploration. And since it has already been explored
(it has already been added to the set \texttt{explored} at
line~\ref{ln:dpor2:add:explored}) we avoid exploring it again.

\begin{figure}
  \centering
  {
    \tt
    \begin{tabular}{l@{\;\;\;}l@{\;\;\;}l}
      thread $P$: & thread $Q$: & thread $R$: \\
      L0: $\areg$ := $\xvar$ & L1: $\xvar$ := 1 & L2: $\xvar$ := 2 \\
    \end{tabular}
  }
  \caption{A small program where one thread $P$ loads from $\xvar$,
    and two threads $Q$ and $R$ store to $\xvar$.}\label{fig:redundant:branches:prog}
\end{figure}

\begin{figure}
  \centering
  \begin{tikzpicture}
    \node (s0) at (0,0) [] {$\initstate$};
    \node (s1a) at ($(s0)+(-1,-1)$) [] {$\astate_1$};
    \node (s1b) at ($(s0)+(0,-1)$) [] {$\astate_4$};
    \node (s1c) at ($(s0)+(2,-1)$) [] {$\astate_7$};
    \draw[->,out=180,in=90] (s0) to node [left,name=ra] {\small{\texttt{L0[$\init{\xvar}$]}}} (s1a);
    \draw[->,line width=1pt] (s0) -- node [right] {\small{\texttt{L1[0]}}} (s1b);
    \draw[->,out=0,in=90,line width=1pt] (s0) to node [right] {\hspace{4pt}\small{\texttt{L2[0]}}} (s1c);
    \node (s2a) at ($(s1a)+(0,-1)$) [] {$\astate_2$};
    \node (s2b) at ($(s1b)+(0,-1)$) [] {$\astate_5$};
    \node (s2c) at ($(s1c)+(0,-1)$) [] {$\astate_8$};
    \draw[->] (s1a) -- node [left,name=w1a] {\small{\texttt{L1[0]}}} (s2a);
    \draw[->,line width=1pt] (s1b) -- node [right,name=rb] {\small{\texttt{L0[L1]}}} (s2b);
    \draw[->,line width=1pt] (s1c) -- node [right,name=rc] {\small{\texttt{L0[L2]}}} (s2c);
    \node (s3a) at ($(s2a)+(0,-1)$) [] {$\astate_3$};
    \node (s3b) at ($(s2b)+(0,-1)$) [circle,inner sep=1pt,draw=black] {$\astate_6$};
    \node (s3c) at ($(s2c)+(0,-1)$) [] {$\astate_9$};
    \draw[->] (s2a) -- node [left,name=w2a] {\small{\texttt{L2[0]}}} (s3a);
    \draw[->] (s2b) -- node [right,name=w2b] {\small{\texttt{L2[0]}}} (s3b);
    \draw[->] (s2c) -- node [right,name=w1c] {\small{\texttt{L1[1]}}} (s3c);
    \node (s2d) at ($(s2c)+(2,0)$) [] {$\astate_{10}$};
    \node (s3d) at ($(s2d)+(0,-1)$) [circle,inner sep=0pt,draw=black] {$\astate_{11}$};
    \draw[->,line width=1pt,out=0,in=90] (s1c) to node [above] {\small{\texttt{L1[1]}}} (s2d);
    \draw[->,line width=1pt] (s2d) -- node [right] {\small{\texttt{L0[L1]}}} (s3d);
    \draw[->,draw=red] (ra) to node [above,rotate=270] {\small{\color{red}race}} (w1a);
    \draw[->,draw=red,out=240,in=110] (ra) to node [above,rotate=75] {\small{\color{red}race}} (w2a.north west);
    % \draw[->,draw=red] (rb) -- node [right] {\small{\color{red}race}} (rb |- w2b.north);
    \draw[->,draw=red] (rc) -- node [right] {\small{\color{red}race}} (rc |- w1c.north);

    \node (exec) at (2,-3.5) [anchor=north,ellipse,draw=black,line width=1pt,inner sep=2pt] {
      \begin{tikzpicture}
        \node (l0) at (0,-5) [] {\texttt{L0}};
        \node (l1) at (1.5,-5) [] {\texttt{L1}};
        \node (l2) at (3,-5) [] {\texttt{L2}};
        \draw[->,line width=1pt] (l1) -- node [above] {\small{\texttt{rf}}} (l0);
        \draw[->,line width=1pt] (l2) -- node [above] {\small{\texttt{co}}} (l1);
      \end{tikzpicture}
    };
    \draw[-] (s3d) -- (exec);
    \draw[-] (s3b) -- (exec);
    \node (expl) at ($(exec.north east)$) [rectangle,draw=black,fill=white,line width=1pt,inner sep=2pt] {
      \scriptsize{$\exec{\astate_6}=\exec{\astate_{11}}$}};
  \end{tikzpicture}
  \caption{Part of a faulty exploration tree for the program above,
    containing redundant branches. The branches ending in
    $\astate_6$ and $\astate_{11}$ correspond to the same
    trace. The \newalgo{} algorithm avoids this redundancy, by
    the mechanism where all branches for read-write races
    from the same load $\anevent_r$ are collected in one set
    \texttt{$\contmap$[$\anevent_r$]}.
  }\label{fig:redundant:branches:tree}
\end{figure}
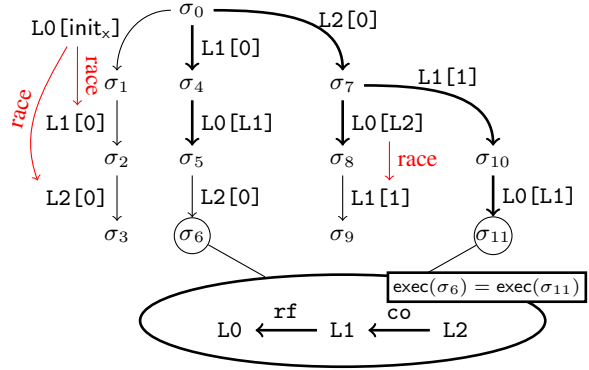

\bjparagraph{Soundness and Optimality.}
\label{sec:dpor-soundness}
We first establish soundness of the RSMC algorithm in
Fig.~\ref{fig:algo:dpor2} for the POWER memory model,
in the sense that it guarantees to  explore all Shasha-Snir traces of a program.
We thereafter establish that RSMC is optimal,
in the sense that it will never explore the same complete trace twice.

\begin{theorem}[Soundness]\label{thm:soundness2}
  Assume that $\fcb$ is valid w.r.t. $\mmodel$, and that $\mmodel$ and
  $\fcb$ are deadlock free.
  Then, for each $\anexec\in\axmodel{\aprog}{\mmodel}$, the evaluation
  of a call to $\newalgo(\emptyseq,\initstate)$ will contain a
  recursive call to $\newalgo(\arun,\astate)$ for some $\arun$,
  $\astate$ such that $\exec{\astate}=\anexec$.  \ifnotextended{\qed}
\end{theorem}

\begin{corollary}\label{cor:soundness2:power}
  RSMC is sound for POWER using $\mmodelpower$ and $\fcbpower$.
\end{corollary}

The proof of Theorem~\ref{thm:soundness2} involves showing that if an
allowed trace exists, then the races detected in previously explored
runs are sufficient to trigger the later exploration of a run
corresponding to that trace.

\begin{theorem}[Optimality for POWER]\label{thm:optimality:power}
  Assume that $\mmodel = \mmodelpower$ and  $\fcb=\fcbpower$.
  Let $\anexec \in \axmodel{\aprog}{\mmodel}$.  Then during the
  evaluation of a call to $\newalgo(\emptyseq,\initstate)$, there will
  be exactly one call $\newalgo(\arun,\astate)$ such that
  $\exec{\astate}=\anexec$.  \ifnotextended{\qed}
\end{theorem}

While the RSMC algorithm is optimal in the sense that it explores
precisely one complete run per Shasha-Snir trace, it may initiate
explorations that block before reaching a complete trace (similarly to
sleep set blocking in classical DPOR).
Such blocking may arise when the RSMC algorithm detects a read-write
race and adds a branch to \texttt{Q}, which upon traversal turns out
to be not allowed under the memory model.
Our experiments in Section~\ref{experimental-section} indicate that
the effect of such blocking is almost negligible, without any blocking
in most benchmarks, and otherwise at most 10\% of explored runs.

\begin{table}[t]
  \begin{tabular}{@{}m{.56\textwidth}@{}m{.42\textwidth}@{}}
  \small{
    \begin{tabular}{@{}l@{\rule{5pt}{0pt}}l@{\rule{5pt}{0pt}}l@{\;}r|r@{\;}r@{\;\;}r@{}}
      \multicolumn{6}{l}{\textbf{Tool running time (s), and trace count}}\\
      \hline
      & & & \multicolumn{1}{@{}r@{}}{\textsf{goto-}} & \multicolumn{3}{@{\rule{3pt}{0pt}}c@{\rule{3pt}{0pt}}}{} \\
      & & & \multicolumn{1}{@{}r@{}}{\textsf{instrument}} & \multicolumn{3}{@{\rule{3pt}{0pt}}c@{\rule{3pt}{0pt}}}{Nidhugg} \\
      \cmidrule(lr){4-4}\cmidrule(lr){5-7}
      & F & LB & \multicolumn{1}{r}{time} & time & SS & \multicolumn{1}{@{}r@{}}{B} \\
      \hline
      dcl\_singleton &  & 7 & *0.40 & *\textbf{0.13} & 3 & 0 \\
dcl\_singleton & y & 7 & 5.05 & \textbf{0.19} & 7 & 0 \\
dekker &  & 10 & *229.39 & *\textbf{0.11} & 5 & 0 \\
dekker & y & 10 & t/o & \textbf{0.76} & 246 & 0 \\
fib\_false &  &  & *\textbf{1.86} & t/o & 109171 & 0 \\
fib\_false\_join &  &  & *\textbf{0.84} & *35.46 & 11938 & 0 \\
fib\_true &  &  & \textbf{7.05} & t/o & 109122 & 0 \\
fib\_true\_join &  &  & \textbf{8.92} & 57.67 & 19404 & 0 \\
indexer &  & 5 & 68.16 & \textbf{1.57} & 19 & 0 \\
lamport &  & 8 & *635.45 & *\textbf{0.12} & 3 & 0 \\
lamport & y & 8 & t/o & \textbf{0.20} & 50 & 2 \\
parker &  & 5 & \sout{1.20} & *\textbf{0.13} & 5 & 0 \\
parker & y & 5 & \textbf{1.24} & 7.44 & 1126 & 0 \\
peterson &  &  & *0.24 & *\textbf{0.11} & 3 & 0 \\
peterson & y &  & 0.19 & \textbf{0.11} & 10 & 1 \\
pgsql &  & 8 & *161.05 & *\textbf{0.11} & 2 & 0 \\
pgsql & y & 8 & t/o & \textbf{0.58} & 16 & 0 \\
pgsql\_bnd &  &  & t/o & *\textbf{0.11} & 2 & 0 \\
pgsql\_bnd & y &  & t/o & t/o & 36211 & 0 \\
stack\_safe &  &  & \textbf{13.84} & 73.86 & 1005 & 0 \\
stack\_unsafe &  &  & *\textbf{1.03} & *3.32 & 20 & 0 \\
szymanski &  &  & *1.02 & *\textbf{0.11} & 17 & 0 \\
szymanski & y &  & 304.87 & \textbf{0.31} & 226 & 0 \\

    \end{tabular}
  }
  &
  \vspace{70pt}

  \caption{A comparison of running times (in seconds) for our
    implementation Nidhugg
    and \textsf{goto-instrument}. The \emph{F} column
    indicates whether fences have been inserted
    code to regain safety. The \emph{LB} column indicates whether the
    tools were instructed to unroll loops up to a certain bound. A
    \emph{t/o} entry means that the tool failed to complete within 900
    seconds. An asterisk (\emph{*}) means that the tool found a safety
    violation. A struck out entry means that the tool gave the wrong
    answer regarding the safety of the benchmark.
    The superior running time for each benchmark is given
    in bold font. The \emph{SS} column indicates the number of complete
    traces explored by
    Nidhugg before detecting an error, exploring all
    traces, or timing out. The \emph{B} (for ``blocking'') column indicates the
    number of incomplete runs that Nidhugg started to
    explore, but that turned out to be invalid.}
  \label{tbl:exp:time}
  \end{tabular}
\end{table}

\section{Experimental Results}\label{sec:experiments}
\label{experimental-section}

In order to evaluate the efficiency of our approach,
we have implemented it as a part of the open source tool Nidhugg~\cite{nidhugggithub},
for stateless model checking of C/pthreads programs under the
relaxed memory. It operates under the restrictions that
\begin{inparaenum}[(i)]
\item
  all executions are bounded
  by loop unrolling, and
\item
  the analysis runs on a given compilation
of the target C code.
\end{inparaenum}
The implementation uses RSMC to explore all allowed program
behaviors under POWER, and detects any assertion violation that can occur.
We validated the correctness of our implementation by
successfully running all 8070
relevant litmus tests published
with~\cite{alglave2014herding}.

The main goals of our experimental evaluation are
\begin{inparaenum}[(i)]
\item to show the feasibility and competitiveness
  of our approach, in particular to show for which programs it performs
  well,
\item
  to compare with
  \textsf{goto-instrument}, which to our knowledge is the only
other tool analyzing C/pthreads programs under POWER\footnote{The
  \textsf{cbmc} tool previously supported POWER~\cite{AlglaveKT13},
  but has withdrawn support in later versions.}, and
\item to show the effectiveness of our approach in terms of wasted
  exploration effort.
\end{inparaenum}

%% The implementation is available as open source at
%% \url{https://github.com/nidhugg/nidhugg}.
Table~\ref{tbl:exp:time} shows running times for Nidhugg and
\textsf{goto-instrument} for several benchmarks in C/pthreads.
All benchmarks were run on an 3.07 GHz Intel Core i7 CPU with 6 GB
RAM. We use \textsf{goto-instrument} version 5.1 with \textsf{cbmc}
version 5.1 as backend.

We note here that the comparison of running time is mainly relevant for
the benchmarks where \emph{no} error is detected (errors are indicated
with a * in Table~\ref{tbl:exp:time}). This is because when an error
is detected, a tool may terminate its analysis without searching the
remaining part of the search space (i.e., the remaining runs in our
case). Therefore the time consumption in such cases, is determined by
whether the search strategy was lucky or not.
This also explains why in e.g. the \textsf{dekker} benchmark, fewer
Shasha-Snir traces are explored in the version \emph{without} fences,
than in the version \emph{with} fences.

\altparagraph{Comparison with \textsf{goto-instrument}.}
\textsf{goto-instrument} employs code-to-code transformation in order
to allow verification tools for SC to work for more relaxed memory
models such as TSO, PSO and POWER~\cite{AlKNT13}.
The results in Table~\ref{tbl:exp:time} show that our technique is
competitive. In many cases Nidhugg significantly outperforms
\textsf{goto-instrument}.
The benchmarks for which
\textsf{goto-instrument} performs better than Nidhugg,
have in common that \textsf{goto-instrument} reports that no trace may
contain a cycle which indicates non-SC behavior.
This allows
\textsf{goto-instrument} to avoid expensive program instrumentation to
capture the extra program behaviors caused by memory consistency
relaxation.
While this treatment is very beneficial in some cases (e.g. for
\textsf{stack\_*} which is data race free and hence has no non-SC executions), it
also leads to false negatives in cases like \textsf{parker}, when
\mbox{\textsf{goto-instrument}} fails to detect
Shasha Snir-cycles that cause safety violations.
In contrast, our technique is precise, and will never miss any
behaviors caused by the memory consistency violation within the
execution length bound.

We remark that our approach is restricted to thread-wisely deterministic programs with fixed input data, whereas the  bounded model-checking used as a backend (CBMC) for \textsf{goto-instrument} can handle both concurrency and data nondeterminism.

\hide{
  \begin{table}
  %% Results 151119, @DeepThought, Nidhugg 0.1 (ddc5c6b, Release, with LLVM-3.6.2:Release)
  \centering
  \begin{tabular}{crr}
    Syncs & \textsf{herd} & Nidhugg \\
    \hline
    No  & 78.09  & 161.36 \\
    Yes & 925.95 & 0.01   \\
  \end{tabular}
  \caption{Running times in seconds under POWER for \textsf{herd} and
    Nidhugg on the litmus test in Fig.~\ref{fig:sb10w},
    with and without \textsf{sync}s.}
  \label{tbl:sb10w:results}
\end{table}
}

\begin{wrapfigure}[27]{r}{.4\linewidth}
  {\fontsize{8pt}{8pt}\selectfont
    \texttt{
      \begin{tabular}{@{}l@{\hspace{2pt}}l@{\;\;\;\;}l@{\hspace{2pt}}l@{}}
        \multicolumn{4}{@{}l}{\xvar = 0 \hspace{10pt} \yvar = 0 \hspace{10pt} \zvar = 0}\\
        \\
        \multicolumn{2}{@{}l}{thread $P$:} & \multicolumn{2}{l}{thread $Q$:}\\
        L0: & $\xvar$ := 1;              & M0: & $\yvar$ := 1;\\
        L1: & \sync;                     & M1: & \sync;\\
        L2: & $\areg_0$ := $\yvar$;      & M2: & $\areg_1$ := $\xvar$;\\
        L3: & if $\areg_0$ = 1           & M3: & if $\areg_1$ = 1\\
            & \hphantom{M}goto L14;     &     & \hphantom{M}goto M14;\\
        L4: & $\zvar$ := 1;          & M4: & $\zvar$ := 1;\\
        L5: & $\zvar$ := 1;          & M5: & $\zvar$ := 1;\\
        L6: & $\zvar$ := 1;          & M6: & $\zvar$ := 1;\\
        L7: & $\zvar$ := 1;          & M7: & $\zvar$ := 1;\\
        L8: & $\zvar$ := 1;          & M8: & $\zvar$ := 1;\\
        L9: & $\zvar$ := 1;          & M9: & $\zvar$ := 1;\\
        L10: & $\zvar$ := 1;         & M10: & $\zvar$ := 1;\\
        L11: & $\zvar$ := 1;         & M11: & $\zvar$ := 1;\\
        L12: & $\zvar$ := 1;         & M12: & $\zvar$ := 1;\\
        L13: & $\zvar$ := 1;         & M13: & $\zvar$ := 1;\\
        L14: & $\areg_0$ := 0;           & M14: & $\areg_1$ := 0;\\
      \end{tabular}
    }
  }
 \caption{\textsf{SB+10W+syncs}: A litmus test based on the idiom
   known as ``Dekker'' or ``SB''. It has 3 allowed Shasha-Snir traces
   under POWER. If the \textsf{sync} fences at lines \texttt{L1} and
   \texttt{M1} are removed, then it has 184759 allowed Shasha-Snir
   traces. This test is designed to have a large difference between
   the \emph{total} number of coherent Shasha-Snir traces and the
   number of \emph{allowed} Shasha-Snir traces.}
 \label{fig:sb10w}
\end{wrapfigure}

\altparagraph{Efficiency of Our Approach.}
While our RSMC algorithm is optimal,
in the sense that it explores
precisely one complete run per Shasha-Snir trace, it
may additionally start to explore runs that then turn out to block
before completing, as described in Section~\ref{sec:dpor}.
The SS and B columns of Table~\ref{tbl:exp:time} indicate that the
effect of such blocking is almost negligible, with no blocking in most
benchmarks, and at most 10\% of the runs.

A costly aspect of our approach is that every time a new event is
committed in a trace, Nidhugg will check which of its
possible parameters are allowed by the axiomatic memory model. This
check is implemented as a search for particular cycles in a graph over
the committed events. The cost is alleviated by the fact that
RSMC is optimal, and avoids exploring unnecessary traces.

To illustrate this tradeoff, we present the small program in
Fig.~\ref{fig:sb10w}. The first three lines of each thread implement
the classical Dekker idiom. It is impossible for both threads to read
the value 0 in the same execution. This property is used to implement
a critical section, containing the lines \texttt{L4}-\texttt{L13} and
\texttt{M4}-\texttt{M13}. However, if the fences at \texttt{L1} and
\texttt{M1} are removed, the mutual exclusion property can be
violated, and the critical sections may execute in an interleaved
manner. The program \emph{with} fences has only three allowed
Shasha-Snir traces, corresponding to the different observable
orderings of the first three instructions of both
threads. \emph{Without} the fences, the number rises to 184759, due to
the many possible interleavings of the repeated stores to \zvar.
The running time of
Nidhugg is $0.01$s with fences and $161.36$s without fences.

We compare this with the results of the litmus test checking tool
\textsf{herd}~\cite{alglave2014herding}, which operates by generating
all possible Shasha-Snir traces, and then checking which are allowed
by the memory model.
The running time of
\textsf{herd} on \textsf{SB+10W+syncs} is $925.95$s with fences and $78.09$s without fences.
Thus \textsf{herd} performs better than Nidhugg on the litmus
test without fences. This is because a large proportion of the
possible Shasha-Snir traces are allowed by the memory model. For each
of them \textsf{herd} needs to check the trace only once. On the other
hand, when the fences are added, the performance of \textsf{herd}
deteriorates. This is because \textsf{herd} still checks every
Shasha-Snir trace against the memory model, and each check becomes
more expensive, since the fences introduce many new dependency edges
into the traces.

We conclude that our approach is particularly superior for application
style programs with control structures, mutual exclusion primitives
etc., where relaxed memory effects are significant, but where most
potential Shasha-Snir traces are forbidden.

\section{Conclusions} \label{sec:conclusion}

We have presented the first framework for efficient application of SMC
to programs running under POWER.
Our framework combines solutions to several challenges.
We developed a scheme for systematically deriving
execution models that are suitable for SMC, from axiomatic ones.
We present RSMC, a novel algorithm for exploring all relaxed-memory traces
of a program, based on our derived execution model. We show that RSMC is sound for POWER,
meaning that it explores all Shasha-Snir
traces of a program, and optimal in the sense that
it explores the same complete trace exactly once.
The RSMC algorithm can
in some situations waste effort by exploring blocked runs, but our
experimental results shows that this is rare in practice.
Our implementation shows that the RSMC approach is competitive relative
to an existing state-of-the-art implementation.
We expect that RSMC will be sound also for other similar
memory models with suitably defined commit-before functions.

\paragraph{Related Work.}

Several SMC techniques have  been  recently developed for programs running under  the memory models TSO and PSO \cite{tacas15:tso,Zhang:pldi15,DBLP:conf/oopsla/DemskyL15}. In this work we  propose a novel and  efficient SMC technique for programs running under  POWER.

In~\cite{alglave2014herding}, a similar execution model was suggested, also based on the axiomatic semantics. However, compared to our semantics, it will lead many spurious executions that will be blocked by the semantics as they are found to be disallowed. This would cause superfluous runs to be explored, if used as a basis for stateless model checking.

Beyond SMC techniques for relaxed memory models, there have been many works related to the verification of programs running under relaxed memory models  (e.g.,~\cite{LNPVY12,KVY10,KVY11,DBLP:conf/sas/DanMVY13,abdulla2012counter,BM08,BurnimSS11,BouajjaniDM13,BAM07,YangGLS04}). Some of these works propose precise analysis techniques for finite-state programs under relaxed memory models (e.g.,~\cite{abdulla2012counter,BouajjaniDM13,DM14}). Others propose algorithms and tools for monitoring and testing programs running under relaxed memory models (e.g.,~\cite{BM08,BurnimSS11,BurSS11,LNPVY12,FlanaganF10}). Different techniques based on explicit state-space exploration  for the verification of programs running under relaxed memory models have also been developed during the last years (e.g.,~\cite{HuynhR07,ParkD99,KVY10,KVY11,DBLP:conf/tacas/LindenW13}).
There are also a number of efforts to design bounded model checking techniques for programs under  relaxed memory models (e.g.,~\cite{AlglaveKT13,YangGLS04,BAM07,TorlakVD10}) which encode the verification problem in SAT/SMT.
Finally, there are code-to-code transformation techniques
(e.g.,~\cite{AtigBP11,AlKNT13,BouajjaniDM13})
which reduce verification of a program under relaxed memory models to
verification of a transformed program under SC.  Most of these works
do not handle POWER.
In~\cite{DM14}, the robustness problem for POWER has been shown to be PSPACE-complete.

The closest works to ours were presented in~\cite{AlglaveKT13,AlKNT13,alglave2014herding}. The work~\cite{AlKNT13} extends  \textsf{cbmc} to work with relaxed memory models (such as TSO, PSO and POWER) using a code-to-code transformation. The work in~\cite{AlglaveKT13} develops a bounded model checking technique that can be applied to different memory models (e.g., TSO, PSO, and POWER). The
  \textsf{cbmc} tool previously supported POWER~\cite{AlglaveKT13}, but has withdrawn support in its later versions.  The  tool \textsf{herd}~\cite{alglave2014herding} operates
by generating all possible Shasha-Snir traces, and then for each one of them checking whether it is allowed by the memory model.
In Section~\ref{experimental-section}, we experimentally compare RSMC with the tools of~\cite{AlKNT13} and~\cite{alglave2014herding}.

%% \acks
%% This work was carried out within the Linnaeus centre of excellence
%% UPMARC (Uppsala Programming for Multicore Architectures Research
%% Center).

\bibliographystyle{splncs03}

\bibliography{biblio-last}

\ifextended{
\vfill\newpage
\appendix

\paragraph{Appendix Overview.}
These appendices contain formal definitions and proofs elided from the
main text.
\begin{description}
\item{Appendix~\ref{app:definitions}} contains formal definitions of
  some concepts used in the semantics (Section~\ref{sec:model}) and
  RSMC algorithm (Section~\ref{sec:dpor}).
\item{Appendix~\ref{app:proofs:model}} contains proofs of theorems
  about the execution model (Section~\ref{sec:model}): In particular
  the proof of equivalence between an execution model and the
  axiomatic model it is derived from, the proof of validity of
  $\fcbpower$, and the deadlock freedom of $\mmodelpower$ and
  $\fcbpower$.
\item{Appendix~\ref{app:rsmc:proofs}} contains proofs of theorems
  about the RSMC algorithm (Section~\ref{sec:dpor}): In particular the
  proof of soundness, and the proof of optimality.
\end{description}

\section{Additional Formal Definitions}\label{app:definitions}

Here we provide some formal definitions that were elided from the main
text.

\subsection{Additional Definitions for Section~\ref{sec:model}}\label{app:semantics}

\begin{figure}
  \begin{center}
    \texttt{
      \begin{tabular}{ll}
        \multicolumn{2}{l}{\xvar = 0}\\
        \multicolumn{2}{l}{\yvar = 0}\\
        \\
        thread $P$:                     & thread $Q$:\\
        L0: $\xvar$ := 1;  & L3: $\areg_1$ := $\yvar$;\\
        L1: \texttt{lwfence}; & L4: \texttt{ffence};\\
        L2: $\yvar$ := 1; & L5: $\areg_2$ := $\xvar$;\\
      \end{tabular}
    }
  \end{center}
  \caption{An example program: MP+lwfence+ffence.}
  \label{fig:prog::mp+lwfence}
\end{figure}

In the following, we introduce some notations and definitions following ~\cite{alglave2014herding} that are needed in order to define dependencies between events. We also give the formal definition of the  partial function $\val{\state}$ which gives the evaluation of an arithmetic expression.

Let $\astate=(\fetched,\events,\po,\co,\rf)\in\states$ be a state.
We define two  partial functions $\val{\astate}$ and $\adeps{\astate}$ over the set of events and arithmetic expression  so that  $\val{\state}(\anevent,\anexpr)$  is the value of the arithmetic expression $\anexpr$
when evaluated at the event $\anevent \in \fetched$ in the state $\state$, and $\adeps{\state}(\anevent,\anexpr)$  is the set of load events in $\fetched$ which are dependencies for the
evaluation of the arithmetic expression $\anexpr$ at the event $\anevent$. Here,  $\val{\astate}(\anevent,\anexpr)$ can be undefined ($\val{\astate}(\anevent,\anexpr)= \perp$)   when the value of $\anexpr$ at the event $\anevent$ depends
on the value of a load which is not yet executed. Formally, we define $\val{\state}(\anevent,\anexpr)$ and $\adeps{\state}(\anevent,\anexpr)$
recursively, depending on the type of arithmetic expression:

\begin{itemize}
\item If $\anexpr$ is
a literal integer $i$, then
$\val{\astate}(\anevent,\anexpr) = i$ and\\ $\adeps{\astate}(\anevent,\anexpr)=\emptyset$.
\item If $\anexpr = f(\anexpr_0,\cdots,\anexpr_n)$ for some arithmetic operator $f$ and
  subexpressions $\anexpr_0,\cdots,\anexpr_n$, then
$\val{\astate}(\anevent,\anexpr) = f(\val{\astate}(\anevent,\anexpr_0),\cdots,\val{\astate}(\anevent,\anexpr_n))$
and
$\adeps{\astate}(\anevent,\anexpr) = \bigcup_{i=0}^n\adeps{\astate}(\anevent,\anexpr_i)$.
\item If $\anexpr = \areg$ for some register \areg, then let
  $\anevent_{\areg}\in\fetched$ be the \po-greatest event such that
  $(\anevent_{\areg},\anevent)\in\po$ and either
  $\instrof{\anevent_{\areg}} = \areg\texttt{:=}\anexpr'$ or
  $\instrof{\anevent_{\areg}} = \areg\texttt{:=[}\anexpr'\texttt{]}$ for some
  expression $\anexpr'$.
  \begin{itemize}
  \item If there is no such event $\anevent_{\areg}$, then
    $\val{\astate}(\anevent,\anexpr) = 0$ and $\adeps{\astate}(\anevent,\anexpr)=\emptyset$.
  \item If $\instrof{\anevent_{\areg}} = \areg\texttt{:=}\anexpr'$, then
    $\val{\astate}(\anevent,\anexpr) = \val{\astate}(\anevent_{\areg},\anexpr')$ and
    $\adeps{\astate}(\anevent,\anexpr) = \adeps{\astate}(\anevent_{\areg},\anexpr')$.
  \item If $\instrof{\anevent_{\areg}} = \areg\texttt{:=[}\anexpr'\texttt{]}$ and
    $\anevent_{\areg}\in\events$ then let $\anevent_w\in\events$ be the event such
    that $(\anevent_w,\anevent_{\areg})\in\rf$. Let $\anexpr''$, $\anexpr'''$ be the arithmetic
    expressions s.t.
    $\instrof{\anevent_w}=\texttt{[}\anexpr''\texttt{]:=}\anexpr'''$. Now we
    define $\val{\state}(\anevent,\anexpr) = \val{\anexec}(\anevent_w,\anexpr''')$ and
    $\adeps{\state}(\anevent,\anexpr) = \{\anevent_{\areg}\}$.
  \item If $\instrof{\anevent_{\areg}} = \areg\texttt{:=[}\anexpr'\texttt{]}$ and
    $\anevent_{\areg}\not\in\events$ then $\val{\astate}(\anevent,\anexpr) = \perp$ and \\
    $\adeps{\astate}(\anevent,\anexpr) = \{\anevent_{\areg}\}$.
  \end{itemize}
\end{itemize}

We overload the function $\adeps{\state}$ for event arguments:

\begin{itemize}
\item If $\instrof{\anevent}=\areg\texttt{:=}\anexpr$, then
  $\adeps{\state}(\anevent)=\adeps{\state}(\anevent,\anexpr)$.
\item If $\instrof{\anevent}=\texttt{if }\anexpr\texttt{ goto }l$, then
  $\adeps{\state}(\anevent)=\adeps{\state}(\anevent,\anexpr)$.
\item If $\instrof{\anevent}=\areg\texttt{:=[}\anexpr\texttt{]}$, then
  $\adeps{\state}(\anevent)=\adeps{\state}(\anevent,\anexpr)$.
\item If $\instrof{\anevent}=\texttt{[}\anexpr\texttt{]:=}\anexpr'$, then
  $\adeps{\state}(\anevent)=\adeps{\state}(\anevent,\anexpr)\cup\adeps{\state}(\anevent,\anexpr')$.
\item If $\instrof{\anevent}\in\{\ffence,\lwfence,\cfence\}$, then
  $\adeps{\state}(\anevent) = \emptyset$.
\end{itemize}

\begin{figure}
  \begin{center}
    \texttt{
      \begin{tabular}{ll}
        \multicolumn{2}{l}{\xvar = 0}\\
        \multicolumn{2}{l}{\yvar = 0}\\
        \\
        thread $P$: & thread Q\\
        L0: $\areg_0$ := $\xvar$; & L3: $\areg_1$ := $\xvar$; \\
        L1: if $\areg_0$ = 1 goto L0; & L4: [$\areg_1$] := 1\\
        L2: $\xvar$ := $1$; & L5: $\areg_2$ := $\yvar$; \\
      \end{tabular}
    }
  \end{center}
  \caption{A program with   address and  control dependencies.}
  \label{fig:prog:term:loop}
\end{figure}

We also define the address  dependency relation $\addrdep{\state}\subseteq(\fetched\times\fetched)$ 
to capture how events depend
on earlier loads for the computation of their address. For  a memory access event $\anevent$ with $\instrof{\anevent} $ is of the form $\texttt{[}\anexpr\texttt{]:=}\anexpr'$ or $\areg\texttt{:=} \texttt{[}\anexpr\texttt{]}$, we have $(\anevent',\anevent)\in\addrdep{\state}$ for any event $\anevent'\in\adeps{\state}(\anevent,\anexpr)$. For instance, in the example described in Figure~\ref{fig:prog:term:loop}, there is an address dependency between   the load \texttt{L3}  and   the store \texttt{L4}.

We define the data  dependency relation
$\datadep{\state}\subseteq(\fetched\times\fetched)$
to capture how events depend
on earlier loads for the computation of their data. 
For  an event $\anevent$ with $\instrof{\anevent} $ is of the form $\areg\texttt{:=}\anexpr$, $\texttt{if }\anexpr\texttt{ goto }l$ or
  $\texttt{[}\anexpr'\texttt{]:=}\anexpr$, we have $(\anevent',\anevent)\in\datadep{\state}$  for any event  $\anevent'\in\adeps{\state}(\anevent,\anexpr)$.  For instance, in the example described in Figure~\ref{fig:prog:LB:data}, there is a data dependency between  the load \texttt{L0}  and  the store \texttt{L1}.

We define the relation $\ctrldep{\state}\subseteq \fetched \times \fetched$ to
capture how the control flow to an event depends on earlier loads. For
two events $\anevent\in\fetched$ and $\anevent'\in\fetched$ we have
$(\anevent,\anevent')\in\ctrldep{\anexec}$ iff $\instrof{\anevent}=\areg\texttt{:=[}\anexpr'\texttt{]}$ (i.e., $\anevent$ is a load event)  and there is a branch event $\anevent_b$ with
$\instrof{\anevent_b}=\texttt{if }\anexpr\texttt{ goto }l$ for some arithmetic
expression $\anexpr$ and label $l$ such that $(\anevent,\anevent_b),(\anevent_b,\anevent')\in\po$ and
$\anevent\in\adeps{\state}(\anevent_b,\anexpr)$. In the example given in Figure~\ref{fig:prog:term:loop}, there is a control  dependency between the load   \texttt{L0}  and   the store \texttt{L2}.

We define the relation $\poloc{\state}\subseteq  \fetched \times \fetched$ to
capture the program order between accesses to the same memory
location:
$\poloc{\state}=\{(\anevent,\anevent')\in\po | \addr{\state}(\anevent) = \addr{\state}(\anevent')\,\wedge \,\addr{\state}(\anevent)\neq \perp\}$.  In the example described in Figure~\ref{fig:prog:term:loop}, the pair (\texttt{L0}, \texttt{L2}) is in   $\poloc{\state}$.

Finally, we define the relations
$\ffencedep{\astate},\lwsyncdep{\astate}\subseteq\fetched\times\fetched$
that contain the set of pairs of events that are separated by the
fence instruction $\ffence$ and $\lwfence$ respectively.
For instance,   in the example described in Figure~\ref{fig:prog::mp+lwfence},  $\ffencedep{\state}$ and $\lwsyncdep{\state}$ will contain the pairs $(\texttt{L3},\texttt{L5})$ and $(\texttt{L0},\texttt{L2})$, respectively.
We then define
$\lwfencedep{\astate}=\{(\anevent,\anevent')\in\lwsyncdep{\astate}| \neg(\textrm{$\anevent$ is a store, and $\anevent'$ is a load})\}$,
corresponding to the intuition that the order between a store and a
later load is not enforced by an $\lwfence$ under POWER.

\subsection{Additional Definitions for Section~\ref{sec:dpor}}\label{app:aux:dpor2}

\paragraph{Definition of the $\cutrun$ Function}

In order to define the $\cutrun$ function, we need to define an
auxiliary function $\cutrun'$. We then define
$\cutrun(\arun,\anevent,\astate)=\cutrun'(\arun,\anevent,\astate,\lambda{}a.\emptyset)$.

The function $\cutrun'(\arun,\anevent,\astate,W)$ works by recursively
traversing $\arun$ and removing each event which is not
$\cb{\astate}$-before $\anevent$. While doing so, for each store
$\inst{\anevent_w}{n}$ that is removed, the parameter $n$ is stored in
$W(\addr{\astate}(\anevent_w))$. When a store $\inst{\anevent_w}{n}$
is retained in the run, its parameter $n$ is updated to reflect that
all the preceding stores with parameters
$W(\addr{\astate}(\anevent_w))$ have disappeared.
Formally, the function $\cutrun'$ is defined as follows:

\noindent
\begin{math}
  \\
  \cutrun'(\emptyseq,\anevent,\astate,W) = \emptyseq\\
  \\
  \cutrun'(\inst{\anevent_0}{\param_0}.\arun,\anevent,\astate,W) =\\
  \left\{
  \begin{array}{@{}l@{\;\;}l@{\;\;}l@{}}
    \inst{\anevent_0}{\param_0}.\cutrun'(\arun,\anevent,\astate,W) & \textrm{if} & \anevent_0\in\R\wedge(\anevent_0,\anevent)\in\cb{\astate}\\
    \cutrun'(\arun,\anevent,\astate,W) & \textrm{if} & \anevent_0\in\R\wedge(\anevent_0,\anevent)\not\in\cb{\astate}\\
    \inst{\anevent_0}{\param_0'}.\cutrun'(\arun,\anevent,\astate,W) & \textrm{if} & \anevent_0\in\W\wedge(\anevent_0,\anevent)\in\cb{\astate} \\
    & \textrm{where} & \param_0' = \param_0-|\{i\in A | i < \param_0\}|\\
    & \textrm{where} & A = W(\addr{\astate}(\anevent_0))\\
    \cutrun'(\arun,\anevent,\astate,W') & \textrm{if} & \anevent_0\in\W\wedge(\anevent_0,\anevent)\not\in\cb{\astate}\\
    & \textrm{where} & W' = W[a\hookleftarrow W(a)\cup\{\param_0\}]\\
    & \textrm{where} & a = \addr{\astate}(\anevent_0)\\
  \end{array}
  \right.
\end{math}

\section{Proofs for Section~\ref{sec:model}}\label{app:proofs:model}

Here we provide proofs for the various theorems appearing in
Section~\ref{sec:model}.

\subsection{Proof of Theorem~\ref{thm:equiv:oper:axiom} (Equivalence of Semantics)}\label{app:proof:equiv:oper:axiom}

\begin{proof}[Proof of theorem~\ref{thm:equiv:oper:axiom}]
  We prove first that
  $\setcomp{\exec{\stateoafter{\exseq}}}{\exseq\in\exmodel{\aprog}{\mmodel}{\fcb}}\subseteq\axmodel{\aprog}{\mmodel}$. This
  follows directly from the fact that every rule in the operational
  semantics checks the new state against $\mmodel$ before allowing the
  transition, and from $\mmodel(\initstate)$.

  We turn instead to proving the other direction
  $\axmodel{\aprog}{\mmodel}\subseteq\setcomp{\exec{\stateoafter{\exseq}}}{\exseq\in\exmodel{\aprog}{\mmodel}{\fcb}}$. Let
  $\anexec=(\events,\po,\co,\rf)$ be an execution in
  $\axmodel{\aprog}{\mmodel}$. Let
  $\astate=(\lblcur,\events,\events,\po,\co,\rf)$ where $\lblcur$ maps
  every thread to its final state be the complete state corresponding
  to $\anexec$. From the assumption that $\fcb$ is valid
  w.r.t. $\mmodel$, we know that $\cb{\astate}$ is acyclic. Let
  $\arun$ be some linearization w.r.t. $\cb{\astate}$ of the memory
  access events in $\events$, instantiated with parameters according
  to $\co$ and $\rf$. We will show that $\arun$ is a run in
  $\exmodel{\aprog}{\mmodel}{\fcb}$ such that
  $\stateoafter{\arun}=\astate$.

  Let
  $\arun=\inst{\anevent_1}{\param_1}.\inst{\anevent_2}{\param_2}\cdots\inst{\anevent_n}{\param_n}$,
  and for every $0\leq{}i\leq{}n$, let $\arun_i$ denote the prefix
  $\inst{\anevent_1}{\param_1}\cdots\inst{\anevent_i}{\param_i}$ of
  $\arun$. In the case that $\arun_i$ is a run (it doesn't block), let
  $\astate^i=(\lblcur_i,\fetched_i,\events_i,\po_i,\co_i,\rf_i)=\stateoafter{\arun_i}$
  for all $0\leq{}i\leq{}n$.

  We will prove by induction that for all $0\leq{}i\leq{}n$ it holds
  that $\arun_i$ is a run, and $\astate^i\cbextends{\fcb}\astate$ and
  the restriction of $\events_i$ to memory access events is the set of
  events in $\arun_i$.

  \paragraph{Base case ($i=0$):}

  From the definition of runs, we see that $\arun_0=\emptyseq$ is a
  run if there is a state $\astate^0$ such that
  $\initstate\xrightarrow{\FLBmax}\astate^0$. This holds vacuously by
  the definition of $\xrightarrow{\FLBmax}$.
  Furthermore, we see from the definition of $\xrightarrow{\FLBmax}$
  that $\fetched_0$ will consist of all events that can be fetched
  without committing any branch which depends on a load. The same
  events must necessarily be fetched in any complete state, and
  therefore we have $\fetched_0\subseteq\fetched$. We see that
  $\events_0$ consists of all local instructions that do not depend on
  any memory access. For the same reason, the same events must also be
  committed in any complete state. And so we have
  $\events_0\subseteq\events$. It follows similarly that
  $\po_0\subseteq\po$. Since no memory access events have been
  committed, $\co_0=\emptyset\subseteq\co$ and
  $\rf_0=\emptyset\subseteq\rf$. Hence we have
  $\astate^0\cbextends{\fcb}\astate$. No memory access events have
  been committed by $\xrightarrow{\FLBmax}$, and no memory access
  events appear in $\arun_0=\emptyseq$, and so the restriction of
  $\events_i$ to memory access events is the set of events in
  $\arun_i$.

  \paragraph{Inductive case ($0<i+1\leq{}n$):}

  We assume as inductive hypothesis that $\arun_i$ is a run, and
  $\astate^i\cbextends{\fcb}\astate$ and the restriction of
  $\events_i$ to memory access events is the set of events in
  $\arun_i$ for some $0\leq{}i<n$.

  We know that $\anevent_{i+1}\in\fetched_i$, since all earlier branch
  events in $\fetched_i$ have been committed (in $\events_i$) by
  $\xrightarrow{\FLBmax}$ and all loads that they depend on have been
  committed (notice
  $(\anevent_l,\anevent_{i+1})\in\ctrldep{\astate}\subseteq\cbzero{\astate}\subseteq\cb{\astate}$
  for all such loads $\anevent_l$) with the same sources as in
  $\astate$. Since the restriction of $\events_i$ to memory access
  events is the set of events in $\arun_i$, we know that
  $\anevent_{i+1}\not\in\events_{i+1}$. To show that
  $\committable_{\astate^i}(\anevent_{i+1})$ holds, it remains to show
  that for all events $\anevent$ such that
  $(\anevent,\anevent_{i+1})\in\cb{\astate^i}$ it holds that
  $\anevent\in\events_i$. This follows from the monotonicity of $\fcb$
  and $\mmodel$ as follows: Since $\astate^i\cbextends{\fcb}\astate$,
  we have $\cb{\astate^i}\subseteq\cb{\astate}$. Since $\arun$ is a
  linearization of $\cb{\astate}$, for any $\anevent$ with
  $(\anevent,\anevent_{i+1})\in\cb{\astate^i}$ it must hold that
  either $\anevent$ is a memory access, and then precedes
  $\anevent_{i+1}$ in $\arun$ and is therefore already committed in
  $\astate^i$, or $\anevent$ is a local event which depends only on
  memory accesses that similarly precede $\anevent_{i+1}$, in which
  case $\anevent$ has been committed by $\xrightarrow{\FLBmax}$. Hence
  we have $\committable_{\astate^i}(\anevent_{i+1})$.

  We now split into cases, depending on whether $\anevent_{i+1}$ is a
  store or a load.

  Assume first that $\anevent_{i+1}$ is a store. From the construction
  of $\arun$, we know that the parameter $\param_{i+1}$ is a coherence
  position chosen such that $\param_{i+1}=\position_{\co_{i+1}}$ for
  some coherence order
  $\co_{i+1}\in\extend_{\astate^i}(\anevent_{i+1})$ such that
  $\anevent_{i+1}$ is ordered with the previous stores in $\events_i$
  in the same order as in $\co$.
  In order to show that the rule
  $\xrightarrow{\inst{\anevent_{i+1}}{\param_{i+1}}}$ applies, we must
  first show that $\committable_{\astate^i}(\anevent_{i+1})$ holds.
  The rule $\xrightarrow{\inst{\anevent_{i+1}}{\param_{i+1}}}$ in the
  operational semantics will produce a state
  $\astate'=(\lblcur_i,\fetched_i,\events_i\cup\{\anevent_{i+1}\},\po_i,\co_{i+1},\rf_i)$,
  and then check whether $\mmodel(\exec{\astate'})$ holds. In order to
  show that $\arun_{i+1}$ is a run, we need to show that
  $\mmodel(\exec{\astate'})$ does indeed hold. Since we have
  $\astate^i\cbextends{\fcb}\astate$, and $\anevent_{i+1}$ was chosen
  for $\arun$ from the committed events in $\astate$, and $\co_{i+1}$
  orders $\anevent_{i+1}$ with $\events_i$ in the same way as $\co$,
  we also have $\astate'\cbextends{\fcb}\astate$. Then by the
  monotonicity of the memory model $\mmodel$, we have
  $\mmodel(\exec{\astate'})$. Hence $\arun_{i+1}$ is a run, and
  $\astate^{i+1}=\stateoafter{\arun_{i+1}}$ for some $\astate^{i+1}$
  such that $\astate'\xrightarrow{\FLBmax}\astate^{i+1}$. By the same
  argument as in the base case, it then follows that
  $\astate^{i+1}\cbextends{\fcb}\astate$ and the restriction of
  $\events_{i+1}$ to memory access events is the set of events in
  $\arun_{i+1}$.

  Assume next that $\anevent_{i+1}$ is a load. From the construction
  of $\arun$, we know that the parameter $\param_{i+1}$ is the store
  event $\anevent_w$ such that $(\anevent_w,\anevent_{i+1})\in\rf$.
  Since $\arun$ is a linearization of $\cb{\astate}$, and
  $(\anevent_w,\anevent_{i+1})\in\rf\subseteq\cb{\astate}$, we know
  that $\anevent_w$ appears before $\anevent_{i+1}$ in
  $\arun_i$. Therefore we know that $\anevent_w$ is already committed
  in $\astate^i$, and that it is therefore available as a parameter
  for the load $\anevent_{i+1}$.
  Since all loads that precede $\anevent_w$ and $\anevent_{i+1}$ in
  $\cb{\astate^i}$ have been committed in the same way as in $\astate$
  we know that the addresses accessed by $\anevent_w$ and
  $\anevent_{i+1}$ are computed in the same way in $\astate^i$ as in
  $\astate$, and therefore we have
  $\addr{\astate^i}(\anevent_w)=\addr{\astate^i}(\anevent_{i+1})$.
  The rule $\xrightarrow{\inst{\anevent_{i+1}}{\param_{i+1}}}$ in the
  operational semantics will produce a state
  $\astate'=(\lblcur_i,\fetched_i,\events_i\cup\{\anevent_{i+1}\},\po_i,\co_i,\rf_i\cup\{(\anevent_w,\anevent_{i+1})\})$,
  and then check whether $\mmodel(\exec{\astate'})$ holds. In order to
  show that $\arun_{i+1}$ is a run, we need to show that
  $\mmodel(\exec{\astate'})$ does indeed hold. Since we have
  $\astate^i\cbextends{\fcb}\astate$, and $\anevent_{i+1}$ was chosen
  for $\arun$ from the committed events in $\astate$, and $\anevent_w$
  was chosen such that $(\anevent_w,\anevent_{i+1})\in\rf$, we also
  have $\astate'\cbextends{\fcb}\astate$. Then by the monotonicity of
  the memory model $\mmodel$, we have
  $\mmodel(\exec{\astate'})$. Hence $\arun_{i+1}$ is a run, and
  $\astate^{i+1}=\stateoafter{\arun_{i+1}}$ for some $\astate^{i+1}$
  such that $\astate'\xrightarrow{\FLBmax}\astate^{i+1}$. By the same
  argument as in the base case, it then follows that
  $\astate^{i+1}\cbextends{\fcb}\astate$ and the restriction of
  $\events_{i+1}$ to memory access events is the set of events in
  $\arun_{i+1}$.

  This concludes the inductive sub-proof.

  Since $\arun_n=\arun$ is a run, and
  $\astate^n=\stateoafter{\arun}\cbextends{\fcb}\astate$, and the
  committed memory access events in $\stateoafter{\arun}$ are the same
  as in $\astate$, and $\astate$ is a complete state, we have that
  $\stateoafter{\arun}=\astate$. Then $\arun$ must also be complete,
  and hence we have $\arun\in\exmodel{\aprog}{\mmodel}{\fcb}$. This
  concludes the proof.\qed
\end{proof}

\subsection{Proof of Theorem~\ref{thm:power:valid} (Validity of $\fcbpower$)}\label{app:power:valid:proof}

\begin{proof}[Proof of Theorem~\ref{thm:power:valid}]
  Monotonicity is proven in
  Lemma~\ref{lemma:power:monotonic}. Acyclicity is proven in
  Lemma~\ref{lemma:power:acyclic}. That
  $\cbzero{\astate}\subseteq\cbpower{\astate}$ for any state $\astate$
  follows directly from the definition of $\fcbpower$.\qed
\end{proof}

\subsubsection{Monotonicity of POWER}

\begin{lemma}\label{lemma:power:monotonic}
  $\fcbpower$ is monotonic w.r.t. $\mmodelpower$.
\end{lemma}

\begin{proof}[Proof of Lemma~\ref{lemma:power:monotonic}]
  Assume that
  $\fcb=\fcbpower$ and $\mmodel=\mmodelpower$. Let
  $\astate=(\lblcur,\fetched,\events,\po,\co,\rf)$ and
  $\astate'=(\lblcur',\fetched',\events',\po',\co',\rf')$ be two
  states such that $\astate\cbextends{\fcb}\astate'$.

  We prove first condition (i): that if $\mmodel(\astate')$ then
  $\mmodel(\astate)$.
  To see this, we need to study the definition of the POWER
  axiomatic memory models as given in~\cite{alglave2014herding}. We
  see that an execution is allowed by the axiomatic memory model,
  unless it contains certain cycles in the relations between
  events. All such forbidden cycles are constructed from some
  combination of the following relations: $\poloc{}$, $\co$, $\rf$,
  $\fr$, $\addrdep{}$, $\datadep{}$, $\fre$, $\rfee$, $\rfi$,
  $\ctrl+\cfence$, $\coe$, $\ctrl$, $\addrdep{};\po$, $\ffence$,
  $\lwfence$. The construction of the forbidden cycles is such that
  adding more relations between events can never cause a forbidden
  cycle to disappear.
  Studying these relations one by one, we see that for each of them,
  the relation in $\astate$ is a subset of the relation in $\astate'$.
  We discuss here only one of the more interesting cases: $\poloc{}$.
  Consider two events $\anevent$ and $\anevent'$ which are
  committed in $\astate$, and where
  $(\anevent,\anevent')\in\poloc{\astate}$. The same events must also
  be committed in $\astate'$, and be ordered in the same way in
  program order in $\astate'$ as in $\astate$. Therefore we must argue
  that $\anevent$ and $\anevent'$ both access the same memory location
  in $\astate'$ as in $\astate$. This follows from the fact that the
  set of committed events $\events$ in $\astate$ is
  $\cb{\astate'}$-closed. Since $\fcbpower$ contains all three of
  $\addrdep{}$, $\datadep{}$ and $\rf$, the computation of the address
  in $\anevent$ and $\anevent'$ must produce the same value in
  $\astate'$ as in $\astate$. Hence we have
  $(\anevent,\anevent')\in\poloc{\astate'}$.
  Since all of the relations participating in forbidden cycles in
  $\astate$ are subsets of the corresponding relations in $\astate'$,
  we know that any forbidden cycle in $\astate$ must also be in
  $\astate'$. Therefore
  $\neg\mmodel(\astate)\Rightarrow\neg\mmodel(\astate')$. The
  contra-positive gives us condition (i).

  We turn now to condition (ii): that
  $\cb{\astate}\subseteq\cb{\astate'}$. We will show that any edge in
  $\cb{\astate}$ is also in $\cb{\astate'}$.
  From the definition of $\fcbpower$, we know that
  $\cb{\astate}$ is the transitive irreflexive closure of the union of
  the following relations: $\addrdep{\astate}$, $\datadep{\astate}$,
  $\ctrldep{\astate}$, $\rf$, $(\addrdep{\astate};\po)$,
  $\poloc{\astate}$, $\ffencedep{\astate}$, $\lwsyncdep{\astate}$.
  We will consider an arbitrary edge
  $(\anevent,\anevent')\in\cb{\astate}$ which is in one of those
  relations, and show that $(\anevent,\anevent')$ is also in the
  corresponding relation in $\astate'$.
  If $(\anevent,\anevent')$ is in $\addrdep{\astate}$ or
  $\datadep{\astate}$, then $\anevent'$ uses the value in a register
  provided by the program order-earlier event $\anevent$. We know that
  in the extended state $\astate'$, the same relation persists.
  This is because any new event $\anevent''$ which might appear in
  $\astate'$ and which breaks the data-flow from $\anevent$ to
  $\anevent'$ must be between $\anevent$ and $\anevent'$ in program
  order. This would contradict the assumption that $\fetched$ is a
  $\po'$-closed subset of $\fetched'$.
  The case when $(\anevent,\anevent')\in(\addrdep{\astate};\po)$
  follows similarly. The case $(\anevent,\anevent')\in\poloc{\astate}$
  was covered in the proof for condition (i) above. In all of the
  remaining cases, $\ctrldep{\astate}$, $\lwsyncdep{\astate}$,
  $\ffencedep{\astate}$, there is some event $\anevent''$ (a branch or
  some fence) which comes between $\anevent$ and $\anevent'$ in
  program order in $\astate$. Since we have
  $\fetched\subseteq\fetched'$ and $\po=\po'|_{\fetched}$, the same
  event must also appear in $\astate$, and cause the same relation
  between $\anevent$ and $\anevent'$.

  Finally we turn to proving condition (iii): that for all
  $\anevent\in\fetched$ such that either
  $\committable_{\astate}(\anevent)$ and $\anevent\not\in\events'$ or
  $\anevent\in\events$, we have
  $(\anevent',\anevent)\in\cb{\astate}\Leftrightarrow(\anevent',\anevent)\in\cb{\astate'}$
  for all $\anevent'\in\fetched'$.
  We have already shown that $\cb{\astate}\subseteq\cb{\astate'}$. So
  we have
  $(\anevent',\anevent)\in\cb{\astate}\Rightarrow(\anevent',\anevent)\in\cb{\astate'}$
  for all $\anevent,\anevent'\in\fetched'$.
  It remains to show that for any event $\anevent$ which is either
  enabled or committed in $\astate$, and which does not become
  committed when extending $\anevent$ to $\anevent'$, we have
  $(\anevent',\anevent)\in\cb{\astate'}\Rightarrow(\anevent',\anevent)\in\cb{\astate}$
  for all $\anevent'\in\fetched'$, i.e., that no additional incoming
  $\fcb$ edges to $\anevent$ appear in $\astate'$ which are not in
  $\astate$. Let $\anevent$ be such an event. If any new incoming
  $\fcb$ edge to $\anevent$ has appeared in $\astate'$, then there
  must be an event $\anevent'\in\fetched'$ such that
  $(\anevent',\anevent)$ is in one of the relations making up
  $\cb{\astate'}$, i.e.: $\addrdep{\astate'}$, $\datadep{\astate'}$,
  $\ctrldep{\astate'}$, $\rf'$, $(\addrdep{\astate'};\po')$,
  $\poloc{\astate'}$, $\ffencedep{\astate'}$, or $\lwsyncdep{\astate'}$.
  We will show that in each case $(\anevent',\anevent)$ is also in the
  corresponding relation in $\astate$.
  Since $\fetched$ is a $\po'$-closed subset of $\fetched'$, all
  events which program order-precede $\anevent$ must be fetched in
  $\astate$. Hence the data-flow forming address or data dependencies
  in $\astate'$ are already visible in $\astate$. So if
  $(\anevent',\anevent)$ is in $\addrdep{\astate'}$ or
  $\datadep{\astate'}$, it must also be in $\addrdep{\astate}$ or
  $\datadep{\astate}$. The cases when $(\anevent',\anevent)$ is in
  $\ctrldep{\astate'}$, $(\addrdep{\astate'};\po')$,
  $\ffencedep{\astate'}$ or $\lwsyncdep{\astate'}$ are similar. If
  $(\anevent',\anevent)\in\rf'$, then $\anevent$ must be committed in
  $\astate'$, since read-from edges are only added upon committing. By
  assumption we have either $\committable_{\astate}(\anevent)$ and
  $\anevent\not\in\events'$ or $\anevent\in\events$. Therefore we must
  have $\anevent\in\events$. We have $\rf=\rf'|_{\events}$ since
  $\astate'$ is a $\fcb$-extension of $\astate$. Therefore we have
  $(\anevent',\anevent)\in\rf$.
  The remaining case is when
  $(\anevent',\anevent)\in\poloc{\astate'}$. Again, since $\fetched$
  is $\po'$-closed, we have $\anevent'\in\fetched$. Since $\anevent$
  is either enabled or committed in $\astate$, its address must be
  computed in $\astate$. It remains to show that the address of
  $\anevent'$ is also computed in $\astate$. If, for a contradiction,
  the address of $\anevent'$ is not computed in $\astate$, then there
  exists another event $\anevent''$ such that
  $(\anevent'',\anevent')\in\addrdep{\astate}$, and
  $\anevent''\not\in\events$. However, then we have
  $(\anevent'',\anevent)\in(\addrdep{\astate};\po)\subseteq\cb{\astate}$. Since
  $\anevent''$ is not committed in $\astate$, this would contradict
  the assumption that $\anevent$ is enabled or committed in $\astate$
  and $\events$ is $\cb{\astate'}$-closed.
  This concludes the proof.\qed
\end{proof}

\subsubsection{Acyclicity Proof for $\fcbzero$ and $\fcbpower$ under POWER.}

In the following we prove that the commit-before functions $\fcbzero$
and $\fcbpower$ are acyclic in states that are allowed under POWER.

\begin{lemma}\label{lemma:power:acyclic}
  For any state $\astate$ such that $\mmodelpower(\exec{\astate})$,
  the relations $\cbzero{\astate}$ and $\cbpower{\astate}$ are
  acyclic.
\end{lemma}

From the definition of the commit-before functions, we see that $(\addrdep{\state} \cup \datadep{\state} \cup \ctrldep{\state}) \subseteq \cbzero{\state} \subseteq \cbpower{\astate}$. Since $\fcbpower$ the strongest one. It is sufficient to only prove the acyclicity of  $\fcbpower$ w.r.t. to POWER.   Here we assume that the POWER  memory model is defined in the way described in the Herding Cats paper \cite{alglave2014herding}. We will assume that the reader is familiar with the notations and definitions used in   \cite{alglave2014herding}.

Define the sets $\R, \W, \M\subseteq\eventtype$ as the sets of load
events, store events and memory accesses events respectively: \\
$\R = \{e\in\eventtype | \exists \areg,a . \instrof{e} = \areg\texttt{:=[}a\texttt{]}\}$ and\\
$\W = \{e\in\eventtype | \exists a,a' . \instrof{e} = \texttt{[}a\texttt{]:=}a'\}$ and
$\M = \R\cup\W$.
Define $\RR=\R\times\R$. Define
$\RW$, $\RM$, $\WR$, $\WW$,
$\WM$, $\MR$, $\MW$, $\MM$ similarly.

The proof of the acyclicity of $\fcbpower$ w.r.t. to POWER  is done by contradiction. Let us assume that a state $\astate=(\fetched,\events,\po,\co,\rf)$ such that  $\mmodel(\exec{\astate})$, with $\mmodel=\mmodelpower$,  holds and $\textsf{acyclic}(\cbpower{{\astate}})$ does not. This implies that there is a  sequence of events $\anevent_0, \anevent_1, \ldots, \anevent_n \in \fetched$  such that $(\anevent_0,\anevent_1), (\anevent_1,\anevent_2),\ldots, (\anevent_{n-1},\anevent_n) ,(\anevent_n,\anevent_0) \in \cbpower{\astate}$ is a cycle. Let $\rfee=\{(\anevent,\anevent')\in \rf\,|\, \tid{(\anevent)}\neq \tid{(\anevent')} \}$, $\rfi=\rf \setminus \rfee$, $\dpp=\addrdep{\state} \cup \datadep{\state}$, $\cc_0= \dpp \cup \ctrldep{\state}\cup (\addrdep{\state};\po) \cup  \poloc{\state}  $, and $\fences=\ffencedep{\state} \cup \lwfencedep{\state}$.  First, we  will  show the acyclicity of the  relation $\cc_0 \cup \fences \cup \rfi$.

\begin{lemma}
\label{acyclicity-lem1}
The  relation $\cc_0 \cup \fences \cup \rfi$ is acyclic .
\end{lemma}

\begin{proof}
This is an immediate consequence of the fact that $\po$ is acyclic by definition and $\cc_0 \cup \fences \cup \rfi \subseteq \po$.  \qed
\end{proof}

 Since $\rfee$ is the only relation in the definition of $\cbpower{\state}$ which relates events of different threads, the cycle  should contains at least two events belonging to two different threads  and related by $\rfee$ (otherwise, we will have a cycle in $\cbpower{\state} \setminus \rfee=\cc_0 \cup \fences \cup \rfi$ contains a cycle and this contradicts Lemma \ref{acyclicity-lem1}).
 We assume w.l.o.g. that $\tid{(\anevent_0)} \neq \tid{(\anevent_n)}$. Since $(\anevent_n,\anevent_0) \in \rfee$, we have that $\anevent_0 \in \R$ and $\anevent_n \in \W$. This implies $(\anevent_0,\anevent_1) \notin \rfee$.

 Let $i_1, i_2, \ldots, i_k\in \{0,\ldots,n\}$ be the maximal sequence of indices  such that for every $j \in \{1,\ldots, k\}$, we have $(\anevent_{i_j},\anevent_{(i_j+1) mod (n+1)}) \in \rfee$.  Let $i_0=-1$. For every $j \in \{1,\ldots,k\}$, we have $\anevent_{i_j} \in \W \cap \events$ and  $\anevent_{i_{j-1}+1} \in \R \cap \events$. In the following, we will show that $(\anevent_{i_{j-1}+1},\anevent_{i_{j}}) \in \ppo \cup \fences$ for all $j \in \{1,\ldots,k\}$. (Observe that $\ppo $ is defined as  in \cite{alglave2014herding}).) This can be seen as an immediate consequence of Lemma \ref{acyclicity-lem2}, since $(\anevent_{i_{j-1}+1},\anevent_{i_{j}}) \in ( \cc_0 \cup \fences \cup \rfi )^* \cap \RW $ by definition. 
 This implies that the sequence of events $\anevent_0, \anevent_{i_1}, \ldots, \anevent_{i_k}$ forms a cycle in $\hbb= \ppo \cup \fences \cup \rfee$. Furthermore $\anevent_0, \anevent_{i_1}, \ldots, \anevent_{i_k}$ are events in \exec{\astate}. This contradicts the POWER axiom ``NO THIN AIR" which requires the acyclicity of $\hbb$ in order that $\mmodel(\exec{\astate}$ holds.
The rest of the proof is dedicated to the proof of the following lemma under the POWER memory model:

\begin{lemma}
\label{acyclicity-lem2}
$( \cc_0 \cup \fences \cup \rfi )^* \cap \RW \subseteq  \ppo \cup \fences$.
\end{lemma}

 \begin{proof}
Assume two events $\anevent, \anevent' \in \eventtype$ such that $(\anevent,\anevent') \in ( \cc_0 \cup \fences \cup \rfi )^* \cap \RW$. We will show that $(\anevent,\anevent') \in \ppo \cup \fences$.

Since $(\anevent,\anevent') \in ( \cc_0 \cup \fences \cup \rfi )^* \cap \RW$ then there is a sequence of events $\anevent_0, \anevent_1, \ldots, \anevent_n \in \eventtype$ such that $\anevent_0=\anevent$, $\anevent_n=\anevent'$, and $(\anevent_{i-1},\anevent_i) \in (\cc_0 \cup \fences \cup \rfi)$ for all $i \in \{1,\ldots,n\}$.

Let us assume first that there is some  $i \in \{1,\ldots,n\}$ such that  $(\anevent_{i-1},\anevent_i) \in \fences$. Then, there is is some fence (\sync or \lwsync) which is program order between $\anevent_{i-1}$ and $\anevent_i$. Since $ \cc_0 \cup \fences \cup \rfi  \subseteq \po$ it also holds that the fence is program order between $\anevent_0$ and $\anevent_n$. Since we know that $(\anevent_0,\anevent_n) \in \RW$, we have $(\anevent_0,\anevent_n) \in \fences$. Thus we conclude that $(\anevent, \anevent') $ is in $ \ppo \cup \fence$.

Let us assume now that there is no $i \in \{1,\ldots,n\}$ such that  $(\anevent_{i-1},\anevent_i) \in \fences$. We will show that for each  $i \in \{1,\ldots,n\}$, we have    $(\anevent_{i-1},\anevent_i) \in  \cc$. First observe that $\rfi \subseteq \poloc{\state} \subseteq \cc_0$.  Then,  $(\anevent_{i-1},\anevent_i) \in \cc $ trivially holds  since $(\anevent_{i-1},\anevent_i) \in \cc_0 \cup \rfi \subseteq\cc_0$ and by  definition of $\cc$, we have  $\cc_0 \subseteq \cc$. Since $\cc$ is transitive by definition we have $(\anevent, \anevent') \in cc$. Finally, from the definition of $\ppo$, we have $\cc \subseteq \ic$ and $(\ic \cap \RW) \subseteq \ppo$. This implies that $(\anevent, \anevent') \in \ppo$ since $(\anevent, \anevent') \in \RW$. \qed
 \end{proof}
 
 This concludes the proof of Lemma~\ref{lemma:power:acyclic}.\qed

\subsection{Proof of Theorem~\ref{thm:power:no:deadlock} (Deadlock Freedom of POWER)}\label{app:power:nodeadlock}

We prove here that our operational semantics instantiated with
$\mmodel=\mmodelpower$ and $\fcb=\fcbpower$ never deadlocks.

We recall what needs to be proven:
Assume that $\mmodel=\mmodelpower$ and $\fcb=\fcbpower$. Let $\arun$
be a run from $\initstate$, with
$\astate_{\arun}=\stateoafter{\arun}$. Assume that $\anevent$ is a
memory access event (load or store) such that
$\committable_{\astate_{\arun}}(\anevent)$. There exists a parameter
$\param$ such that $\arun.\inst{\anevent}{\param}$ is a run from
$\initstate$ with
$\astate_{\arun}'=\stateoafter{\arun.\inst{\anevent}{\param}}$.

\newcommand{\dt}{.}
\newcommand{\allowed}[1]{\textrm{$ARM(#1)$}\xspace}
\newcommand{\xra}[1]{\xrightarrow{#1}}

\newcommand{\rfe}{\textsf{rfe}}
\newcommand{\epo}[1]{\textsf{po}_{#1}}
\newcommand{\eco}[1]{\textsf{co}_{#1}}
\newcommand{\ecoe}[1]{\textsf{coe}_{#1}}
\newcommand{\erf}[1]{\textsf{rf}_{#1}}
\newcommand{\erfe}[1]{\textsf{rfe}_{#1}}
\newcommand{\efr}[1]{\textsf{fr}_{#1}}
\newcommand{\efre}[1]{\textsf{fre}_{#1}}
\newcommand{\ecom}[1]{\textsf{com}_{#1}}
\newcommand{\epoloc}[1]{\textsf{po-loc}_{#1}}
\newcommand{\eprop}[1]{\textsf{prop}_{#1}}
\newcommand{\epropbase}[1]{\textsf{prop-base}_{#1}}
\newcommand{\effence}[1]{\textsf{ffence}_{#1}}
\newcommand{\efences}[1]{\textsf{fences}_{#1}}
\newcommand{\ehb}[1]{\textsf{hb}_{#1}}
\newcommand{\ehbz}[1]{\textsf{hb}_{\textsf{0}\;#1}}
\newcommand{\edp}[1]{\textsf{dp}_{#1}}
\newcommand{\erdw}[1]{\textsf{rdw}_{#1}}
\newcommand{\erfi}[1]{\textsf{rfi}_{#1}}
\newcommand{\ectrlcfence}[1]{\textsf{ctrl+cfence}_{#1}}
\newcommand{\edetour}[1]{\textsf{detour}_{#1}}
\newcommand{\ectrl}[1]{\textsf{ctrl}_{#1}}
\newcommand{\eaddr}[1]{\textsf{addr}_{#1}}
\newcommand{\eii}[1]{\textsf{ii}_{#1}}
\newcommand{\eiiz}[1]{\textsf{ii}_{\textsf{0}\;#1}}
\newcommand{\eci}[1]{\textsf{ci}_{#1}}
\newcommand{\eciz}[1]{\textsf{ci}_{\textsf{0}\;#1}}
\newcommand{\eic}[1]{\textsf{ic}_{#1}}
\newcommand{\ecc}[1]{\textsf{cc}_{#1}}
\newcommand{\eccz}[1]{\textsf{cc}_{\textsf{0}\;#1}}
\newcommand{\eppo}[1]{\textsf{ppo}_{#1}}

\newcommand{\poA}{\epo{\anexec_w}}
\newcommand{\coA}{\eco{\anexec_w}}
\newcommand{\coeA}{\ecoe{\anexec_w}}
\newcommand{\rfA}{\erf{\anexec_w}}
\newcommand{\rfeA}{\erfe{\anexec_w}}
\newcommand{\frA}{\efr{\anexec_w}}
\newcommand{\freA}{\efre{\anexec_w}}
\newcommand{\comA}{\ecom{\anexec_w}}
\newcommand{\polocA}{\epoloc{\anexec_w}}
\newcommand{\propA}{\eprop{\anexec_w}}
\newcommand{\ffenceA}{\effence{\anexec_w}}
\newcommand{\hbA}{\ehb{\anexec_w}}
\newcommand{\hbzA}{\ehbz{\anexec_w}}
\newcommand{\dpA}{\edp{\anexec_w}}
\newcommand{\rdwA}{\erdw{\anexec_w}}
\newcommand{\rfiA}{\erfi{\anexec_w}}
\newcommand{\ctrlcfenceA}{\ectrlcfence{\anexec_w}}
\newcommand{\detourA}{\edetour{\anexec_w}}
\newcommand{\ctrlA}{\ectrl{\anexec_w}}
\newcommand{\addrA}{\eaddr{\anexec_w}}
\newcommand{\iiA}{\eii{\anexec_w}}
\newcommand{\iizA}{\eiiz{\anexec_w}}
\newcommand{\ciA}{\eci{\anexec_w}}
\newcommand{\cizA}{\eciz{\anexec_w}}
\newcommand{\icA}{\eic{\anexec_w}}
\newcommand{\ccA}{\ecc{\anexec_w}}
\newcommand{\cczA}{\eccz{\anexec_w}}
\newcommand{\ppoA}{\eppo{\anexec_w}}

\newcommand{\poB}{\epo{\anexec_w'}}
\newcommand{\coB}{\eco{\anexec_w'}}
\newcommand{\coeB}{\ecoe{\anexec_w'}}
\newcommand{\rfB}{\erf{\anexec_w'}}
\newcommand{\rfeB}{\erfe{\anexec_w'}}
\newcommand{\frB}{\efr{\anexec_w'}}
\newcommand{\freB}{\efre{\anexec_w'}}
\newcommand{\comB}{\ecom{\anexec_w'}}
\newcommand{\polocB}{\epoloc{\anexec_w'}}
\newcommand{\propB}{\eprop{\anexec_w'}}
\newcommand{\ffenceB}{\effence{\anexec_w'}}
\newcommand{\hbB}{\ehb{\anexec_w'}}
\newcommand{\hbzB}{\ehbz{\anexec_w'}}
\newcommand{\dpB}{\edp{\anexec_w'}}
\newcommand{\rdwB}{\erdw{\anexec_w'}}
\newcommand{\rfiB}{\erfi{\anexec_w'}}
\newcommand{\ctrlcfenceB}{\ectrlcfence{\anexec_w'}}
\newcommand{\detourB}{\edetour{\anexec_w'}}
\newcommand{\ctrlB}{\ectrl{\anexec_w'}}
\newcommand{\addrB}{\eaddr{\anexec_w'}}
\newcommand{\iiB}{\eii{\anexec_w'}}
\newcommand{\iizB}{\eiiz{\anexec_w'}}
\newcommand{\ciB}{\eci{\anexec_w'}}
\newcommand{\cizB}{\eciz{\anexec_w'}}
\newcommand{\icB}{\eic{\anexec_w'}}
\newcommand{\ccB}{\ecc{\anexec_w'}}
\newcommand{\cczB}{\eccz{\anexec_w'}}
\newcommand{\ppoB}{\eppo{\anexec_w'}}

\newcommand{\er}{\anevent_r}
\newcommand{\ew}{\anevent_w}
\newcommand{\ewp}{\anevent_w'}

{

\newcommand{\sgm}{\astate_{\arun}}
\newcommand{\sgmp}{\sgm'}
\newcommand{\exc}{\anexec_{\arun}}
\newcommand{\excp}{\exc'}

\begin{proof}[Proof of Theorem~\ref{thm:power:no:deadlock}]
  If $\anevent$ is a store, then let $\param$ be the number of
  committed stores in $\sgm$ to the same address as $\anevent$, so
  that $\anevent$ becomes $\co$-last in $\sgmp$. If $\anevent$ is a
  load, then let $\param$ be the $\co$-last store to
  $\addr{\sgm}(\anevent)$.

  We will now investigate the new edges in various inter-event
  relations in $\sgmp$. Let
  $\exc=(\events_{\exc},\epo{\exc},\eco{\exc},\erf{\exc})=\exec{\sgm}$ and
  $\excp=(\events_{\excp},\epo{\excp},\eco{\excp},\erf{\excp})=\exec{\sgmp}$.

  \begin{description}
  \item{\textsf{E}: } We have $\events_{\excp} = \events_{\exc}\cup\{\anevent\}$.
  \item{\textsf{po-loc}: } We have $\epoloc{\excp} \subseteq \epoloc{\exc}\cup\{(\anevent',\anevent) | \anevent'\in\eventtype\}$.
  \item{\textsf{co}: } We have $\eco{\excp}\subseteq\eco{\exc}\cup\{(\anevent',\anevent) | \anevent'\in\eventtype\}$.
  \item{\textsf{fr}: } We have $\efr{\excp}\subseteq\efr{\exc}\cup\{(\anevent',\anevent) | \anevent'\in\eventtype\}$.
  \item{\textsf{fre}: } We have $\efre{\excp}\subseteq\efre{\exc}\cup\{(\anevent',\anevent) | \anevent'\in\eventtype\}$.
  \item{\textsf{rf}: } We have $\erf{\excp}\subseteq\erf{\exc}\cup\{(\anevent',\anevent) | \anevent'\in\eventtype\}$.
  \item{\textsf{rfi}: } We have $\erfi{\excp}\subseteq\erfi{\exc}\cup\{(\anevent',\anevent) | \anevent'\in\eventtype\}$.
  \item{\textsf{rfe}: } We have $\erfe{\excp}\subseteq\erfe{\exc}\cup\{(\anevent',\anevent) | \anevent'\in\eventtype\}$.
  \item{\textsf{com}: } We have $\ecom{\excp}\subseteq\ecom{\exc}\cup\{(\anevent',\anevent) | \anevent'\in\eventtype\}$.
  \item{\textsf{fences}: } We have $\efences{\excp}\subseteq\efences{\exc}\cup\{(\anevent',\anevent) | \anevent'\in\eventtype\}$\\
    Since, for each new edge
    $\anevent_0\xra{\efences{\excp}}\anevent_1$ with
    $(\anevent_0,\anevent_1)\not\in\efences{\exc}$, we must have
    $\anevent_0=\anevent$ or $\anevent_1=\anevent$. Furthermore, we
    cannot have $\anevent\xra{\efences{\excp}}\anevent_1$, since
    $\efences{\excp}\subseteq\cbpower{\excp}$, and $\anevent$ is
    $\cbpower{\excp}$-last.
  \item{\textsf{ffence}: } We have $\effence{\excp}\subseteq\effence{\exc}\cup\{(\anevent',\anevent) | \anevent'\in\eventtype\}$\\
    Same motivation as for \textsf{fences}.
  \item{\textsf{dp}: } We have $\edp{\excp}\subseteq\edp{\exc}\cup\{(\anevent',\anevent) | \anevent'\in\eventtype\}$\\
    Same motivation as for \textsf{fences}.
  \item{\textsf{rdw}: } We have $\erdw{\excp}\subseteq\erdw{\exc}\cup\{(\anevent',\anevent) | \anevent'\in\eventtype\}$\\
    Same motivation as for \textsf{fences}, where we notice that $\erdw{\excp}\subseteq\epoloc{\excp}\subseteq\cbpower{\excp}$.
  \item{\textsf{ctrl+cfence}: } We have $\ectrlcfence{\excp}\subseteq\ectrlcfence{\exc}\cup\{(\anevent',\anevent) | \anevent'\in\eventtype\}$\\
    Same motivation as for \textsf{fences}, where we notice that $\ectrlcfence{\excp}\subseteq\ectrl{\excp}\subseteq\cbpower{\excp}$.
  \item{\textsf{detour}: } We have $\edetour{\excp}\subseteq\edetour{\exc}\cup\{(\anevent',\anevent) | \anevent'\in\eventtype\}$\\
    Same motivation as for \textsf{fences}, where we notice that $\edetour{\excp}\subseteq\epoloc{\excp}\subseteq\cbpower{\excp}$.
  \item{\textsf{po-loc}: } We have $\epoloc{\excp}\subseteq\epoloc{\exc}\cup\{(\anevent',\anevent) | \anevent'\in\eventtype\}$\\
    Same motivation as for \textsf{fences}.
  \item{\textsf{ctrl}: } We have $\ectrl{\excp}\subseteq\ectrl{\exc}\cup\{(\anevent',\anevent) | \anevent'\in\eventtype\}$\\
    Same motivation as for \textsf{fences}.
  \item{\textsf{addr;po}: } We have $\eaddr{\excp};\epo{\excp}\subseteq\eaddr{\exc};\epo{\exc}\cup\{(\anevent',\anevent) | \anevent'\in\eventtype\}$\\
    Same motivation as for \textsf{fences}.
  \item{\textsf{ppo}: } We have $\eppo{\excp}\subseteq\eppo{\exc}\cup\{(\anevent',\anevent) | \anevent'\in\eventtype\}$\\
    To see this, let $(\anevent_0,\anevent_1)$ be any edge in
    $\eppo{\excp}\setminus\eppo{\exc}$. From the definition of
    \textsf{ppo}, we have that
    $\anevent_0\xra{(\eiiz{\excp}\cup\eciz{\excp}\cup\eccz{\excp})^+}\anevent_1$. Furthermore
    there must be at least one edge along that chain which is not in
    $\eiiz{\exc}\cup\eciz{\exc}\cup\eccz{\exc}$. We have seen above
    that any edge which is in
    $\eiiz{\excp}\cup\eciz{\excp}\cup\eccz{\excp}$ but not in
    $\eiiz{\exc}\cup\eciz{\exc}\cup\eccz{\exc}$ must be of the form
    $(\anevent_2,\anevent)$ for some $\anevent_2$. Hence we have

    \noindent
    \begin{math}
      \anevent_0\xra{(\eiiz{\excp}\cup\eciz{\excp}\cup\eccz{\excp})*}\\
      \anevent_2\xra{(\eiiz{\excp}\cup\eciz{\excp}\cup\eccz{\excp})}\\
      \anevent\xra{(\eiiz{\excp}\cup\eciz{\excp}\cup\eccz{\excp})*}\anevent_1
    \end{math}

    But there is no edge going from $\anevent$ in
    $\eiiz{\excp}\cup\eciz{\excp}\cup\eccz{\excp}$, so it must be the
    case that $\anevent = \anevent_1$.
  \item{\textsf{hb}: } We have
    $\ehb{\excp}\subseteq\ehb{\exc}\cup\{(\anevent',\anevent) | \anevent'\in\eventtype\}$.\\
    By the above, we have

    \noindent
    \begin{math}
      \ehb{\excp} =\\
      \efences{\excp}\cup\erfe{\excp}\cup\eppo{\excp} \subseteq\\
      \efences{\exc}\cup\erfe{\exc}\cup\eppo{\exc}\cup\{(\anevent',\anevent) | \anevent'\in\eventtype\} =\\
      \ehb{\excp}\cup\{(\anevent',\anevent) | \anevent'\in\eventtype\}
    \end{math}
  \end{description}

  Notice that none of the $\exc$-relations contain links to or from
  $\anevent$, since $\anevent$ is not in $\events_{\exc}$.

  We will now show that $POWER(\excp)$. We need to show that each of
  the four POWER axioms~\cite{alglave2014herding} holds for $\excp$:

  \paragraph{Subproof \ref{thm:power:no:deadlock}.1: Show \normalfont{$\textsf{acyclic}(\epoloc{\excp}\cup\ecom{\excp})$}}\mbox{}\\

  By the assumption $POWER(\exc)$, we have
  $\textsf{acyclic}(\epoloc{\exc}\cup\epoloc{\exc})$. Since
  $(\epoloc{\excp}\cup\epoloc{\excp})\setminus(\epoloc{\exc}\cup\epoloc{\exc})$
  only contains edges leading to $\anevent$, and
  $\epoloc{\excp}\cup\epoloc{\excp}$ contains no edges leading from
  $\anevent$, we also have
  $\textsf{acyclic}(\epoloc{\excp}\cup\epoloc{\excp})$.

  \paragraph{Subproof \ref{thm:power:no:deadlock}.2: Show \normalfont{$\textsf{acyclic}(\ehb{\excp})$}}\mbox{}\\

  The proof is analogue to that in
  Subproof~\ref{thm:power:no:deadlock}.1.

  \paragraph{Subproof \ref{thm:power:no:deadlock}.3: Show \normalfont{$\textsf{irreflexive}(\efre{\excp};\eprop{\excp};\ehb{\excp}*)$}}\mbox{}\\

  By the assumption $POWER(\exc)$, we have
  $\textsf{irreflexive}(\efre{\exc};\eprop{\exc};\ehb{\exc}*)$. Assume
  for a contradiction that we have $(\anevent_0,\anevent_0) \in
  \efre{\excp};\eprop{\excp};\ehb{\excp}*$. By examining the
  definition of \textsf{prop}, we see that every edge building up the
  chain from $\anevent_0$ to $\anevent_0$ through
  $\efre{\excp};\eprop{\excp};\ehb{\excp}*$, must be in one of
  $\efre{\excp}$, $\erfe{\excp}$, $\efences{\excp}$, $\ehb{\excp}$,
  $\ecom{\excp}$ or $\effence{\excp}$. Since at least one edge must
  not be in the corresponding $\exc$-relation, the chain of relations
  must go through $\anevent$. But we have seen above that there is no
  edge going out from $\anevent$ in any of the above mentioned
  relations. Hence there can be no such cycle.

  \paragraph{Subproof \ref{thm:power:no:deadlock}.4: Show \normalfont{$\textsf{acyclic}(\eco{\excp}\cup\eprop{\excp})$}}\mbox{}\\

  The proof is analogue to that in
  Subproof~\ref{thm:power:no:deadlock}.3.

  This concludes the proof.\qed
\end{proof}

}

\section{Proofs for Section~\ref{sec:dpor}}
\label{app:rsmc:proofs}

Here we provide proofs for the theorems appearing in
Section~\ref{sec:dpor}.

\subsection{Proof of Theorem~\ref{thm:soundness2} (Soundness of RSMC)}
\label{app:dpor2:soundness:proof}

\begin{lemma}\label{lemma:promote:load}
  Assume that $\fcb$ is valid w.r.t. $\mmodel$, and that $\mmodel$ and
  $\fcb$ are deadlock free.
  Assume that
  $\arun_A=\arun.\inst{\anevent_w}{n_w}.\inst{\anevent_r}{\anevent_w}$
  is a run from $\initstate$. Further assume that
  $\committable_{\stateoafter{\arun}}(\anevent_r)$. Then there is a
  run $\arun_B=\arun.\inst{\anevent_r}{\anevent_w'}$ from $\initstate$
  such that $\committable_{\stateoafter{\arun_B}}(\anevent_w)$.
\end{lemma}

\begin{proof}[Proof of Lemma~\ref{lemma:promote:load}]
  The lemma follows from deadlock freedom and monotonicity.\qed
\end{proof}

\begin{lemma}\label{lemma:promote:load:split}
  Assume that $\fcb$ is valid w.r.t. $\mmodel$, and that $\mmodel$ and
  $\fcb$ are deadlock free.
   Assume that
  $\arun_A = \arun.\inst{\anevent_r}{\anevent_{w0}}$ is a run
  from $\initstate$. Assume that
  $\arun_B=\arun.\arun'.\inst{\anevent_w}{n_w}.\inst{\anevent_r}{\anevent_w}$
  is a run from $\initstate$. Let
  $\astate_B = \stateoafter{\arun_B}$. Then either
  \begin{itemize}
  \item
    $\arun_C=\arun.\inst{\anevent_r}{\anevent_{w1}}.\arun'$ is a run
    from $\initstate$ and
    $\committable_{\stateoafter{\arun_C}}(\anevent_w)$, or
  \item there is an event $\anevent_w'\in\arun'$ such that
    $\arun_D=\arun.\arun''.\inst{\anevent_w'}{n_w'}.\inst{\anevent_r}{\anevent_w'}.\arun'''$
    is a run from $\initstate$ with
    $\arun''.\inst{\anevent_w'}{n_w'}.\arun''' = \arun'$ and
    $\committable_{\stateoafter{\arun_D}}(\anevent_w)$.
  \end{itemize}
\end{lemma}

\begin{proof}[Proof of Lemma~\ref{lemma:promote:load:split}]
  Monotonicity of $\fcb$ and Lemma~\ref{lemma:promote:load}
  give that there exists a run
  $\arun_B'=\arun.\arun'.\inst{\anevent_r}{\param}$ from $\initstate$
  for some store $\param$ with
  $\committable_{\stateoafter{\arun_B'}}(\anevent_w)$. We case split
  on whether or not $\param\in\arun'$:

  Assume first that $\param\not\in\arun'$. Then since
  $\committable_{\stateoafter{\arun}}(\anevent_r)$, we know that
  $\anevent_r$ is not $\fcb$-related to any event in
  $\arun'$. Therefore, we can rearrange $\arun_B'$ into
  $\arun_C=\arun.\inst{\anevent_r}{\param}.\arun'$, which is a run
  from $\initstate$. Furthermore, since the committed
  events are the same in $\arun_C$ as in $\arun_B'$, we have
  $\committable_{\stateoafter{\arun_C}}(\anevent_w)$.

  Assume instead that $\param\in\arun'$. Then
  $\arun_B'=\arun.\arun''.\inst{\anevent_w'}{n_w'}.\arun'''.\inst{\anevent_r}{\anevent_w'}$
  for some $\arun''$, $n_w'$, $\arun'''$ and $\anevent_w'=\param$. Then
  by the same argument as above, we can reorder this run to form
  $\arun_D=\arun.\arun''.\inst{\anevent_w'}{n_w'}.\inst{\anevent_r}{\anevent_w'}.\arun'''$
  which is a run from $\initstate$ and
  $\arun''.\inst{\anevent_w'}{n_w'}.\arun''' = \arun'$ and
  $\committable_{\stateoafter{\arun_D}}(\anevent_w)$.\qed
\end{proof}

We are now ready to state and prove the main lemma, from which the
soundness theorem directly follows:

\begin{lemma}\label{lemma:soundness2}
  Assume that $\fcb$ is valid w.r.t. $\mmodel$, and that $\mmodel$ and
  $\fcb$ are deadlock free.
  Let $\arun$, $\astate$, $\anexec$ be such that
  $\stateoafter{\arun} = \astate$ and $\exec{\astate} = \anexec$ and
  $M(\anexec)$.
  Then for all $\anexec'$ s.t. $M(\anexec')$, and $\anexec'$ is a
  complete $\fcb$-extension of $\anexec$, the evaluation of
  $\newalgo(\arun,\astate)$ will contain a recursive call to
  $\newalgo(\arun',\astate')$ for some $\arun'$, $\astate'$ such
  that $\exec{\astate'} = \anexec'$.
\end{lemma}

\begin{proof}[Proof of Lemma~\ref{lemma:soundness2}]
  By assumption there is an upper bound $B$ on the length of any run
  of the fixed program. Therefore, we can perform the proof by total
  induction on $B$ minus the length of $\arun$.

  Fix arbitrary $\arun$,
  $\astate=(\lblcur,\fetched,\events,\po,\co,\rf)$ and $\anexec$. We
  will show that for all $\anexec'$ s.t. $M(\anexec')$, and $\anexec'$
  is an complete $\fcb$-extension of $\anexec$, the evaluation of
  $\newalgo(\arun,\astate)$ will contain a recursive call to
  $\newalgo(\arun',\astate')$ for some $\arun'$, $\astate'$ such that
  $\exec{\astate'} = \anexec'$.
  Our inductive hypothesis states that for all $\arun''$, $\astate''$, $\anexec''$ such that
  $|\arun|<|\arun''|$ and $\stateoafter{\arun''}=\astate''$ and $\exec{\astate''}=\anexec''$
  and $M(\anexec'')$,
  it holds that for all $\anexec'''$ s.t. $M(\anexec''')$, and
  $\anexec'''$ is a complete $\fcb$-extension of $\anexec''$, the evaluation
  of $\newalgo(\arun'',\astate'')$ will contain a recursive call to
  $\newalgo(\arun''',\astate''')$ for some $\arun'''$, $\astate'''$
  such that $\exec{\astate'''} = \anexec'''$.

  Consider the evaluation of $\newalgo(\arun,\astate)$. If there is no
  enabled event on line~\ref{ln:dpor2:select:e:disarm}, then $\astate$
  is complete, and there are no (non-trivial) $\fcb$-extensions of
  $\anexec$. The Lemma is then trivially satisfied. Assume therefore
  instead that a enabled event $\anevent$ is selected on
  line~\ref{ln:dpor2:select:e:disarm}.

  We notice first that for all executions
  $\anexec'=(\events',\po',\co',\rf')$ s.t. $M(\anexec')$, and
  $\anexec'$ is a complete $\fcb$-extension of $\anexec$, it must be the case
  that $\anevent\in\events'$. This is because $\anexec'$ is an
  extension of $\anexec$, and $\anevent$ is enabled after $\anexec$
  and $\anexec'$ is complete. Since $\anevent$ is enabled in
  $\anexec$, it must be executed at some point before the execution
  can become complete.

  The event $\anevent$ is either a store or a load. Assume first that
  $\anevent$ is a store.
  Let $\anexec'=(\events',\po',\co',\rf')$ be an arbitrary execution
  s.t. $M(\anexec')$, and $\anexec'$ is a complete $\fcb$-extension of
  $\anexec$.
  There exists some parameter (natural number) $n$ for $\anevent$
  which inserts $\anevent$ in the same position in the coherence order
  relative to the other stores in $\events$ as $\anevent$ has in
  $\anexec'$.
  Let $\astate_n = \stateoafter{\arun.\inst{\anevent}{n}}$ and
  $\anexec_n = \exec{\astate_n}$. Notice that $\anexec'$ is an
  extension of $\anexec_n$. We have $M(\anexec')$, and by
  the monotonicity of the memory model we then also have that
  the parameter $n$ is allowed for the event $\anevent$ by the memory
  model, i.e., $M(\anexec_n)$.
  Since the parameter $n$ is allowed for $e$ in $\astate$ by the
  memory model, $(n,\astate_n)$ will be in $\newbnc$ on
  line~\ref{ln:dpor2:store:comp:S}, and so there will be a recursive
  call $\newalgo(\arun.\inst{\anevent}{n},\astate_n)$ on
  line~\ref{ln:dpor2:store:rec:call}. The run
  $\arun.\inst{\anevent}{n}$ is a longer run than $\arun$. Hence the
  inductive hypothesis can be applied, and yields that the sought
  recursive call $\newalgo(\arun',\astate')$ will be made. This
  concludes the case when $\anevent$ is a store.

  Next assume instead that $\anevent$ is a load.
  Let $\mathfrak{E}$ be the set of all executions $\anexec'$
  s.t. $M(\anexec')$, and $\anexec'$ is a complete $\fcb$-extension of
  $\anexec$. Now define the set $\mathfrak{E}_0\subseteq\mathfrak{E}$
  s.t. $\mathfrak{E}_0$ contains precisely the executions
  $\anexec'=(\events',\po',\co',\rf')\in\mathfrak{E}$ where
  $(\anevent_w,\anevent)\in\rf'$ for some
  $\anevent_w\in\events$. I.e. we define $\mathfrak{E}_0$ to be the
  executions in $\mathfrak{E}$ where the read-from source for
  $\anevent$ is already committed in $\anexec$.
  Let $\mathfrak{E}_1 = \mathfrak{E}\setminus\mathfrak{E}_0$.
  We will show that the lemma holds, first for all executions in
  $\mathfrak{E}_0$, and then for all executions in $\mathfrak{E}_1$.

  Let $\anexec'$ be an arbitrary execution in $\mathfrak{E}_0$. We can
  now apply a reasoning analogue to the reasoning for the case when
  $\anevent$ is a store to show that there will be a recursive call
  $\newalgo(\arun',\astate')$ with $\exec{\astate'} = \anexec'$.

  We consider instead the executions in $\mathfrak{E}_1$. Assume for a
  proof by contradiction that there are some executions in
  $\mathfrak{E}_1$ that will not be explored.
  Let $\mathfrak{E}_2\subseteq\mathfrak{E}_1$ be the set of executions
  in $\mathfrak{E}_1$ that we fail to explore. I.e., let
  $\mathfrak{E}_2\subseteq\mathfrak{E}_1$ be the set of executions
  $\anexec'$ such that there are no $\arun'$, $\astate'$ where
  $\stateoafter{\arun'} = \astate'$ and $\exec{\astate'} = \anexec'$
  and there is a recursive call $\newalgo(\arun',\astate')$ made
  during the evaluation of $\newalgo(\arun,\astate)$.
  We will now define a ranking function $R$ over executions in
  $\mathfrak{E}_2$, and then investigate one of the executions that
  minimize $R$.
  For a run $\arun'$ of the form
  $\arun.\arun_0.\inst{\anevent_w}{n_w}.\inst{\anevent}{\anevent_w}.\arun_1$
  (notice the fixed run $\arun$ and the fixed load event $\anevent$),
  define $R(\arun') = |\arun_0|$. For an execution
  $\anexec'\in\mathfrak{E}_2$, let $T(\anexec')$ be the set of runs
  $\arun'$ of the form
  $\arun.\arun_0.\inst{\anevent_w}{n_w}.\inst{\anevent}{\anevent_w}.\arun_1$
  such that $\exec{\stateoafter{\arun'}} = \anexec'$.
  Now define $R(\anexec') = R(\arun')$ for a run $\arun'\in
  T(\anexec')$ that minimizes $R$ within $T(\anexec')$.

  Now let $\anexec'\in\mathfrak{E}_2$ be an execution minimizing $R$
  in $\mathfrak{E}_2$ and
  $\arun'=\arun.\arun_0.\inst{\anevent_w}{n_w}.\inst{\anevent}{\anevent_w}.\arun_1$
  be a run in $T(\anexec')$ minimizing $R$ in $T(\anexec')$. Let
  $\astate' = \stateoafter{\arun'}$.
  Lemma~\ref{lemma:promote:load:split} tells us that either
  \begin{itemize}
  \item
    $\arun_C=\arun.\inst{\anevent}{\anevent_{w1}}.\arun_0$ is a run
    from $\initstate$ and
    $\committable_{\stateoafter{\arun_C}}(\anevent_w)$, or
  \item there is an event $\anevent_w'\in\arun_0$ such that
    $\arun_D=\arun.\arun'.\inst{\anevent_w'}{n_w'}.\inst{\anevent}{\anevent_w'}.\arun''$
    is a run from $\initstate$ and
    $\arun'.\inst{\anevent_w'}{n_w'}.\arun'' = \arun_0$ and
    $\committable_{\stateoafter{\arun_D}}(\anevent_w)$.
  \end{itemize}

  Consider first the case when
  $\arun_C=\arun.\inst{\anevent}{\anevent_{w1}}.\arun_0$ is a run from
  $\initstate$ and $\committable_{\stateoafter{\arun_C}}(\anevent_w)$.
  Since $\arun_C$ is a run, we know that $\anevent_{w1}$ is an allowed
  parameter for $\anevent$ after $\arun$. So
  $(\anevent_{w1},\astate_{w1})$ will be added to $\newbnc$ for some
  $\astate_{w1}$ on line~\ref{ln:dpor2:load:comp:S} in the call
  $\newalgo(\arun,\astate)$, and a recursive call
  $\newalgo(\arun.\inst{\anevent}{\anevent_{w1}},\astate_{w1})$ will
  be made on line~\ref{ln:dpor2:load:rec:call}. Since
  $\arun.\inst{\anevent}{\anevent_{w1}}$ is a longer run than $\arun$,
  the inductive hypothesis tells us that all complete
  $\fcb$-extensions of $\exec{\astate_{w1}}$ will be explored. Let
  $\anexec_C'$ be any complete $\fcb$-extension of
  $\exec{\stateoafter{\arun_C}}$. Notice that $\anexec_C'$ is also a
  complete $\fcb$-extension of $\exec{\astate_{w1}}$. Since we have
  $\committable_{\stateoafter{\arun_C}}(\anevent_w)$ it must be the
  case that $\anevent_w$ is committed in $\anexec_C'$. Then the race
  detection code in $\newalgodetectrace$ is executed for $\anevent_w$
  on line~\ref{ln:dpor2:store:call:detect:race} in the call to
  $\newalgo$ where $\anevent_w$ is committed. When the race detection
  code is run for $\anevent_w$, the $R\rightarrow{}W$ race from
  $\anevent$ to $\anevent_w$ will be detected. To see this, notice
  that $\anexec_C'$ is an extension of
  $\exec{\stateoafter{\arun_C}}$. Hence we know that the events that
  are $\fcb$-before $\anevent_w$ in $\anexec_C'$ are the same as the
  events that are $\fcb$-before $\anevent_w$ in $\arun_C$ and also in
  $\arun'$. Therefore $\anevent_w$ targets the same memory location in
  $\anexec_C'$ as it does in $\arun'$. Furthermore, $\anevent$ cannot
  be $\fcb$-before $\anevent_w$, since $\anevent$ appears after
  $\anevent_w$ in $\arun'$. Therefore, the race from $\anevent$ to
  $\anevent_w$ is detected, and the branch
  $\arun_0.\inst{\anevent_w}{\dporstar}.\inst{\anevent}{\anevent_w}$
  is added to \texttt{$\contmap$[$\anevent$]}. When the
  lines~\ref{ln:dpor2:load:cont:begin}-\ref{ln:dpor2:load:cont:end}
  are executed in the call $\newalgo(\arun,\astate)$, that branch will
  be traversed. During the traversal, all parameters for $\anevent_w$
  will be explored, and in particular the following call will be made:
  $\newalgo(\arun.\arun_0.\inst{\anevent_w}{n_w}.\inst{\anevent}{\anevent_w},\astate_w)$
  for some $\astate_w$. Since the run
  $\arun.\arun_0.\inst{\anevent_w}{n_w}.\inst{\anevent}{\anevent_w}$
  is longer than $\arun$, the inductive hypothesis tells us that all
  its complete $\fcb$-extensions will be explored. These include
  $\anexec'$, contradicting our assumption that $\anexec'$ is never
  explored.

  Consider then the case when there is an event
  $\anevent_w'\in\arun_0$ such that
  $\arun_D=\arun.\arun'.\inst{\anevent_w'}{n_w'}.\inst{\anevent}{\anevent_w'}.\arun''$
  is a run from $\initstate$ and
  $\arun'.\inst{\anevent_w'}{n_w'}.\arun'' = \arun_0$ and
  $\committable_{\stateoafter{\arun_D}}(\anevent_w)$.
  Let $\anexec_D'$ be any execution which is a complete $\fcb$-extension of
  $\exec{\stateoafter{\arun_D}}$. Since $\anexec_D'$ can be reached by
  a run which extends $\arun_D$, it must be the case that $\anexec_D'$
  has a lower rank than $\anexec'$, i.e.,
  $R(\anexec_D')<R(\anexec')$. Since, $\anexec'$ has a minimal rank in
  $\mathfrak{E}_2$, it must be the case that
  $\anexec_D'\in\mathfrak{E}_1$, and so we know that $\anexec_D'$ is
  explored by some call $\newalgo(\arun_D',\astate_D')$, made
  recursively from $\newalgo(\arun,\astate)$, with
  $\stateoafter{\arun_D'}=\astate_D'$ and
  $\exec{\astate_D'}=\anexec_D'$. By a reasoning analogue to that in
  the previous case, the store $\anevent_w$ must be committed
  in $\anexec_D'$, and the $R\rightarrow{}W$ race from $\anevent$ to
  $\anevent_w$ is detected by $\newalgodetectrace$. Again,
  the branch
  $\arun_0.\inst{\anevent_w}{\dporstar}.\inst{\anevent}{\anevent_w}$
  is added to \texttt{$\contmap$[$\anevent$]}. So there will be a
  recursive call $\newalgo(\arun_w,\astate_w)$ for
  $\arun_w=\arun_0.\inst{\anevent_w}{n_w}.\inst{\anevent}{\anevent_w}$
  and $\stateoafter{\arun_w}=\astate_w$ where the parameter
  $\dporstar$ for $\anevent_w$ has been instantiated with $n_w$. Since
  $\arun_w$ is a longer run than $\arun$, the inductive hypothesis
  tells us that all extensions of $\arun_w$ will be explored. The
  sought execution $\anexec'$ is an extension of $\exec{\stateoafter{\arun_w}}$, and
  so it will be explored. This again contradicts our assumption that
  $\anexec'$ is never explored. This concludes the proof.\qed
\end{proof}

\begin{proof}[Proof of Theorem~\ref{thm:soundness2}]
  The theorem follows directly from Lemma~\ref{lemma:soundness2},
  since each complete execution a $\fcb$-extension of
  $\exec{\initstate}$.\qed
\end{proof}

\subsection{Proof of Theorem~\ref{thm:optimality:power} (Optimality of RSMC for POWER)}
\label{app:dpor2:optimality:proof}
\begin{lemma}\label{lemma:diff:cuts:diff:params}
  Assume that $\mmodel = \mmodelpower$ and $\fcb=\fcbpower$.
  Let $\arun$, and $\arun.\arun_A.\inst{\anevent}{\param_A}$ and
  $\arun.\arun_B.\inst{\anevent}{\param_B}$ be runs. Let
  $\astate_A=\stateoafter{\arun.\arun_A}$ and
  $\astate_B=\stateoafter{\arun.\arun_B}$. Let
  $\arun_a=\normalizerun(\cutrun(\arun_A,\anevent,\astate_A))$ and
  $\arun_b=\normalizerun(\cutrun(\arun_B,\anevent,\astate_B))$. Assume
  $\arun_a\neq\arun_b$. Then there is some event $\anevent'$ and
  parameters $\param_a\neq\param_b$ such that
  $\inst{\anevent'}{\param_a}\in\arun_a$ and
  $\inst{\anevent'}{\param_b}\in\arun_b$.
\end{lemma}

\begin{proof}[Proof of Lemma~\ref{lemma:diff:cuts:diff:params}]
  We start by considering the control flow leading to $\anevent$ in
  the thread $\tid(\anevent)$. Either the control flow to $\anevent$
  is the same in both $\arun_A$ and $\arun_B$, or it differs. Assume
  first that the control flow differs. Then there is some branch
  instruction which program order-precedes $\anevent$ which evaluates
  differently in $\arun_A$ and $\arun_B$. That means that the
  arithmetic expression which is the condition for a conditional
  branch evaluates differently in $\arun_A$ and $\arun_B$. Our program
  semantics does not allow data nondeterminism. The only way for an
  expression to evaluate differently is for it to depend on the value
  of a program order-earlier load $\anevent_l$. This loaded value must
  differ in $\arun_A$ and $\arun_B$. This can only happen if
  $\anevent_l$ gets its value from some chain (possibly empty) of
  read-from and data dependencies, which starts in a load
  $\anevent_l'$ (possibly equal to $\anevent_l$) which takes different
  parameters in $\arun_A$ and $\arun_B$. However, since both read-from
  and data dependencies are a part of $\fcbpower$, this gives us
  $(\anevent_l',\anevent_l)\in\cb{\astate_A}^*$ and
  $(\anevent_l',\anevent_l)\in\cb{\astate_B}^*$. Furthermore, since
  $\anevent_l$ provides a value used in a branch that program
  order-precedes $\anevent$, we have a control dependency between
  $\anevent_l$ and $\anevent$. Control dependencies are also part of
  $\fcbpower$, so we have
  $(\anevent_l',\anevent)\in\cb{\astate_A}^*$ and
  $(\anevent_l',\anevent)\in\cb{\astate_B}^*$. Then $\anevent_l'$ is
  the sought witness event.

  Now assume instead that the control flow leading to $\anevent$ is
  the same in $\arun_A$ and $\arun_B$.

  Consider the sets
  $A=\{\anevent\in\eventtype|\exists\param.\inst{\anevent}{\param}\in\arun_a\}$
  and
  $B=\{\anevent\in\eventtype|\exists\param.\inst{\anevent}{\param}\in\arun_b\}$.

  Assume first that $A=B$. Then if all events in $A$ (or equivalently
  in $B$) have the same parameters in $\arun_a$ as in $\arun_b$, then
  $\arun_a$ and $\arun_b$ consist of exactly the same parameterized
  events. Since all the events have the same parameters in $\arun_a$
  and $\arun_b$, they must also be related in the same way by the
  commit-before relation. But then since $\arun_a$ and $\arun_b$ are
  normalized, (by the function $\normalizerun$) it must hold that
  $\arun_a=\arun_b$, which contradicts our assumption
  $\arun_a\neq\arun_b$. Hence there must be some event which appears
  in both $\arun_a$ and $\arun_b$ which takes different parameters in
  $\arun_a$ and $\arun_b$.

  Assume next that $A\neq{}B$. Assume without loss of generality that
  $A$ contains some event which is not in $B$. Let $C$ be the largest
  subset of $A$ such that $C\cap{}B=\emptyset$. Then by the
  construction of $\arun_a$, at least one of the events
  $\anevent_c\in{}C$ must precede some event
  $\anevent_a\in{}(A\cap{}B)\cup\{\anevent\}$ in
  $\cb{\astate_A}$. Since $\anevent_c\not\in{}B$ we also have
  $(\anevent_c,\anevent_a)\not\in\cb{\astate_B}$. We now look to the
  definition of $\fcbpower$, and consider the relation
  between $\anevent_c$ and $\anevent_a$ in $\astate_A$. From
  $(\anevent_c,\anevent_a)\in\cb{\astate_A}$ we know that
  $(\anevent_c,\anevent_a)\in\addrdep{\astate_A}\cup\datadep{\astate_A}\cup\ctrldep{\astate_A}\cup(\addrdep{\astate_A};\po_A)\cup\ffencedep{\astate_A}\cup\lwsyncdep{\astate_A}\cup\poloc{\astate_A}\cup\rf_A$. However,
  all the relations $\addrdep{\astate_A}$, $\datadep{\astate_A}$,
  $\ctrldep{\astate_A}$, $(\addrdep{\astate_A};\po_A)$,
  $\ffencedep{\astate_A}$, $\lwsyncdep{\astate_A}$ are given by the
  program (and the control flow, which is fixed by assumption). So if
  $(\anevent_c,\anevent_a)\in\addrdep{\astate_A}\cup\datadep{\astate_A}\cup\ctrldep{\astate_A}\cup(\addrdep{\astate_A};\po_A)\cup\ffencedep{\astate_A}\cup\lwsyncdep{\astate_A}$,
  then the same relation holds in $\arun_B$, and then we would have
  $(\anevent_c,\anevent_a)\in\cb{\astate_B}$ contradicting our
  previous assumption. Hence it must be the case that
  $(\anevent_c,\anevent_a)\in\poloc{\astate_A}$ or
  $(\anevent_c,\anevent_a)\in\rf_A$.
  If $(\anevent_c,\anevent_a)\in\rf_A$ then $\anevent_a\neq\anevent$,
  since $\anevent$ is not committed in $\astate_A$, and therefore
  cannot have a read-from edge. Hence $\anevent_a$ must then be a load
  which appears both in $\arun_a$ and $\arun_b$. Furthermore, since
  $(\anevent_c,\anevent_a)\not\in\rf_B$ the load $\anevent_a$ must
  have different parameters in $\arun_a$ and $\arun_b$. Then
  $\anevent_a$ is the sought witness.
  The final case that we need to consider is when
  $(\anevent_c,\anevent_a)\in\poloc{\astate_A}$. Since we have
  $(\anevent_c,\anevent_a)\in\poloc{\astate_A}$ and
  $(\anevent_c,\anevent_a)\not\in\poloc{\astate_B}$, the address of
  either $\anevent_c$ or $\anevent_a$ must be computed differently in
  $\arun_a$ and $\arun_b$. Therefore, the differing address must
  depend on the value read by some earlier load $\anevent_l$, which
  reads different values in $\arun_a$ and $\arun_b$. If the differing
  address is in $\anevent_a$, then we have
  $(\anevent_l,\anevent_a)\in\addrdep{\astate_A}$ and
  $(\anevent_l,\anevent_a)\in\addrdep{\astate_B}$. If the differing
  address is in $\anevent_c$, then we have
  $(\anevent_l,\anevent_a)\in(\addrdep{\astate_A};\po_A)$ and
  $(\anevent_l,\anevent_a)\in(\addrdep{\astate_B};\po_B)$. In both
  cases we have $(\anevent_l,\anevent_a)\in\cb{\astate_A}$ and
  $(\anevent_l,\anevent_a)\in\cb{\astate_B}$. Hence
  $\anevent_l\in{}A\cap{}B$. Now, by the same reasoning as above, in
  the case for differing control flows, we know that there is a load
  $\anevent_l'\in{}A\cap{}B$ which has different parameters in
  $\arun_a$ and $\arun_b$. This concludes the proof.\qed
\end{proof}

We recall the statement of Theorem~\ref{thm:optimality:power} from
Section~\ref{sec:dpor}:

{
  \renewcommand{\thetheorem}{\ref{thm:optimality:power}}
  \begin{theorem}[Optimality for POWER]
    Assume that $\mmodel = \mmodelpower$ and  $\fcb=\fcbpower$.
    Let $\anexec \in \axmodel{\aprog}{\mmodel}$.
    Then during the evaluation of a call to
    $\newalgo(\emptyseq,\initstate)$, there will be exactly one call
    $\newalgo(\arun,\astate)$ such that $\exec{\astate}=\anexec$.
  \end{theorem}
  \addtocounter{theorem}{-1}
}

\begin{proof}[Proof of Theorem~\ref{thm:optimality:power}]
  It follows from Corollary~\ref{cor:soundness2:power} that at least
  one call $\newalgo(\arun,\astate)$ such that
  $\exec{\astate}=\anexec$ is made. It remains to show that at most
  one such call is made.

  Assume for a proof of contradiction that two separate calls
  $\newalgo(\arun_a,\astate_a)$ and $\newalgo(\arun_b,\astate_b)$ are
  made such that $\exec{\astate_a}=\exec{\astate_b}=\anexec$.

  Let $\newalgo(\arun_c,\astate_c)$ be the latest call to $\newalgo$,
  which is an ancestor to both of the calls
  $\newalgo(\arun_a,\astate_a)$ and $\newalgo(\arun_b,\astate_b)$.
  Since at least two calls were made from
  $\newalgo(\arun_c,\astate_c)$ to $\newalgo$ or $\newalgorun$, we
  know that an event $\anevent = \anevent_c$ must have been chosen on
  line~\ref{ln:dpor2:select:e:disarm} in the call
  $\newalgo(\arun_c,\astate_c)$.

  The event $\anevent_c$ is a store or a load. Assume first that
  $\anevent_c$ is a store. We see that $\newalgo(\arun_c,\astate_c)$
  may make calls to $\newalgo$, but not to $\newalgorun$. Hence the
  calls $\newalgo(\arun_a,\astate_a)$ and
  $\newalgo(\arun_b,\astate_b)$ must be reached from two
  \emph{different} calls to $\newalgo$ on
  line~\ref{ln:dpor2:store:rec:call} in the call
  $\newalgo(\arun_c,\astate_c)$. However, we see that the different
  calls made to $\newalgo$ from $\newalgo(\arun_c,\astate_c)$ will all
  fix different parameters for $\anevent_c$. Therefore there will be
  two stores which are ordered differently in the coherence order of
  the different calls to $\newalgo$. As we proceed deeper in the
  recursive evaluation of $\newalgo$, new coherence edges may
  appear. But coherence edges can never disappear. Therefore it cannot
  be the case that $\exec{\astate_a}=\exec{\astate_b}$, which is the
  sought contradiction.

  Next we assume instead that $\anevent_c$ is a load. In this case the
  calls $\newalgo(\arun_a,\astate_a)$ and
  $\newalgo(\arun_b,\astate_b)$ will be reached via calls to either
  $\newalgo$ on line~\ref{ln:dpor2:load:rec:call} or $\newalgorun$ on
  line~\ref{ln:dpor2:call:explore:run}. If both are reached through
  calls to $\newalgo$, then the contradiction is reached in an way
  analogue to the case when $\anevent_c$ is a store above.

  Assume instead that the call $\newalgo(\arun_a,\astate_a)$ is
  reached via a recursive call to
  $\newalgo(\arun_c.\inst{\anevent_c}{\anevent_w},\astate_d)$ from the
  call $\newalgo(\arun_c,\astate_c)$, and that
  $\newalgo(\arun_b,\astate_b)$ is reached via a call to
  $\newalgorun(\arun_c,\astate_c,\arun_c')$ from the call
  $\newalgo(\arun_c,\astate_c)$. Notice from the way that subruns in
  \texttt{$\contmap$[$\anevent_c$]} are constructed that $\arun_c'$
  must have the form
  $\arun_c''.\inst{\anevent_w'}{\dporstar}.\inst{\anevent_c}{\anevent_w'}$
  for some subrun $\arun_c''$ and some store $\anevent_w'$ which is
  not committed in $\astate_c$. In the evaluation of
  $\newalgorun(\arun_c,\astate_c,\arun_c')$, the subrun $\arun_c'$
  will be traversed, and for any call to $\newalgo(\arun_e,\astate_e)$
  made during that evaluation it will hold that $\anevent_c$ loads
  from the store $\anevent_w'$. However, in the call
  $\newalgo(\arun_c.\inst{\anevent_c}{\anevent_w},\astate_d)$ it holds
  that $\anevent_c$ loads from $\anevent_w$. Since $\anevent_w$ is
  committed in $\astate_c$, it must be that
  $\anevent_w\neq\anevent_w'$. And so by the same reasoning as above,
  we derive the contradiction $\exec{\astate_a}\neq\exec{\astate_b}$.

  Next we assume that $\newalgo(\arun_a,\astate_a)$ and
  $\newalgo(\arun_b,\astate_b)$ are reached via calls from
  $\newalgo(\arun_c,\astate_c)$ to
  $\newalgorun(\arun_c,\astate_c,\arun_c^a.\inst{\anevent_w^a}{\dporstar}.\inst{\anevent_c}{\anevent_w^a})$
  and
  $\newalgorun(\arun_c,\astate_c,\arun_c^b.\inst{\anevent_w^b}{\dporstar}.\inst{\anevent_c}{\anevent_w^b})$
  respectively. If $\anevent_w^a\neq\anevent_w^b$, then the
  contradiction is derived as above. Assume therefore that
  $\anevent_w^a=\anevent_w^b$.

  If $\arun_c^a=\arun_c^b$, then the entire new branches are equal:
  $\arun_c^a.\inst{\anevent_w^a}{\dporstar}.\inst{\anevent_c}{\anevent_w^a}=\arun_c^b.\inst{\anevent_w^b}{\dporstar}.\inst{\anevent_c}{\anevent_w^b}$. By
  the mechanism on
  lines~\ref{ln:dpor2:load:cont:begin}-\ref{ln:dpor2:load:cont:end}
  using the set \texttt{explored}, we know that the calls
  $\newalgorun(\arun_c,\astate_c,\arun_c^a.\inst{\anevent_w^a}{\dporstar}.\inst{\anevent_c}{\anevent_w^a})$
  and
  $\newalgorun(\arun_c,\astate_c,\arun_c^b.\inst{\anevent_w^b}{\dporstar}.\inst{\anevent_c}{\anevent_w^b})$
  must then be the same call, since $\newalgorun$ is not called twice
  with the same new branch. The call to $\newalgorun$ will eventually
  call $\newalgo$ after traversing the new branch. From the assumption
  that the call $\newalgo(\arun_c,\astate_c)$ is the last one that is
  an ancestor to both calls $\newalgo(\arun_a,\astate_a)$ and
  $\newalgo(\arun_b,\astate_b)$, we know that the call to
  $\newalgorun$ must perform two different calls to $\newalgo$, which
  will eventually lead to the calls $\newalgo(\arun_a,\astate_a)$ and
  $\newalgo(\arun_b,\astate_b)$ respectively. However, when the
  function $\newalgorun$ traverses a branch, all events have fixed
  parameters, except the last store (i.e. $\anevent_w^a$ or
  $\anevent_w^b$). So if different calls to $\newalgo$ from
  $\newalgorun$ lead to the calls $\newalgo(\arun_a,\astate_a)$ and
  $\newalgo(\arun_b,\astate_b)$, then $\anevent_w^a$ (which is the
  same as $\anevent_w^b$) must have different coherence positions in
  $\astate_a$ and $\astate_b$. This leads to the usual contradiction.

  Hence we know that $\arun_c^a\neq\arun_c^b$. From the way new
  branches are constructed in $\newalgodetectrace$, we know that the
  branches
  $\arun_c^a.\inst{\anevent_w^a}{\dporstar}.\inst{\anevent_c}{\anevent_w^a}$
  and
  $\arun_c^b.\inst{\anevent_w^b}{\dporstar}.\inst{\anevent_c}{\anevent_w^b}$
  must have been constructed and added to
  \texttt{$\contmap$[$\anevent_c$]} during the exploration of some
  continuation of $\arun_c$. So there must exist runs
  $\arun_c.\arun_c^A.\inst{\anevent_w^a}{n^a}$ and
  $\arun_c.\arun_c^B.\inst{\anevent_w^b}{n^b}$ ending in the states
  $\astate_A$ and $\astate_B$ respectively. Furthermore $\arun_c^a$ is
  the restriction of $\arun_c^A$ to events that precede $\anevent_w^a$
  in $\cb{\astate_A}$, and $\arun_c^b$ is the restriction of
  $\arun_c^B$ to events that precede $\anevent_w^b$ in
  $\cb{\astate_B}$.
  From $\arun_c^a\neq\arun_c^b$ and
  Lemma~\ref{lemma:diff:cuts:diff:params}, it now follows that there
  is an event $\anevent_d$ which appears both in $\arun_c^a$ and
  $\arun_c^b$ but which has different parameters in the two
  subruns. This difference will also be reflected in $\astate_a$ and
  $\astate_b$. Hence $\exec{\astate_a}\neq\exec{\astate_b}$, which
  gives a contradiction and concludes the proof.\qed
\end{proof}

%% \input cb-illustration

%% \todo{Here are earlier sections, that have been shortened in the current paper}
%% \input cl-overview
%% \input op-semantics
%% \input sem-framework
%% \input bj-framework
%% \input det-dpor
%% \input det-correctness
%% \input source-correctness
}

\end{document}

%                       Revision History
%                       -------- -------
%  Date         Person  Ver.    Change
%  ----         ------  ----    ------

%  2013.06.29   TU      0.1--4  comments on permission/copyright notices